\documentclass[11pt]{article}
\usepackage[utf8]{inputenc}
\usepackage[english]{babel}
\usepackage{amsmath,amsfonts,amsthm,graphicx,float}
\usepackage{natbib}

\DeclareMathAlphabet{\mathpzc}{OT1}{pzc}{m}{it}

\newtheorem{rem}{Remark}[section]
\newcommand{\Var}  {\mbox{Var}}

\newcommand{\tr} {\mbox{tr}}

\usepackage{algorithm}

\usepackage[noend]{algpseudocode}

\def\bm{{\mathbf m}}

\def\b1{{\mathbf 1}}

\usepackage[font=small]{caption}

\newtheorem{theorem}{Theorem}
\newtheorem{lem}{Lemma}

\usepackage{titlesec}

\titleformat*{\section}{\normalfont\fontsize{14}{17}\bfseries}
\titleformat*{\subsection}{\normalfont\fontsize{12}{15}\bfseries}

\usepackage{bm}
\usepackage{enumerate}
\providecommand{\abs}[1]{\lvert#1\rvert}
\providecommand{\norm}[1]{\lVert#1\rVert}

\usepackage{soul}
\usepackage[usenames, dvipsnames]{color}

\date{}

\begin{document}

\title{\LARGE A goodness-of-fit test for regression models with spatially correlated errors}\normalsize
\author{Andrea Meil\'an-Vila \\
Carlos III University of Madrid\thanks{%
Department of Statistics, Carlos III University of Madrid, Av. de la Universidad 30, Legan\'es, 28911, Spain}
\and
Jean D. Opsomer \\
Westat\thanks{Westat, 1600 Research Boulevard
		Rockville, MD, 20850, Maryland, USA}
\and
Mario Francisco-Fern\'andez\\
Universidade da Coru\~{n}a\thanks{%
Research group MODES, CITIC, Department of Mathematics, Faculty of Computer Science, Universidade da Coru\~na, Campus de Elvi\~na s/n, 15071,
A Coru\~na, Spain}
\and
Rosa M. Crujeiras \\
Universidade de Santiago de Compostela\thanks{Department of Statistics, Mathematical Analysis and Optimization, Faculty of Mathematics, Universidade de Santiago de Compostela, R\'ua Lope G\'omez de Marzoa s/n,
	15782, Santiago de Compostela, Spain}
\and %
}
\maketitle


\begin{abstract}
The problem of assessing a parametric regression model in the presence of spatial correlation is addressed in this work. For that purpose, a goodness-of-fit test based on a $L_2$-distance comparing a parametric and a nonparametric regression estimators is proposed. Asymptotic properties of the test statistic, both under the null hypothesis and under local alternatives, are derived. Additionally, a bootstrap procedure is designed to calibrate the test in practice. Finite sample performance of the test is analyzed through a si\-mu\-la\-tion study, and its applicability is illustrated using a real data example.
\end{abstract}	


\section{Introduction}
\label{intro}
The problem of testing a parametric regression model, confronting a parametric estimator of the regression function with a smooth alternative estimated by a nonparametric method, has been approached by several authors in the statistical literature \citep[see, for example][]{azzalini1989use, eubank1990testing}. For instance, \cite{weihrather1993testing} and \cite{eubank2005testing} described tests based on an overall distance between parametric and nonparametric regression fits,  giving some strategies on bandwidth selection. \cite{hardle1993comparing} proposed a testing procedure to check if a regression function belongs to a class of parametric models by measuring a $L_2$-distance between parametric and nonparametric estimates. Specifically, the Nadaraya-Watson estimator \citep{nadaraya1964estimating, watson1964smooth} was considered for the nonparametric approach. The same type of study was performed by \cite{alcala1999goodness}, but using a local polynomial regression estimator \citep{fan1996local}. Following similar ideas, a local test  for a univariate parametric
model checking was proposed by \cite{opsomer2010finding}, while \cite{li2005using} assessed the lack of fit
of a nonlinear regression model, comparing a local linear smoother and parametric fits. 

The previous testing procedures, all of them formulated with independent errors, have been also adapted for scenarios where data exhibit correlation in time. For example, \cite{park2015using} considered a model specification test based on a kernel for a nonparametric regression model with an equally-spaced fixed design and correlated errors. 
Also in the context of time series, goodness-of-fit tests for linear regression models with correlated errors have been studied
by \cite{gonzalez1995testing}, also considering an equispaced fixed design. \cite{biedermann2000testing} extended
the previous results under fixed alternatives, considering a regression model with explanatory variables $x_i$, $i=1,\ldots,n$, being fixed and given by $i/n= \int _0^{x_i} f(t)dt$, where $f$ is a positive density on the interval $[0,1]$. For further discussion
and examples of nonparametric specification tests for regression models, see the comprehensive review by
\cite{gonzalez2013updated}.

Although for time dependent errors, the problem of assessing a pa\-ra\-me\-tric regression model has been widely studied, this is not the case for spatial (or even spatio-temporal) correlated data. Observations from spatially varying processes are quite frequent in applied sciences such as ecology, environmental and soil sciences. In order to gain some insight in the process evolution across space, a regression model where the regression function captures the first-order structure, whereas the error term collects the second-order structure, can be formulated in the previous contexts. Usually, parametric models are considered for the regression function, e.g. polynomial models on latitude and longitude \citep[see][]{cressie1993statistics,ribeiro2007model}, and estimation is accomplished by  least squares methods, providing reliable inferences if the model is correctly specified. As an example, a classical dataset which is analyzed under this scope is the Wolfcamp aquifer data presented by \cite{harper1986geostatistical}, collecting $85$ measurements of levels of piezometric-head. In this example, several parametric trend models are considered after performing different analyses, concluding that a linear trend seems to be a reasonable model (see Figure~\ref{map}). However, to determine if this linear model (or in general, any parametric fit) is an appropriate representation of a dataset, it would be advisable to carry out a statistical test in order to assess the goodness-of-fit of the selected model. 
In this context, the statistical literature initially focused on the assessment of independence \citep{diblasi2001use} and on testing a parametric correlation model \citep{maglione2004exploring}, considering the variogram as the function describing the spatial dependence pattern. Also taking the variogram as the target function, \cite{bowman2013inference} proposed some testing methods for simplifying hypothesis (namely, stationarity and isotropy). Although these proposals investigate the dependence structure of the data (a nuisance when the primary goal is the regression or trend function), the ideas which inspired these methods are common to the goodness-of-fit tests for regression models. 


A new proposal for testing a parametric regression model (with univariate responses and possibly $d$-dimensional covariates), in the presence of spatial correlation, is presented in this work. Following similar ideas as those of \cite{hardle1993comparing}, the test statistic is based on a comparison between a smooth version of a parametric fit and a nonparametric estimator of the regression function, using a weighted $L_2$-distance. The null hypothesis that the regression function follows a parametric model is rejected if the distance exceeds a certain threshold. 
To perform the parametric estimation, an iterative procedure based on generalized least squares  is used \citep[see][]{ribeiro2007model}, although other fitting techniques such as maximum likelihood me\-thods could be employed. For the nonparametric alternative, the multivariate local linear regression estimator is used \citep{liu2001kernel, francisco2005smoothing, hallin2004local}, generalizing in some way the results of \cite{alcala1999goodness} for the univariate case with independent errors.


This paper is organized as follows. Section \ref{sec:1} introduces the regression model, as well as the nonparametric and parametric estimators of the regression function used in our approach. Assumptions and the asymptotic distribution of the proposed test statistic, jointly with a bootstrap procedure to calibrate the test are presented in Section \ref{sec:test}. A simulation study for assessing the final performance of the test is provided in Section \ref{Simulations}. Finally, Section \ref{examples} shows how to apply the testing procedure to the Wolfcamp aquifer dataset introduced above. 
Appendix B contains further simulation results.

\section{Statistical model}
\label{sec:1}
Denote by $\{(\mathbf{X}_i,Z_i)\}_{i=1}^n$ a random sample of $(d+1)$-valued random vectors, where $Z_i$ denotes a scalar response which depends on a $d$-dimensional covariate $\mathbf{X}$, with support $D\subset\mathbb R^d$, through the following regression model:
\begin{equation}\label{processdecomposed}
Z_i=m(\mathbf{X}_i)+\varepsilon_i, \quad i=1,\dots,n,
\end{equation} 
where $m$ is the regression function and $\varepsilon$ denotes a spatially correlated error process, which is assumed to be second order stationary, where  $$\mathbb{E}[\varepsilon_i]=0, \quad\mbox{Cov}(\varepsilon_i,\varepsilon_j)=\Sigma(i,j)=\sigma^2\rho_n(\mathbf{X}_i-\mathbf{X}_j),\quad i,j=1,\dots,n,$$
	with $\sigma^2$ being the point variance and $\rho_n$ a continuous stationary correlation function satisfying $\rho_n(0)=1$, $\rho_n(\mathbf{x})=\rho_n(-\mathbf{x})$, and $\abs{\rho_n(\mathbf{x})}\le1$, $\forall \mathbf{x}$. The subscript $n$ in $\rho_n$ allows the correlation function to shrink as $n\to\infty$ (this will be made more precise below). 
Under these assumptions, the semivariogram function $\gamma_n$ satisfies that $\gamma_n(\mathbf{u})=\sigma^2(1-\rho_n(\mathbf{u}))$, $\forall\mathbf{u}\in \mathbb{R}^d$.
For simplicity, the subscript $n$ will be sometimes omitted. It should be noted that the previous expression for the covariance of the errors is correct if the nugget effect,  denoted by  $c_0$, is equal to zero. If $c_0 \neq 0$, then $\mbox{Cov}(\varepsilon_i,\varepsilon_j)=c_1\rho_n(\mathbf{X}_i-\mathbf{X}_j),$ if $i\neq j$, where $c_1=\sigma^2-c_0$ is the partial sill. In what follows, only the case of $c_0=0$ is considered. However, the case of considering a nugget effect has also been analyzed through simulations.

The goal of this work is to propose and study a testing procedure to assess the goodness-of-fit of a parametric regression model, that is:
\begin{equation}\label{test}
H_0:m\in \mathcal{M}_{\bm{\beta}}=\{m_{\bm{\beta}},{\bm{\beta}}\in\mathcal{B}\}, \hspace{1.2cm}\text{vs.} \hspace{1.2cm} H_a:m\notin \mathcal{M}_{\bm{\beta}},
\end{equation}
where $\mathcal{B}\subset\mathbb{R}^p$ is a compact set, and $p$ denotes the dimension of the parameter space $\mathcal{B}$. For example, in the bidimensional case ($d=2$), considering that $\mathcal{M}_{\bm{\beta}}$ is the family of linear models, then $p=3$. In addition, $m_{\bm{\beta}}$ denotes a $d$-variate parametric function with parameter vector $\bm{\beta}$.  Note that $m_{\bm{\beta}}$ is not restricted to be polynomial, although that is a common choice in practice.


As pointed out in the Introduction, the goodness-of-fit test is based on a weighted $L_2$-distance which measures the discrepancy between a smooth version of a parametric estimator and a nonparametric estimator of the regression function. $H_0$ is rejected if the distance between both fits exceeds a critical value. The estimation methods (parametric and nonparametric) considered in this proposal will be described below. 
As it will be seen in Section 3, the pa\-ra\-me\-tric estimator which is used in the test must satisfy a $\sqrt{n}-$consistency property. As an example, an iterative least squares estimator will be also presented.

A note of caution should be made about regression estimation in this context: for spatially correlated data, when just a single realization of the process F$\{z_1,\dots,z_n\}$ is a\-vai\-la\-ble, additional stationarity assumptions on the process are required in order to enable statistical inference. In addition, it should be also noted that, from a single realization, it may be difficult to disentangle the regression and error components, especially if the dependence is strong.

\subsection{Nonparametric regression estimation}
\label{trend:est}
For the nonparametric estimation of $m $ in model  (\ref{processdecomposed}), the multivariate local linear estimator \citep{fan1996local} is employed. This non\-pa\-ra\-me\-tric a\-pproach presents some advantages over other kernel-type methods \citep{hallin2004local}. For example, it adapts to a broad class of design densities. Moreover, unlike other kernel-type smoothers, this estimator does not suffer from boundary effects. 
In the spatial framework, the local linear estimator for
$m $ at a location $\mathbf{x}$ can be explicitly written as:
\begin{equation}
\label{LL}
\hat{m}^{LL}_{\mathbf{H}}(\mathbf{x})=\mathbf{e}_1'(X'_xW_xX_x)^{-1}X_x'W_x\mathbf{Z},
\end{equation}
where $\mathbf{e}_1$ is a vector of length $(d+1)$ with value 1 in the first entry and all other entries 0, $X_x$ is a $n\times(d+1)$ matrix with $i$-th row equal to $(1, (\mathbf{X}_i-\mathbf{x})')$, $W_x=\mbox{diag}\{K_\mathbf{H}(\mathbf{X}_1-\mathbf{x}),\dots,K_\mathbf{H}(\mathbf{X}_n-\mathbf{x})\}$, with $K_\mathbf{H}(\mathbf{x})=\abs{\mathbf{H}}^{-1}K(\mathbf{H}^{-1}\mathbf{x})$, being $K$ a $d$-dimensional kernel function and $\mathbf{H}$  a $d\times d$ symmetric positive definite matrix, and  $\mathbf{Z}=(Z_1,\ldots,Z_n)'$.

For the case of uncorrelated data with a random design, \cite{ruppert1994multivariate} derived the asymptotic mean squared error (AMSE) formula for the multivariate local linear estimator, while \cite{liu2001kernel} generalized those results when the errors are correlated. 
The bandwidth matrix $\mathbf{H}$ controls the shape and the size
of the local neighborhood used to estimate $m(\mathbf{x})$ and its selection plays an important role in the estimation process. If $\mathbf{H}$ is ``small" an undersmoothed estimator is obtained with high variability and, on the other hand, if  $\mathbf{H}$ is ``large", the resulting estimator will be very smooth and possible with larger bias. Cross-validation procedures for bandwidth selection are the usual ones in this context, but this type of methods derived under independence should not be used directly when data exhibit dependence given that its expectation is severely affected by the correlation \citep{liu2001kernel}. In that case, the dependence of the observations should be taken into account in some way in the bandwidth selection method  to estimate ``optimal" smoothing parameters \citep{liu2001kernel, francisco2005smoothing}. 

\subsection{Parametric regression estimation}
\label{trend:est_par}
As pointed out previously, the goodness-of-fit test proposed in this paper also requires a parametric estimation of the regression function. As it will be remarked in the next section, the test statistic can be applied taking any parametric estimator, as long as it satisfies a consistency property. Specifically, if $m_{{{\bm{\beta}}}_0}$ denotes the ``true'' regression function under the null hypothesis, and $m_{\hat{{\bm{\beta}}}}$ the corresponding parametric estimator, it is needed that the difference  $m_{\hat{{\bm{\beta}}}}(\mathbf{x})-m_{{{\bm{\beta}}}_0}(\mathbf{x})=O_p(n^{-1/2})$ uniformly in $\mathbf{x}$.
A suitable parametric estimator satisfying this property is, for example, the one considered by \cite{rosa_csda}, and this is the parametric method employed for the practical application of the test. 

The parametric estimator studied by \cite{rosa_csda} is obtained using an iterative least squares algorithm. A feasible version of this method includes an approximation of the variance-covariance matrix of the errors. However, for estimating the covariance structure, an initial estimation of the regression is required. This feature leads to the design of iterative estimation procedures in this setting. Following these ideas, this parametric regression estimator is computed as follows: 
\begin{enumerate}
\item Get an initial estimator of $\bm{{\bm{\beta}}}$ by  least squares regression:
\begin{equation}\label{OLS}
\tilde{\bm{\beta}}=\mbox{arg}\min_{\bm{{\bm{\beta}}}}({\mathbf{Z}}-\mathbf{m}_{\bm{\beta}})'({\mathbf{Z}}-\mathbf{m}_{\bm{\beta}}), 
\end{equation}
where ${\mathbf{m}}_{\bm{\beta}}=({m}_{\bm{\beta}}(\mathbf{X}_1),\dots,{m}_{\bm{\beta}}(\mathbf{X}_n))'$ is the regression function evaluated at the explicative variables.
\item Using the residuals $\tilde{\varepsilon}_i=Z_i-{m}_{\tilde{\bm{\beta}}}(\mathbf{X}_i)$, $i=1,\dots,n$, and assuming that the variogram belongs to a valid parametric family $\{2\gamma_{\bm{\theta}},\;\bm{\theta}\in\bm{\Theta}\subset\mathbb{R}^q\}$ (usually $q=3$, with the vector $\bm{\theta}$ made up of the nugget effect, the partial sill, and the practical range), obtain a parameter estimate $\hat{\bm{\theta}}$ of $\bm{\theta}$.
Fo\-llo\-wing a classical approach, $\bm{\theta}$ is approximated by fitting the parametric model considered for the variogram to a set of empirical semivariogram estimates, computed using the residuals $\tilde{\varepsilon}_i$, applying the weighted  least squares method \citep{cressie1985fitting}.  
Under this parametric assumption, the variance-covariance matrix of the errors can be denoted by $\Sigma_{\bm{\theta}}$, with elements  $\Sigma_{\bm{\theta}}(i,j)$, $i,j=1\dots,n$. Then, replacing $\bm{\theta}$ by $\hat{\bm{\theta}}$ in these elements, a parametric estimation of $\Sigma_{\bm{\theta}}$ (denoted by $\Sigma_{\hat{\bm{\theta}}}$) is obtained.   
\item Using $\Sigma_{\hat{\bm{\theta}}}$, estimate the regression parameter $\bm{{\bm{\beta}}}$ applying the weighted  least squares method:
\begin{equation}\label{IGLS}
\hat{\bm{\beta}}=\mbox{arg}\min_{\bm{{\bm{\beta}}}}({\mathbf{Z}}-\mathbf{m}_{\bm{\beta}})'\Sigma_{\hat{\bm{\theta}}}^{-1}({\mathbf{Z}}-\mathbf{m}_{\bm{\beta}}).
\end{equation}
Finally, the parametric estimator of $m $ considered is given by $m_{\hat{\bm{\beta}}} $.
\end{enumerate}

\section{Test statistic}
\label{sec:test}

As pointed out in Section \ref{sec:1}, 
the aim of this paper is to propose a goodness-of-fit test to check if the regression function in model (\ref{processdecomposed}) can be assumed to belong to a certain parametric family, $\{m_{\bm{\beta}},{\bm{\beta}}\in\mathcal{B}\}$. To tackle this problem, a natural approach consists in comparing a parametric estimator of the re\-gre\-ssion function with a nonparametric one. The question arises if the differences between both fits can be explained by small stochastic fluctuations or if such differences suggest that the parametric assumption is not correct and it is more reasonable to use nonparametric methods to approximate the regression function. Using these ideas, one way to proceed is to measure the distance between both fits and to employ this distance as the test statistic for checking the parametric model. 

The approach followed in this work to solve this problem, as in \cite{alcala1999goodness}, considers a test statistic given by a weighted $L_2$-distance between the nonparametric and parametric fits to address the testing problem (\ref{test}): 
\begin{equation}
\label{statistic}
T_n=n\abs{\mathbf{H}}^{1/2}\int_{D}^{}(\hat{m}^{LL}_{\mathbf{H}}(\mathbf{x})-\hat{m}^{LL}_{\mathbf{H},\hat{\bm{\beta}}}(\mathbf{x}))^2w(\mathbf{x})d\mathbf{x},
\end{equation}
where $w$ is a weight function that helps in mitigating possible edge effects.
The use of a weight function is quite frequent in this type of tests, both for density and regression \citep{gonzalez2013updated}.
Moreover, $\hat{m}^{LL}_{\mathbf{H},\hat{\bm{\beta}}}$ is a smooth version of the parametric estimator ${m}_{\hat{{\bm{\beta}}}}$ which is defined by
\begin{equation}
\label{par_smooth}
\hat{m}^{LL}_{\mathbf{H},\hat{\bm{\beta}}}(\mathbf{x})=\mathbf{e}_1'(X'_xW_xX_x)^{-1}X_x'W_x\mathbf{\mathbf{m}}_{\hat{\bm{\beta}}},
\end{equation}
with ${\mathbf{m}}_{\hat{\bm{\beta}}}=({m}_{\hat{\bm{\beta}}}(\mathbf{X}_1),\dots,{m}_{\hat{\bm{\beta}}}(\mathbf{X}_n))'$.

In the particular situation that the parametric family $\mathcal{M}_{\bm{\beta}}$ in (\ref{test}) is the class of polynomials of degree less or equal than $k$, it could be more reasonable to use, as the nonparametric fit,  the multivariate local polynomial estimator of degree $l$, with $l \geq k$, and considering the $L_2$-distance between this estimator and ${m}_{\hat{{\bm{\beta}}}}$. In that case, it would not be necessary to employ a smooth version of ${m}_{\hat{{\bm{\beta}}}}$, because both are consistent unbiased estimators of the regression function, under the null hypothesis. However, for a general parametric fa\-mi\-ly $\mathcal{M}_{\bm{\beta}}$, this is not true, and using the simpler local linear estimator, given that  $\mathbb{E}[\hat{m}^{LL}_{\mathbf{H}}(\mathbf{x})]=\mathbf{e}_1'(X'_xW_xX_x)^{-1}X_x'W_xm(\mathbf{x})$, it is convenient to smooth the parametric estimator so that the parametric term in (\ref{statistic}) has the same expected value as the nonparametric term, under $H_0$. This fact also justifies the use of the same bandwidth matrix $\mathbf{H}$ in $\hat{m}^{LL}_{\mathbf{H}}$ and in $\hat{m}^{LL}_{\mathbf{H},\hat{\bm{\beta}}}$ \citep[see][p. 1928]{hardle1993comparing}. It is clear that the statistic $T_n$ will be large when the parametric and nonparametric fits, evaluated on the domain $D$, differ substantially.

For example, considering the Wolfcamp aquifer dataset described in the Introduction, Figure \ref{fits} shows the smooth version of the parametric (left) and the nonparametric (right) regression estimators  for the level of piezometric-head in the area of study. In this case, a linear model is considered for the parametric fit, while the local linear estimator (\ref{LL}) is employed to perform the nonparametric fit (specific details on the estimation
procedures and the fits will be discussed later). Given that both surfaces are very similar, the value of the test statistic $T_n$ will be \emph{small}, and there may be no evidences against the assumption of a linear trend. This feature will indeed be confirmed with the statistical illustration of (\ref{statistic}) presented in Section \ref{examples}.

The types of model deviations
that can be captured by this test are of the form $m(\mathbf{x})=m_{{\bm{\beta}}_0}(\mathbf{x})+ c_n g(\mathbf{x})$, where $c_n$ is a sequence, such that $c_n\to0$ and $g$ is a deterministic function collecting the deviation direction from the null model. In the following section, the asymptotic distribution of the test statistic (\ref{statistic}) is derived under the null hypothesis, and also under local alternatives converging to the null hypothesis at a certain rate controlled by $c_n$. Specifically, it is assumed that the function $g$ is bounded (uniformly in $\mathbf{x}$ and $n$) and $c_n=n^{-1/2}\abs{\mathbf{H}}^{-1/4}$. In particular, this contains the null hypothesis corresponding to $g(\mathbf{x})=0$. 

It is clear from expression (\ref{statistic}) that $T_n$ depends on the bandwidth matrix $\mathbf{H}$. While the bandwidth selection problem has been well studied in the regression estimation framework, it is still an open issue in goodness-of-fit studies relying on nonparametric methods. In this paper, the smoothing parameter selection problem is not investigated further. Instead, the performance of the test statistic $T_n$ is analyzed for a range of bandwidths in the numerical studies, allowing to check how sensitive the results are to variations in $\mathbf{H}$. 
Note that although technically it is  possible to consider different bandwidth matrices in $\hat{m}^{LL}_{\mathbf{H}}$ and $\hat{m}^{LL}_{\mathbf{H},\hat{\bm{\beta}}}$, the use of just one bandwidth matrix simplifies the application of the test in practice.

Note that the test statistic (\ref{statistic}) generalizes to the framework of spatial correlated data (with a $d$-dimensional covariate) the statistic proposed for independent data by \cite{hardle1993comparing}, using the Nadaraya-Watson estimator, and that of \cite{alcala1999goodness} using the local polynomial estimator and considering a single covariate.


\subsection{Main result}

Next, the asymptotic distribution of $T_n$ is derived. 
The following assumptions  on the stochastic nature of the observations, and on the  nonparametric estimator of the regression function are needed:

\begin{enumerate}[{(A}1)]
	\item The regression and the density functions $m$ and $f$, respectively, are twice continuously differentiable.
	\item The weight function $w$ is continuously differentiable. 
	\item The marginal density $f$ is continuous, bounded away from zero and $f(\mathbf{x})>0$ for all $\mathbf{x}\in D$.  
	
	
	\item For the correlation function $\rho_n$, there exist constants $\rho_{M}$ and $\rho_{c}$ such that
	$n\int \abs{\rho_n(\mathbf{x})}d\mathbf{x}<\rho_{M}$ and $\lim_{n\to\infty}n\int {\rho_n(\mathbf{x})}d\mathbf{x}=\rho_{c}.$
	For any sequence $\epsilon_n>0$ satisfying $n^{1/d}\epsilon_n\to\infty$,
	$$n\int_{\norm{\mathbf{x}}\ge\epsilon_n} \abs{\rho_n(\mathbf{x})}d\mathbf{x}\to 0\quad \text{as}\quad n\to\infty.$$
	
	\item For any $i$, $j$, $k$, $l$,
	$$
		\mbox{Cov}({\varepsilon}_i{\varepsilon}_j,{\varepsilon}_k{\varepsilon}_l )=\mbox{Cov}({\varepsilon}_i,{\varepsilon}_k)\mbox{Cov}({\varepsilon}_j,{\varepsilon}_l)+\mbox{Cov}({\varepsilon}_i,{\varepsilon}_l)\mbox{Cov}({\varepsilon}_j,{\varepsilon}_k).$$
	\item It is assumed that errors are a geometrically strong mixing sequence with mean zero and $\mathbb{E}\abs{\varepsilon(\mathbf{x})}^r<\infty$ for all $r> 4$.
	\item The kernel $K$ is a spherically symmetric density function, twice continuously differentiable and with compact support (for simplicity with a nonzero value only if $\norm{\mathbf{u}}\le 1$). Moreover, $\int \mathbf{u}\mathbf{u}' K(\mathbf{u})d\mathbf{u}=\mu_2(K)\mathbf{I}_d$, where $\mu_2(K)\neq 0$ is scalar and $\mathbf{I}_d$ is the $d\times d$ identity matrix.
	
	\item $K$ is Lipschitz continuous. That is, there exists $\mathfrak{L}>0$, such that 
	$$\abs{K(\mathbf{X}_1)-K(\mathbf{X}_2)}\le \mathfrak{L}\norm{\mathbf{X}_1-\mathbf{X}_2}, \quad \forall \mathbf{X}_1,\mathbf{X}_2\in D.$$

\item The bandwidth matrix $\mathbf{H}$ is symmetric and positive definite, with $\mathbf{H}\to 0$ and  $n\abs{\mathbf{H}}\lambda^2_{\min}(\mathbf{H})\to\infty$, when $n\to\infty$. 
The ratio $\lambda_{{\max}}(\mathbf{H})/\lambda_{{\min}}(\mathbf{H})$ is bounded above, where $\lambda_{{\max}}(\mathbf{H})$ and $\lambda_{{\min}}(\mathbf{H})$ are the maximum and minimum eigenvalues of $\mathbf{H}$, respectively.

	\end{enumerate}

As for the parametric estimator, just the assumption of being a $\sqrt{n}$-con\-sis\-tent estimator is required. This is guaranteed if the estimator $m_{\hat{{\bm{\beta}}}}$ des\-cri\-bed in Section \ref{trend:est_par} is employed in the statistic (\ref{statistic}). Anyway, as pointed out in the previous section, a different parametric estimator of the regression function
could be used in the test statistic (\ref{statistic}) as long as this property was fulfilled. 

Assumption (A4) implies that the correlation function depends on $n$, and  the integral $\int \abs{\rho_n(\mathbf{x})}d\mathbf{x}$ should vanish  as $n\to\infty$. The vanishing speed should not be slower than $O(n^{-1})$. This assumption also  implies that the integral of $\abs{\rho_n(\mathbf{x})}$  is essentially dominated by the values of $\rho_n(\mathbf{x})$ near to the origin $\bm{0}$. Hence, the correlation
is short-range and decreases as $n\to\infty$. Arguing somewhat loosely, this can be considered as a
case of increasing-domain spatial asymptotics \citep[see][]{cressie1993statistics}, since this setup can 
immediately be transformed to one in which the correlation function $\rho_{n}$ is fixed with respect to the sample
size, but the support  $D$
for $\mathbf{x}$ expands. The current setup with fixed domain $D$ and shrinking $\rho_n$ is more natural to consider when the primary purpose of the estimation is a fixed regression function $m$ defined over a spatial domain, not the correlation function itself.

Two examples of commonly used correlation functions that satisfy the conditions of assumption (A4) are the exponential model
$$\rho_n(\mathbf{x})=\mbox{exp}(-a n\norm{\mathbf{x}}),$$
and the rational quadratic model
$$\rho_n(\mathbf{x})=\dfrac{1}{1+a(n\norm{\mathbf{x}})^2},$$
with $a>0$ in both cases  \citep[see][]{cressie1993statistics}.
In general, if $\rho_n(\mathbf{x})=\rho(n^{1/d}\mathbf{x})$ and $\rho(\mathbf{x})$ is a fixed valid correlation function, which is continuous everywhere except at a finite number of points and absolutely integrable in $\mathbb{R}^d$, then it is easy to check that $\rho_n(\mathbf{x})$ satisfies assumption (A4).

Assumption (A5)  is satisfied, for example, when the errors follow a Gaussian distribution. As for (A6), if $\mathcal{M}^b_a$ is the $\sigma-$field generated by $\{\xi(t):a\le t\le b\}$, then $\{\xi(t):  t\in \mathbb{R}\}$ is geometrically strong mixing if the mixing coefficients  verify
\begin{equation}
\alpha(\tau)=\mbox{sup}\{\abs{\mathbb{P}(A\cap B)-\mathbb{P}(A)\mathbb{P}(B)}: A\in \mathcal{M}_{-\infty}^0 \quad \mbox{and} \quad B\in \mathcal{M}^{\infty}_\tau\}=O(\zeta^\tau),
\label{mixing}
\end{equation}
for some $0<\zeta<1$, when $\tau\to\infty$. This assumption is needed to apply the central limit theorem for reduced U-statistics under dependence given by \cite{kim2013central}. 
Note that if a random variable is a real Gaussian process, the strong mixing coefficient and the correlation function are equivalent \citep[][p. 181]{rozanov67}. Therefore, hypotheses (A4)-(A6) could be satisfied by Gaussian error processes with exponential or rational quadratic (among others) correlation functions, having a decay rate larger than or equal to that indicated in (\ref{mixing}).

In assumption (A9), $\mathbf{H}\to 0$ means that every entry of $\mathbf{H}$ goes to $0$. Since $\mathbf{H}$ is symmetric and positive definite, $\mathbf{H}\to 0$ is equivalent to $\lambda_{{\max}}(\mathbf{H})\to 0$. $\abs{\mathbf{H}}$ is a quantity of order $O(\lambda_{{\max}}^d(\mathbf{H}))$ because $\abs{\mathbf{H}}$ is equal to the product of all eigenvalues of $\mathbf{H}$.

The following theorem shows the asymptotic distribution of the test statistic (\ref{statistic}). The detailed proof is provided in Appendix A.
\begin{theorem}
\label{teo1}
	Under Assumptions \textnormal{(A1)-(A9)}, and if $0<V<\infty$, it can be proved that
	$$V^{-1/2}(T_n-b_{0\mathbf{H}}-b_{1\mathbf{H}})\to_{\mathcal{L}} N(0,1) \text{ as } n\to\infty,$$
	where $\to_{\mathcal{L}}$ denotes convergence in distribution, with
	\begin{eqnarray*}
		b_{0\mathbf{H}}&=& \abs{\mathbf{H}}^{-1/2}\sigma^2K^{(2)}(\bm{0})\bigg[\int \dfrac{w(\mathbf{x})}{f(\mathbf{x})}d\mathbf{x}+\rho_{c}\int {w(\mathbf{x})}d\mathbf{x}\bigg],\\
		b_{1\mathbf{H}}&=& \int (K_{\mathbf{H}}\ast g(\mathbf{x}))^2w(\mathbf{x})d\mathbf{x},\\
	\end{eqnarray*}
	and
	\begin{eqnarray*}
		V&=&2\sigma^4 K^{(4)}(\bm{0})\bigg[\int \dfrac{w^2(\mathbf{x})}{f^2(\mathbf{x})}d\mathbf{x}+2\rho_{c}\int \dfrac{w^2(\mathbf{x})}{f(\mathbf{x})}d\mathbf{x}+4\rho^2_{c}\int {w^2(\mathbf{x})}d\mathbf{x}\bigg], 	\end{eqnarray*}
	where $K^{(j)}$ denotes the $j$-times convolution product of $K$ with itself.
\end{theorem}

This result generalizes Theorem 2.1 of \cite{alcala1999goodness} in the univariate case and with independent errors   (corresponding to $\rho_c=0$), considering the local polynomial regression estimator.

\begin{rem}
The asymptotic distribution of the test statistic (\ref{statistic}) can be also obtained under a geostatistical spatial trend model. In this scenario, model ({\ref{processdecomposed}}) can be viewed as an additive decomposition of the spatial process: the regression function $m$ corresponds to the first-order moment of the process and captures the large-scale variability, whereas the error term collects the second-order structure, reflecting the small-scale variation. The covariates in this setting are given by the spatial locations (latitude and longitude), which are usually fixed in a geostatistical setting. In this case, considering assumptions (A1)-(A9), except the ones relative to $f$ (given that we are under a fixed design scheme), and following similar steps to those employed in the proof of Theorem \ref{teo1}, but using Riemann approximations of sums by integrals, the asymptotic distribution of $T_n$ is given by: 
$$V^{-1/2}(T_n-b_{0\mathbf{H}}-b_{1\mathbf{H}})\to_{\mathcal{L}} N(0,1) \text{ as } n\to\infty,$$
	with
	\begin{eqnarray*}
		b_{0\mathbf{H}}&=& \abs{\mathbf{H}}^{-1/2}\sigma^2K^{(2)}(\bm{0})\bigg[\int w(\mathbf{x})d\mathbf{x}+\rho_{c}\int {w(\mathbf{x})}d\mathbf{x}\bigg],\\
		b_{1\mathbf{H}}&=& \int (K_{\mathbf{H}}\ast g(\mathbf{x}))^2w(\mathbf{x})d\mathbf{x},\\
	\end{eqnarray*}
	and
	\begin{eqnarray*}
		V&=&2\sigma^4 K^{(4)}(\bm{0})\bigg[\int w^2(\mathbf{x})d\mathbf{x}+2\rho_{c}\int w^2(\mathbf{x})d\mathbf{x}+4\rho^2_{c}\int {w^2(\mathbf{x})}d\mathbf{x}\bigg]. 	\end{eqnarray*}
\end{rem}

%

\subsection{Calibration in practice}
\label{sec:boot}
Once a suitable test statistic is available, a crucial task is the calibration of
the critical value for a given level $\alpha$, namely $t_\alpha$. Usually, the determination of the critical
value $t_\alpha$, such that $\mathbb{P}_{H_0}(T_n\ge t_\alpha)=\alpha$ (denoting by $\mathbb{P}_{H_0}$ the pro\-ba\-bi\-li\-ty under $H_0$), can be done by means of the asymptotic
distribution of $T_n$. However, as noted in other nonparametric testing contexts, the asymptotic distribution obtained in Theorem \ref{teo1} is often not sufficiently precise for constructing a practical test in small-to-medium sample size situation. Moreover, to use the asymptotic expression of $T_n$ in practice, it is necessary to
estimate some nuisance functions. 
The poor performance of the normal approximation for moderate sample sizes was observed in some simulation studies. A simple example, taking $f$ and $\sigma^2$ as known, is included in Appendix B.1.

Under these circumstances, calibration can be done
by means of resampling procedures, such as bootstrap \citep[see, for example,][]{francisco2006nonparametric}. The bootstrap procedure considered (detailed below) extends to the case of spatially correlated data the parametric bootstrap discussed in \cite{vilar1996bootstrap}.
The specific steps are the
following:

\begin{enumerate}
	\item  Obtain, using (\ref{IGLS}), the parametric regression estimator  $\hat{\bm{{\bm{\beta}}}}$.
	
	\item Compute the estimated variance-covariance matrix  of the errors, $\hat{\Sigma}$, using the residuals $\bm{\hat{\varepsilon}}=(\hat{\varepsilon}_1,\dots,\hat{\varepsilon}_n)'$,   where  $\hat{\varepsilon}_i=Z_i-m_{\hat{\bm{\beta}}}(\mathbf{X}_i)$, $i=1,\dots,n$.
	
	\item Find the matrix $L$, such that  $\hat{\Sigma}=LL'$, using Cholesky decomposition.
	
	\item Compute the ``independent" variables, $\mathbf{e}=(e_1,\dots,e_n)'$, given by $\mathbf{e}=L^{-1}\bm{\hat{\varepsilon}}$.
	
	\item The previous independent variables are centered and an independent bootstrap sample of size $n$, denoted by $\mathbf{e}^*=(e^*_1,\dots,e^*_n)$, is obtained.
	
	\item Finally, the bootstrap errors  $\bm{\varepsilon}^*=(\varepsilon^*_1,\dots,\varepsilon^*_n)$ are $\bm{\varepsilon}^*=L\mathbf{e}^*$, and the bootstrap samples are  $Z^*(\mathbf{X}_i)=m_{\hat{\bm{\beta}}}(\mathbf{X}_i)+{\varepsilon}^*_i$.
	
\end{enumerate}

Using the bootstrap sample $\{Z^*_i, i=1,\dots,n\}$, the bootstrap test statistic $T_n^*$ is computed as in (\ref{statistic}), by the weighted $L_2$-distance between the bootstrap versions of the  smooth  parametric fit (\ref{par_smooth}) and the nonparametric estimator (\ref{LL}). Once the bootstrap statistic is obtained, the distribution of $T_n^*$  can be approximated by Monte Carlo, and the $(1-\alpha)$ quantile $t_\alpha^*$ easily computed. Finally, the null hypothesis is rejected if $T_n>t_\alpha^*$.

\section{Simulations}
\label{Simulations}

The finite sample performance of the proposed test, proceeding with a bootstrap calibration, is illustrated in this section with a simulation study. For this purpose, a linear parametric regression surface is chosen,
$
m_{\bm{\beta}}(X_1,X_2)=\beta_0+\beta_1 X_1+\beta_2 X_2,
$ being $\mathbf{X}=(X_1,X_2)$, 
and for different values of $c$ the mean function
\begin{equation}
m(X_1,X_2)=2+X_1+X_2+cX_1^3
\label{trend_sim}
\end{equation}
is considered. Therefore, the parameter $c$ controls whether the null ($c=0$) or the alternative ($c\neq 0$) hypotheses are assumed. Values $c=0$, $3$, and $5$  are considered in the study. 

For each value of $c$, 500 samples of sizes  $n=225$ and $400$ are generated on a bidimensional regular grid in the unit square, following model (\ref{processdecomposed}), with regression function (\ref{trend_sim}) and 
random errors $\varepsilon_i$ normally distributed with  zero mean and isotropic exponential covariance function:  
\begin{equation}
\mbox{Cov}({\varepsilon}_i,{\varepsilon}_j)=   
\sigma^2\{\exp(-\norm{\mathbf{X}_i-\mathbf{X}_j}/a_e)\},
\label{expo}
\end{equation}
 where  $\sigma^2$ is the variance and $a_e$ is the practical
range. Different degrees of spatial dependence were
studied, considering values of  $\sigma=0.4$, $0.6$, and $0.8$, and   $a_e=0.1$ (weak correlation), $a_e=0.2$ (medium correlation) and $a_e=0.4$ (strong correlation). Note that no nugget effect is considered in this scenario.

To analyze the behavior of the test statistic given in (\ref{statistic}) in the different scenarios, the bootstrap procedure described in Section \ref{sec:boot} was applied, using  $B=500$ replications. The weight function used was taken constant with value 1. The nonparametric fit used for constructing (\ref{statistic}) was obtained using the multivariate local linear estimator, described in Section \ref{trend:est}, with a multiplicative triweight kernel. The parametric one was computed using the iterative least squares procedure described in Section \ref{trend:est_par}, for a linear model.  The bandwidth selection
problem was addressed by using the same classical procedure as the one used in \cite{hardle1993comparing}, \cite{alcala1999goodness}, or \cite{opsomer2010finding}, among others. The
test was run in a grid of several bandwidths to check how it is affected by the bandwidth choice. In order to simplify the calculations, the bandwidth matrix was restricted to a class of  diagonal matrices with both equal elements (scalar matrices). To give a reasonable grid, the optimal bandwidth obtained by minimizing the mean average squared error (MASE) of the multivariate local linear estimator \citep[see][p. 288]{francisco2005smoothing} was calculated for each scenario.  These bandwidths were in the interval $[0.6, 1]$, therefore, the bandwidth was taken as a diagonal matrix $\mathbf{H}=\text{diag}(h,h)$, and different  values of $h$ were chosen, $h=0.6, 0.7, 0.8, 0.9, 1.$

Rejection proportions of the null hypothesis, for a significance level $\alpha=0.05$, are displayed in Table~\ref{Simu}, where it can be observed that the test has a reasonable behavior. If $c=0$ (null hypothesis), the rejection proportions are similar to the theoretical level, although these proportions are affected by the value of  $h$.  In fact, in most of the cases, the rejection proportions are smaller when the bandwidth value is larger. As expected, considering a larger sample size, the bandwidth value should be smaller. For alternative assumptions ($c=3$ and $c=5$), a decreasing power of the test when the values of $h$ increase is observed.  For all the scenarios, the power of the test becomes larger as the value of $c$ increases.  As expected, large values of the variance $\sigma^2$ lead to a decrease in power. Regarding the effect of the range $a_e$, when this parameter is larger, the power of the test increases, which justifies the correct performance of the bootstrap procedure for dependent data considered. It can be also noticed that, for large values of $a_e$, the bandwidth values providing an effective calibration of the test are also large. 

Additional simulation studies with other regression functions, selecting bandwidth matrices with different values in the main diagonal, including a nugget effect and considering random designs were also performed, obtaining similar results to those shown in Table \ref{Simu}. These experiments are reported in Appendix B.

\begin{table}
\scriptsize
	\centering
	\begin{tabular}{cccccccccc}
		&&	&&&&	 &$h$&&  \\
		\cline{6-10}
		$\sigma$	&$a_e$&$c$&$n$&&0.6&$0.7$&0.8&0.9&1    \\
		\hline
		$0.4$ &0.1&0&225&&0.092& 0.068& 0.050& 0.038& 0.024 \\
		&&&400&&  0.050& 0.036& 0.024& 0.022& 0.020 \\  
		$0.4$ &0.1&3&225&& 0.522& 0.480& 0.458& 0.446& 0.458  \\
		&&&400& &0.438& 0.396& 0.360& 0.360 &0.368   \\			
		$0.4$ &0.1&5&225&&0.988&0.984&0.978&0.980&0.984  \\
		&&&400&&  1.000&1.000&1.000&1.000&1.000  \\
		\hline
		$0.4$ &0.2&0&225&& 0.082&0.062&0.048&0.032&0.022   \\
		&&&400&&0.078& 0.050& 0.032& 0.028& 0.014   \\  
		$0.4$ &0.2&3&225&&0.902&0.876&0.854&0.840&0.834 \\
		&&&400& &0.896& 0.870& 0.832& 0.818& 0.806 \\			
		$0.4$ &0.2&5&225&&1.000&1.000&1.000&1.000&1.000  \\
		&&&400&&1.000&1.000&1.000&1.000&1.000 \\
		\hline
		$0.4$ &0.4&0&225& &0.162& 0.126& 0.084& 0.074& 0.068 \\
		&&&400&&  0.164& 0.126& 0.098& 0.076& 0.058  \\  
		$0.4$ &0.4&3&225&& 0.978& 0.976& 0.974& 0.970& 0.970 \\
		&&&400&& 0.990& 0.990 &0.988& 0.986& 0.986   \\			
		$0.4$ &0.4&5&225&&1.000&1.000&1.000&1.000&1.000  \\
		&&&400&& 1.000&1.000&1.000&1.000&1.000  \\
		\hline	
		$0.6$ &0.1&0&225&& 0.090& 0.068& 0.054& 0.038& 0.026  \\
		&&&400&& 0.050& 0.036& 0.022& 0.022& 0.020 \\  
		$0.6$ &0.1&3&225& &0.096& 0.084& 0.062& 0.056& 0.066 \\
		&&&400& &0.082& 0.058& 0.046& 0.034& 0.036 \\			
		$0.6$ &0.1&5&225&& 0.684& 0.652& 0.624& 0.602& 0.608  \\
		&&&400&&   0.630& 0.576& 0.538& 0.532& 0.536\\
		\hline
		$0.6$ &0.2&0&225&& 0.082&0.060&0.046&0.034&0.024   \\
		&&&400& &0.074&0.050&0.032&0.028&0.014   \\  
		$0.6$ &0.2&3&225&&0.492&0.430&0.370&0.332&0.322 \\
		&&&400& &0.466& 0.408& 0.362& 0.334& 0.330  \\			
		$0.6$ &0.2&5&225&&0.964&0.958&0.942&0.930&0.920  \\
		&&&400&& 0.962& 0.948& 0.930& 0.916 &0.912 \\
		\hline
		$0.6$ &0.4&0&225&&0.158 &0.126& 0.084& 0.074& 0.068 \\
		&&&400&&  0.164& 0.126& 0.100& 0.076& 0.058 \\  
		$0.6$ &0.4&3&225& & 0.766& 0.742& 0.716& 0.694& 0.684 \\
		&&&400&&  0.818& 0.784& 0.744& 0.714& 0.704  \\			
		$0.6$ &0.4&5&225& &0.998& 0.998& 0.998& 0.998& 0.998  \\
		&&&400&&  0.996& 0.996& 0.994& 0.994& 0.994 \\
		\hline
		$0.8$ & 0.1&0&225&& 0.088& 0.066& 0.052& 0.038& 0.026  \\ 
		&&&400& &0.050& 0.036& 0.022& 0.022& 0.020 \\  
		$0.8$ & 0.1&3&225&&  0.080& 0.052& 0.036& 0.030& 0.026  \\
		&&&400& &0.046& 0.018& 0.008& 0.006& 0.006 \\			
		$0.8$ & 0.1&5&225&& 0.282& 0.240& 0.204& 0.196& 0.204  \\
		&&&400&& 0.190 &0.158& 0.128& 0.120& 0.126\\
		\hline
		$0.8$ & 0.2&0&225&& 0.082&0.060&0.046&0.032&0.024  \\ 
		&&&400&& 0.076& 0.050& 0.032& 0.028& 0.014 \\  
		$0.8$ & 0.2&3&225&& 0.282&0.212&0.174&0.142&0.146  \\
		&&&400&&  0.256& 0.202& 0.164& 0.144& 0.144  \\			
		$0.8$ & 0.2&5&225&& 0.716&0.654&0.614&0.588&0.574  \\
		&&&400& &0.704& 0.654& 0.628 &0.600& 0.572 \\
		\hline
		$0.8$ & 0.4&0&225&&  0.158& 0.124& 0.084& 0.074& 0.068  \\ 
		&&&400& &0.164& 0.126& 0.100& 0.074& 0.058 \\  
		$0.8$ & 0.4&3&225&& 0.556& 0.496& 0.458& 0.434& 0.426 \\
		&&&400& &0.580& 0.532& 0.484& 0.450& 0.430  \\			
		$0.8$ & 0.4&5&225&& 0.928 &0.920 &0.906 &0.888 &0.874 \\
		&&&400& &0.952& 0.940& 0.930& 0.920& 0.904 \\
		\hline
	\end{tabular}
			\caption{Rejection proportions of the null hypothesis for $\alpha=0.05$.}
	\label{Simu}
\end{table}

\section{Illustration with real data}\label{examples}

In order to illustrate the performance in practice of the test statistic $T_n$, given in (\ref{statistic}), the Wolfcamp aquifer dataset briefly mentioned in the Introduction is considered. These data were reported and geostatistically analyzed in \cite{harper1986geostatistical} and \cite{cressie1993statistics}, and are  available in the \texttt{R} package \texttt{npsp} \citep{npsp}.

\begin{figure}
	\centering
\includegraphics[width=6.5cm]{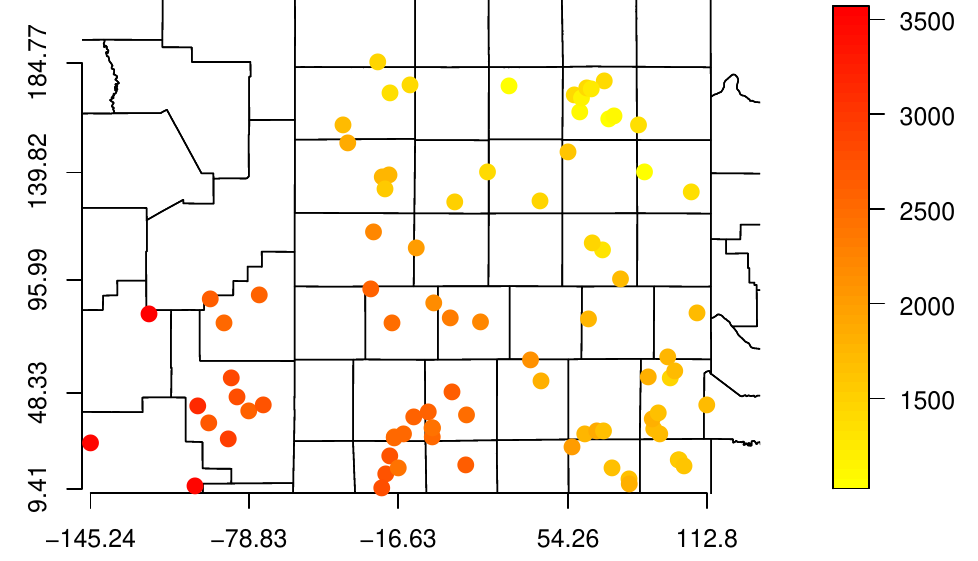}\hspace{0.4cm}
\includegraphics[width=4.8cm]{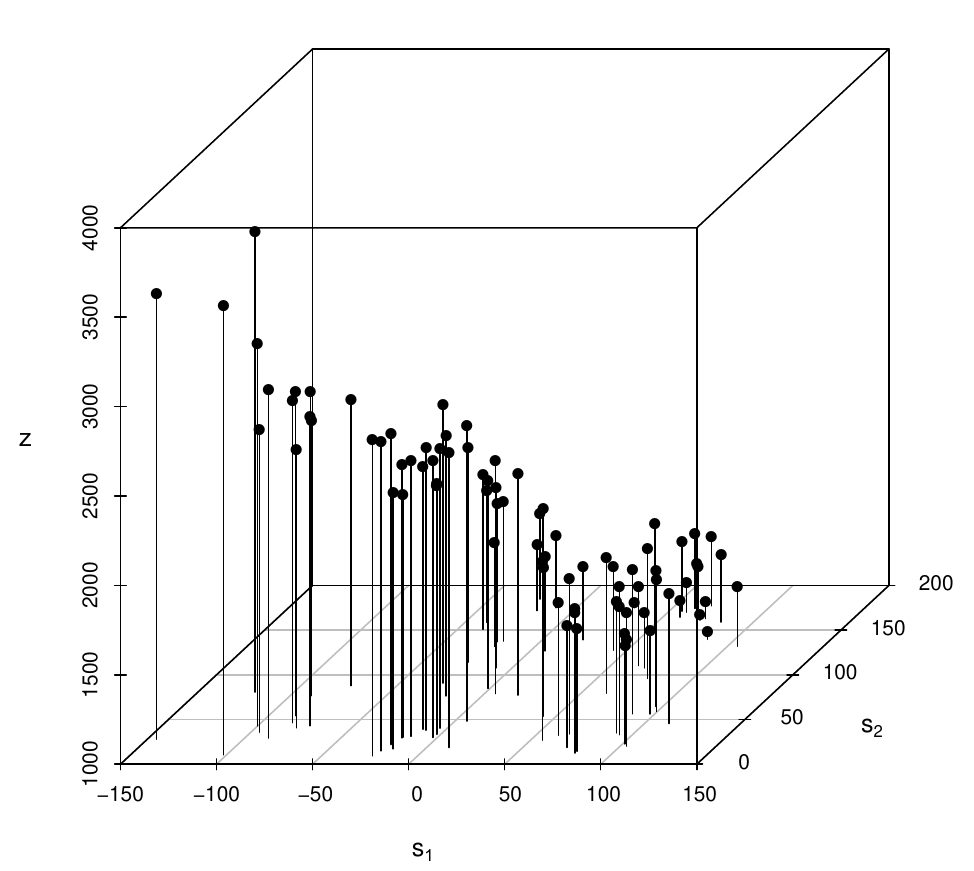}
\vspace{0.4cm}
	\caption{\footnotesize Locations with the levels of piezometric-head for the Wolfcamp Aquifer (left) and its own $3$-dimensional representation (right).}
	\label{map}
\end{figure}

The Deaf Smith County (Texas, bordering New Mexico) was selected as an alternate site for a possible nuclear waste disposal repository in the 1980s. This site was later dropped on grounds of contamination of the aquifer, the source of much of the water supply for west Texas. In a study conducted by the U.S. Department of Energy, piezometric-head levels  were obtained irregularly at 85 locations, shown in the left panel of Figure  \ref{map},  by drilling a narrow pipe through the aquifer \citep[see][]{harper1986geostatistical}.  With higher values generally in the lower left (southwest) and lower values in the upper
right (northwest), the groundwater gradient would cause water to flow in a northeasterly direction
from the repository in Deaf Smith County toward Amarillo in lower Potter county.

Figure \ref{map} (right panel) displays  the $3$-dimensional scatterplot of the piezometric heads levels (feet above sea level) against the coordinates (miles, from a reference point). This plot shows a clear downwards
trend from south-west to north-east. \cite{cressie1993statistics} used the median polish
approach to model this trend, whereas \cite{harper1986geostatistical} considered a linear trend surface, that is, a linear regression model on latitude and longitude. 
In order to check if a linear model is plausible, the test $T_n$, using the bootstrap procedure described in Section \ref{sec:boot} with $B=1000$ replications, was applied considering a linear parametric model, $
m_{\bm{\beta}}(X_1,X_2)=\beta_0+\beta_1 X_1+\beta_2 X_2,
$
as the null hypothesis, being $X_1$ and $X_2$ the spatial coordinates of the points where the process is observed. It should be noted that the (nonparametric) detrended data were also tested for isotropy and stationarity, following the proposals by \cite{bowman2013inference}, obtaining $p$-values of 0.838 for isotropy and 0.031 for stationarity.

To apply the test (\ref{statistic}), the parametric fit was carried out using the iterative least squares estimator described in Section \ref{trend:est_par}, assuming a linear regression model. After analyzing the initial residuals obtained by  least squares re\-gre\-ssion,  a spherical correlation model  \citep[as it was suggested by][]{harper1986geostatistical}
was considered to estimate the variance-covariance matrix of the errors, needed to obtain a feasible estimate of $\bm{\beta}$. As for the nonparametric fit in (\ref{statistic}), the local linear estimator (\ref{LL}) with a multiplicative triweight kernel was con\-si\-de\-red. The bandwidth was taken as a diagonal matrix $\mathbf{H}=\text{diag}(h_1,h_2)$, being the values of $h_1$ and $h_2$ different. Note that the corrected generalized cross-validation bandwidth \citep{francisco2005smoothing, francisco2012nonparametric} is $\mathbf{H}=\mbox{diag}(403.19, 226.20)$.

\begin{figure}
	\centering
	\includegraphics[width=0.46\textwidth]{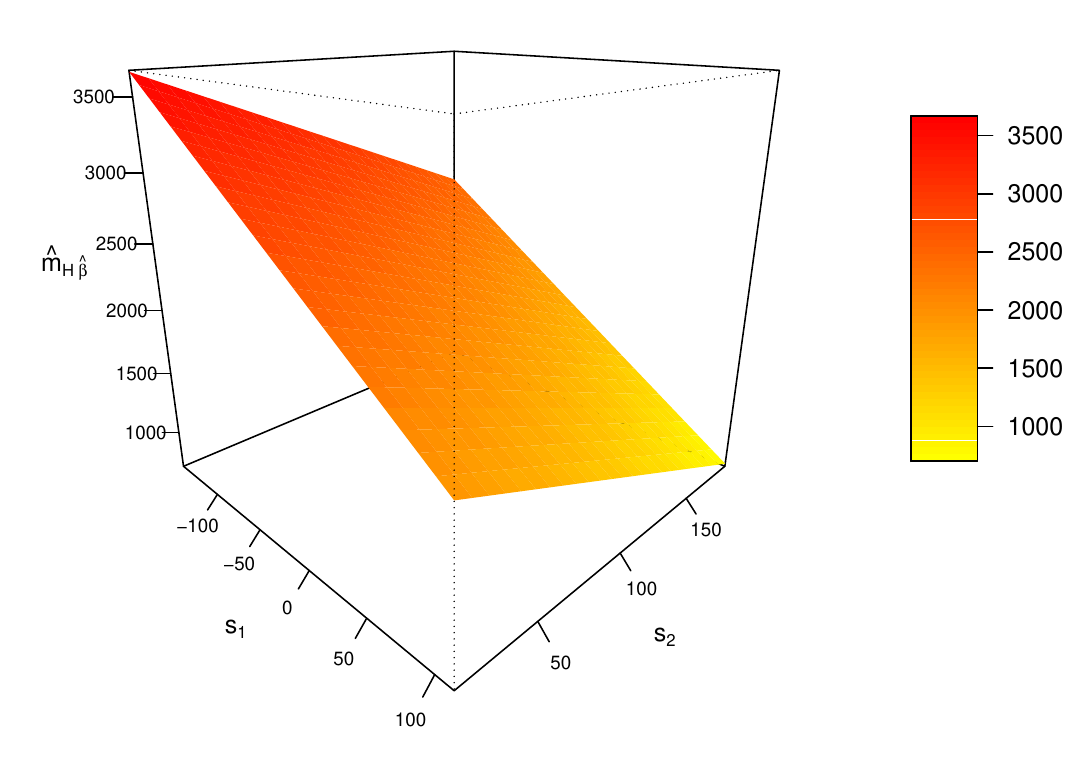}\hspace{0.5cm}
\includegraphics[width=0.46\textwidth]{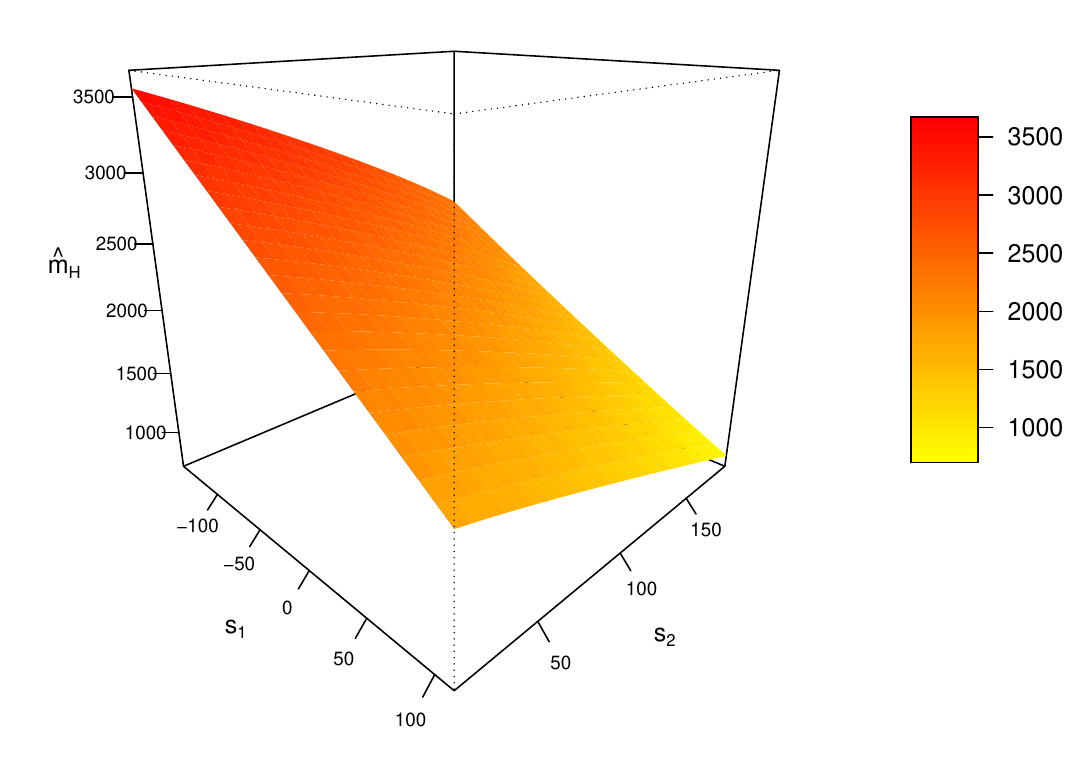}
	\caption{\footnotesize  Smooth version of the parametric fit (left) and nonparametric estimator of the regression (right)  using the corrected generalized cross-validation bandwidth for the Wolfcamp Aquifer. }
	\label{fits}
\end{figure}

\begin{figure}
	\centering
	\includegraphics[width=0.79\textwidth]{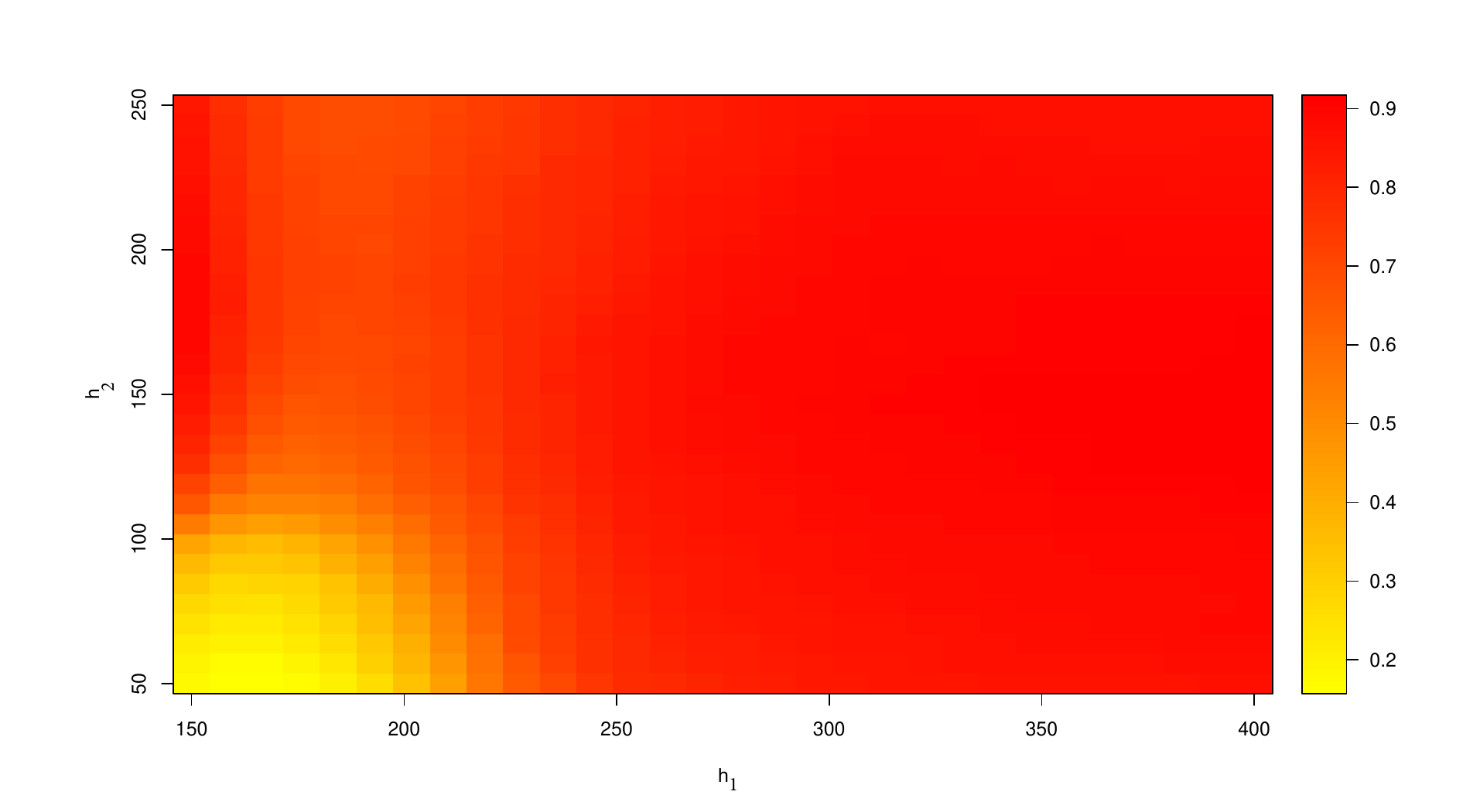}
	\caption{\footnotesize  Significance trace of the test  for $\alpha=0.05$ for the Wolfcamp aquifer dataset. }
	\label{trace}
\end{figure}

Figure \ref{fits} shows the smooth version of the parametric (left panel) and the nonparametric (right panel) regression estimators using the corrected ge\-ne\-ra\-li\-zed cross-validation bandwidth for the level of piezometric-head in the area of study. These regression surfaces are compared in the proposed test statistic. Figure \ref{trace} shows the $p$-values of the test using the so-called significance trace \citep{bowman1997applied}, that is, the proportions of empirical rejections for different
bandwidths.  Taking into account this plot,  there are no evidences against a linear spatial regression. Note that smaller bandwidths than those considered should not be taken to avoid boundary problems.

\section{Discussion}
\label{sec:discussion}
A goodness-of-fit test for a parametric regression model with correlated errors is presented in this work, based on a $L_2$-distance between a parametric and a nonparametric fits. A least squares procedure has been considered as a parametric approach, given its efficiency, but other methods such as maximum likelihood methods, could be also used, as long as a $\sqrt{n}-$consistency property is satisfied. In this case, it should be noted that both the regression function and the dependence structure of the errors are jointly estimated, but usually restricted to a (multivariate) Gaussian distribution of the process rea\-li\-za\-tion. In both cases (least squares and maximum likelihood), a parametric form for the correlation is considered. Considering spatial correlation as a nuisance (which should certainly be accounted for in inference, but it is not of primary interest), it is expected that the proposed goodness-of-fit test has a good performance even when the correlation is misspecified as long as it can be reasonably well approximated. Testing approaches as those proposed in \cite{maglione2004exploring} can be useful for this task. Regarding the nonparametric counterpart in the test statistic, other kernel estimators such as Priestley-Chao or Nadaraya-Watson estimators could be used.

Asymptotic results, under the null and under local alternatives, support
the proposal but due to the slow convergence to the limit distribution, a
bootstrap procedure is presented. Simulation results confirm that the bootstrap
algorithm works, facilitating the practical application of the test, with no other competitor (up to our knowledge). It may be argued
that this simulation study was limited to bidimensional linear regression models, but it could be extended to any parametric family. It  should be noted \citep{cressie1993statistics,ribeiro2007model}  that, in the geostatistical context, simple parametric models
are usually preferred in order to preserve interpretability. If one would be interested
in a more sophisticated regression structure, then a nonparametric fit could provide an appealing alternative. In any case, the bandwidth matrix needed to apply (\ref{statistic}) can be selected by
cross-validation but recall that this bandwidth is not necessarily a good one for
testing. With this purpose, it is advisable to explore a range of bandwidths,
taking a data-driven one as a reference.

Although a homoscedastic regression model has been considered in this paper, under suitable assumptions, the asymptotic results of the test statistic could be also derived for certain heteroscedastic regression models. In such a context, the bootstrap method to calibrate the test, described in Section \ref{sec:boot}, could be also modified, using an appropriate route to estimate the dependence of the model. To do this, the nonparametric approach described by \cite{fernandez2017nonparametric} could be used. Note that in that case, due to heteroscedasticity, the use of a wild bootstrap procedure in the resampling process could be more convenient. The design of this type of resampling approach in this context is, indeed, an interesting issue for a future research.

The procedures used in the simulation study as well as in the illustration with real data were implemented in
the statistical environment \texttt{R} \citep{Rsoft}, using functions included in the  \texttt{geoR}  and \texttt{npsp} packages  \citep{geor,npsp} to estimate the variogram and the spatial regression functions.

 \section*{Acknowledgements}\label{acknowledgements}
The authors acknowledge the support from the Xunta de Galicia and the
European Union (European Social Fund - ESF). This research has been partially supported by MINECO grants MTM2014-52876-R, MTM2016-76969-P and MTM2017-82724-R, and by the Xunta de Galicia (Grupos de Referencia Competitiva ED431C-2016-015 and ED431C- 2017-38, and Centro Singular de Investigaci\'on de Galicia ED431G/01), all of them through the ERDF. 

\section*{Appendix A. Proof of Theorem 1}
\label{app_proof}

In what follows, $\mathbf{1}_d$  and  $\mathbf{1}_{d\times d}$ are used to denote the $d \times 1$ vector and the $d\times d$ matrix with all entries equal to 1, respectively. Moreover, if $\mathbf{U}_n$ is a random matrix, then $O_p(\mathbf{U}_n)$ and $o_p(\mathbf{U}_n)$ are to be taken componentwise.

Before deriving the proof of Theorem 1, some auxiliary lemmas are required.

\begin{lem}\label{L1}
	Let $$W_{1n}(\mathbf{x})=\dfrac{1}{n}\sum_{i=1}^n K_{\mathbf{H}}\left({\mathbf{X}_i-\mathbf{x}}\right)g(\mathbf{X}_i),$$
	where $g$ is a bounded function uniformly  at $\mathbf{x}$.
	For any $\mathbf{x}\in D$, under assumptions (A1),(A3), (A7) and (A9), one gets that
	$$
	W_{1n}(\mathbf{x})=\int K\left(\mathbf{p}\right)g(\mathbf{x}+\mathbf{Hp})f(\mathbf{x}+\mathbf{Hp})d\mathbf{p}+o_p(1).$$
	\begin{proof}[Proof of Lemma 1] For any $\mathbf{x}\in D,$ under assumptions (A1),(A3), (A7) and (A9), it follows that
		\begin{eqnarray*}
			E(W_{1n}(\mathbf{x}))&=&\int K_{\mathbf{H}}\left({\mathbf{u}-\mathbf{x}}\right)g(\mathbf{u})f(\mathbf{u})d\mathbf{u}\\&=&\int K\left(\mathbf{p}\right)g(\mathbf{x}+\mathbf{Hp})f(\mathbf{x}+\mathbf{Hp})d\mathbf{p}
		\end{eqnarray*}
		and
		\begin{eqnarray*}
			\Var(W_{1n}(\mathbf{x}))&\le&\dfrac{1}{n}\int K^2_{\mathbf{H}}\left({\mathbf{u}-\mathbf{x}}\right)g^2(\mathbf{u})f(\mathbf{u})d\mathbf{u}\\&=&\dfrac{1}{n\abs{\mathbf{H}}}\int K^2\left({\mathbf{p}}\right)g^2(\mathbf{x}+\mathbf{Hp})f(\mathbf{x}+\mathbf{Hp})d\mathbf{p}\\&=&o_p(1)
		\end{eqnarray*}	
	\end{proof}
\end{lem}

	\begin{lem}\label{L2}
	Let $$W_{2n}(\mathbf{x},\mathbf{t})=\dfrac{1}{n}\sum_{i=1}^n K_{\mathbf{H}}\left({\mathbf{X}_i-\mathbf{x}}\right)K_{\mathbf{H}}\left({\mathbf{X}_i-\mathbf{t}}\right).$$ For any $\mathbf{x}\in D$, under assumptions (A1), (A3), (A7) and (A9), then 
	$$W_{2n}(\mathbf{x},\mathbf{t})=\abs{\mathbf{H}}^{-1}K^{(2)}(\mathbf{H}^{-1}({\mathbf{x}}-\mathbf{t}))f(\mathbf{t})\cdot\{1+o_p(1)\}.$$
	
	\begin{proof}[Proof of Lemma 2]
		 For any $\mathbf{x},\mathbf{t}\in D$
		\begin{eqnarray*}
			\mathbb{E}(\abs{\mathbf{H}}W_{2n}(\mathbf{x},\mathbf{t}))&=&\abs{\mathbf{H}}\int K_{\mathbf{H}}\left({\mathbf{u}-\mathbf{x}}\right)K_{\mathbf{H}}\left({\mathbf{u}-\mathbf{t}}\right)f(\mathbf{u})d\mathbf{u}\\&=&\int K\left({\mathbf{p}}\right)K\left(\mathbf{p}-\mathbf{H}^{-1}({\mathbf{x}}-\mathbf{t})\right)f(\mathbf{t}+\mathbf{Hp})d\mathbf{p}\\&=&K^{(2)}(\mathbf{H}^{-1}({\mathbf{x}}-\mathbf{t}))\{f(\mathbf{t})+o(1)\}.
		\end{eqnarray*}
		
	Moreover,
		\begin{eqnarray*}
			\mbox{Var}(\abs{\mathbf{H}}W_{2n}(\mathbf{x},\mathbf{t}))&\le&\dfrac{\abs{\mathbf{H}}^2}{n}\int K^2_{\mathbf{H}}\left({\mathbf{u}-\mathbf{x}}\right)K^2_{\mathbf{H}}\left({\mathbf{u}-\mathbf{t}}\right)f(\mathbf{u})d\mathbf{u}\\&=&\dfrac{1}{n\abs{\mathbf{H}}}\int K^2\left({\mathbf{p}}\right)K^2\left(\mathbf{p}-\mathbf{H}^{-1}({\mathbf{x}}-\mathbf{t})\right)f(\mathbf{t}+\mathbf{Hp})d\mathbf{p}\\&=&o_p(1).
		\end{eqnarray*}	
	\end{proof}
\end{lem}

	\begin{lem}\label{L3}
	Let $$W_{3n}(\mathbf{x},\mathbf{t})=\dfrac{1}{n^2}\sum_{i=1}^n\sum_{j=1}^n K^2_{\mathbf{H}}\left({\mathbf{X}_i-\mathbf{x}}\right)K^2_{\mathbf{H}}\left({\mathbf{X}_j-\mathbf{t}}\right)\rho^2_n(\mathbf{X}_i-\mathbf{X}_j).$$ For any $\mathbf{x},\mathbf{t}\in D$, under assumptions (A1), (A3) and (A9), then
	$$\mathbb{E}(W_{3n}(\mathbf{x},\mathbf{t}))=\abs{\mathbf{H}}^{-2}f(\mathbf{x})f(\mathbf{t})\int \int K^2(\mathbf{p})K^2(\mathbf{q})\rho^2_n(\mathbf{x}-\mathbf{t}+\mathbf{H}(\mathbf{p}-\mathbf{q}))d\mathbf{p}d\mathbf{q}\cdot\{1+o(1)\}.$$
	\begin{proof}[Proof of Lemma 3]
	For any $\mathbf{x},\mathbf{t}\in D$
		\begin{eqnarray*}
			\mathbb{E}(W_{3n}(\mathbf{x},\mathbf{t}))&=&\int \int K^2_{\mathbf{H}}\left({\mathbf{u}-\mathbf{x}}\right)K^2_{\mathbf{H}}\left({\mathbf{v}-\mathbf{t}}\right)\rho^2_n(\mathbf{u}-\mathbf{v})f(\mathbf{u})f(\mathbf{v})d\mathbf{u}d\mathbf{v}\\&=&\abs{\mathbf{H}}^{-2}\int \int K^2(\mathbf{p})K^2(\mathbf{q})\rho^2_n(\mathbf{x}-\mathbf{t}+\mathbf{H}(\mathbf{p}-\mathbf{q}))f(\mathbf{x}+\mathbf{Hp})f(\mathbf{t}+\mathbf{Hq})d\mathbf{p}d\mathbf{q}\\&=&\abs{\mathbf{H}}^{-2}f(\mathbf{x})f(\mathbf{t})\int \int K^2(\mathbf{p})K^2(\mathbf{q})\rho^2_n(\mathbf{x}-\mathbf{t}+\mathbf{H}(\mathbf{p}-\mathbf{q}))d\mathbf{p}d\mathbf{q}\cdot\{1+o(1)\}.
		\end{eqnarray*}			
	\end{proof}
\end{lem}

\begin{lem}\label{L4}
	Let $$W_{4n}(\mathbf{x},\mathbf{t})=\dfrac{1}{n^2}\sum_{i=1}^n\sum_{j=1}^n K_{\mathbf{H}}\left({\mathbf{X}_i-\mathbf{x}}\right)K_{\mathbf{H}}\left({\mathbf{X}_j-\mathbf{t}}\right)\rho_n(\mathbf{X}_i-\mathbf{X}_j).$$ For any $\mathbf{x},\mathbf{t}\in D$,  under assumptions (A1), (A3) and (A9), then 
	\begin{equation*}\label{W3n}
	\mathbb{E}(	W_{4n}(\mathbf{x},\mathbf{t}))=f(\mathbf{x})f(\mathbf{t})\int \int K\left(\mathbf{p}\right)K\left(\mathbf{q}\right)\rho_n(\mathbf{x}-\mathbf{t}+\mathbf{H}(\mathbf{p}-\mathbf{q}))d\mathbf{p}d\mathbf{q}+o(1).	
	\end{equation*}
	\begin{proof}[Proof of Lemma 4]
	 For any $\mathbf{x},\mathbf{t}\in D$,
		\begin{eqnarray*}
			\mathbb{E}(W_{4n}(\mathbf{x},\mathbf{t}))&=&\int \int K_{\mathbf{H}}\left({\mathbf{u}-\mathbf{x}}\right)K_{\mathbf{H}}\left({\mathbf{v}-\mathbf{t}}\right)\rho_n(\mathbf{u}-\mathbf{v})f(\mathbf{u})f(\mathbf{v})d\mathbf{u}d\mathbf{v}\\&=&\int \int K\left(\mathbf{p}\right)K\left(\mathbf{q}\right)\rho_n(\mathbf{x}-\mathbf{t}+\mathbf{H}(\mathbf{p}-\mathbf{q}))f(\mathbf{x}+\mathbf{Hp})f(\mathbf{t}+\mathbf{Hq})d\mathbf{p}d\mathbf{q}\\&=&f(\mathbf{x})f(\mathbf{t})\int \int K\left(\mathbf{p}\right)K\left(\mathbf{q}\right)\rho_n(\mathbf{x}-\mathbf{t}+\mathbf{H}(\mathbf{p}-\mathbf{q}))d\mathbf{p}d\mathbf{q}+o(1).
		\end{eqnarray*}
	\end{proof}
\end{lem}

\begin{lem}\label{L5}
	Let $$W_{5n}(\mathbf{x},\mathbf{t})=\dfrac{1}{n^2}\sum_{i=1}^n\sum_{j=1}^n K_{\mathbf{H}}\left({\mathbf{X}_i-\mathbf{x}}\right)K_{\mathbf{H}}\left({\mathbf{X}_i-\mathbf{t}}\right)K_{\mathbf{H}}\left({\mathbf{X}_j-\mathbf{x}}\right)K_{\mathbf{H}}\left({\mathbf{X}_j-\mathbf{t}}\right)\rho^2_n(\mathbf{X}_i-\mathbf{X}_j).$$ For any $\mathbf{x},\mathbf{t}\in D$,  under assumptions (A1), (A3) and (A9), then
		\begin{eqnarray*}\label{W3n}
	\mathbb{E}(W_{5n}(\mathbf{x},\mathbf{t}))&=&\abs{\mathbf{H}}^{-2}f^2(\mathbf{t})\int\int K\left(-\mathbf{p}+\mathbf{H}^{-1}({\mathbf{x}-\mathbf{t}})\right)K\left(-\mathbf{q}+\mathbf{H}^{-1}({\mathbf{x}-\mathbf{t}})\right)\\&\nonumber\cdot&K(\mathbf{p})K(\mathbf{q})\rho^2_n(\mathbf{H}(\mathbf{p}-\mathbf{q}))d\mathbf{p}d\mathbf{q}\cdot\{1+o(1)\}.	
	\end{eqnarray*}
	\begin{proof}[Proof of Lemma 5]
		 For any $\mathbf{x},\mathbf{t}\in D$,
		\begin{eqnarray*}
			\mathbb{E}(W_{5n}(\mathbf{x},\mathbf{t}))&=&\int \int K_{\mathbf{H}}\left({\mathbf{u}-\mathbf{x}}\right)K_{\mathbf{H}}\left({\mathbf{u}-\mathbf{t}}\right)K_{\mathbf{H}}\left({\mathbf{v}-\mathbf{x}}\right)K_{\mathbf{H}}\left({\mathbf{v}-\mathbf{t}}\right)\rho^2_n(\mathbf{u}-\mathbf{v})\\&\cdot&f(\mathbf{u})f(\mathbf{v})d\mathbf{u}d\mathbf{v}\\&=&\abs{\mathbf{H}}^{-2}\int\int K\left(-\mathbf{p}+\mathbf{H}^{-1}({\mathbf{x}-\mathbf{t}})\right)K\left(-\mathbf{q}+\mathbf{H}^{-1}({\mathbf{x}-\mathbf{t}})\right)K\left(\mathbf{p}\right)K\left(\mathbf{q}\right)\\&\nonumber\cdot&f(\mathbf{t}+\mathbf{Hp})f(\mathbf{t}+\mathbf{Hq})\rho^2_n(\mathbf{H}(\mathbf{p}-\mathbf{q}))d\mathbf{p}d\mathbf{q}\\&=&\abs{\mathbf{H}}^{-2}f^2(\mathbf{t})\int\int K\left(-\mathbf{p}+\mathbf{H}^{-1}({\mathbf{x}-\mathbf{t}})\right)K\left(-\mathbf{q}+\mathbf{H}^{-1}({\mathbf{x}-\mathbf{t}})\right)\\&\nonumber\cdot&K\left(\mathbf{p}\right)K\left(\mathbf{q}\right)\rho^2_n(\mathbf{H}(\mathbf{p}-\mathbf{q}))d\mathbf{p}d\mathbf{q}\cdot\{1+o(1)\}.
		\end{eqnarray*}
	\end{proof}
\end{lem}

\begin{lem}\label{L6}
	Let $$W_{6n}(\mathbf{x},\mathbf{t})=\dfrac{1}{n^3}\sum_{i\neq j}\sum_{k\neq i,j} K_{\mathbf{H}}\left({\mathbf{X}_i-\mathbf{x}}\right)K_{\mathbf{H}}\left({\mathbf{X}_j-\mathbf{x}}\right)K_{\mathbf{H}}({\mathbf{X}_i-\mathbf{t}})K_{\mathbf{H}}({\mathbf{X}_k-\mathbf{t}})\rho_n(\mathbf{X}_i-\mathbf{X}_k)\rho_n(\mathbf{X}_j-\mathbf{X}_i).$$ For any $\mathbf{x},\mathbf{t}\in D$,  under assumptions (A1), (A3) and (A9), then
	\begin{eqnarray*}\label{W3n}
		\mathbb{E}(W_{6n}(\mathbf{x},\mathbf{t}))&=&\abs{\mathbf{H}}^{-1}f^2(\mathbf{x})f(\mathbf{t})\int\int\int K(\mathbf{p}) K\left(\mathbf{q}\right)K\left(-\mathbf{p}+\mathbf{H}^{-1}({\mathbf{x}-\mathbf{t}})\right)K\left(\mathbf{r}\right)\\&\nonumber\cdot&\rho_n(\mathbf{H}(\mathbf{p}-\mathbf{q}))\rho_n(\mathbf{x}-\mathbf{t}+\mathbf{H}(\mathbf{p}-\mathbf{r}))d\mathbf{p}d\mathbf{q}d\mathbf{r}\cdot\{1+o(1)\}.	
	\end{eqnarray*}
	\begin{proof}[Proof of Lemma 6]
		 For any $\mathbf{x},\mathbf{t}\in D$,
		\begin{eqnarray*}
			\mathbb{E}(W_{6n}(\mathbf{x},\mathbf{t}))&=&\int \int \int K_{\mathbf{H}}\left({\mathbf{u}-\mathbf{x}}\right)K_{\mathbf{H}}\left({\mathbf{v}-\mathbf{x}}\right)K_{\mathbf{H}}\left({\mathbf{u}-\mathbf{t}}\right)K_{\mathbf{H}}\left({\mathbf{y}-\mathbf{t}}\right)\\&\cdot&\rho_n(\mathbf{u}-\mathbf{y})\rho_n(\mathbf{u}-\mathbf{v})f(\mathbf{u})f(\mathbf{v})f(\mathbf{y})d\mathbf{u}d\mathbf{v}d\mathbf{y}\\&=&\abs{\mathbf{H}}^{-1}\int\int\int K(\mathbf{p}) K\left(\mathbf{q}\right)K\left(-\mathbf{p}+\mathbf{H}^{-1}({\mathbf{x}-\mathbf{t}})\right)K\left(\mathbf{r}\right)\\&\nonumber\cdot&f(\mathbf{x}+\mathbf{Hp})f(\mathbf{x}+\mathbf{Hq})f(\mathbf{t}+\mathbf{Hr})\rho_n(\mathbf{H}(\mathbf{p}-\mathbf{q}))\rho_n(\mathbf{x}-\mathbf{t}+\mathbf{H}(\mathbf{p}-\mathbf{r}))d\mathbf{p}d\mathbf{q}d\mathbf{r}\\&=&\abs{\mathbf{H}}^{-1}f^2(\mathbf{x})f(\mathbf{t})\int\int\int K(\mathbf{p}) K\left(\mathbf{q}\right)K\left(-\mathbf{p}+\mathbf{H}^{-1}({\mathbf{x}-\mathbf{t}})\right)K\left(\mathbf{r}\right)\\&\nonumber\cdot&\rho_n(\mathbf{H}(\mathbf{p}-\mathbf{q}))\rho_n(\mathbf{x}-\mathbf{t}+\mathbf{H}(\mathbf{p}-\mathbf{r}))d\mathbf{p}d\mathbf{q}d\mathbf{r}\cdot\{1+o(1)\}.
		\end{eqnarray*}
	\end{proof}
\end{lem}

Next, the proof of Theorem 1 is presented. 

\begin{proof}[Proof of Theorem 1]

The test statistic (\ref{statistic}) can be written as		\begin{eqnarray*}
		T_n&=&n\abs{\mathbf{H}}^{1/2}\int (\hat{m}^{LL}_{\mathbf{H}}(\mathbf{x})-\hat{m}^{LL}_{\mathbf{H},\hat{\bm{\beta}}}(\mathbf{x}))^2w(\mathbf{x})d\mathbf{x}\\&=&n\abs{\mathbf{H}}^{1/2}\int \bigg[e_1'\left(\frac{1}{n}X_x'W_xX_x\right)^{-1}\frac{1}{n}\sum_{i=1}^n (1, (\mathbf{X}_i-\mathbf{x})') K_{\mathbf{H}}({\mathbf{X}_i-\mathbf{x}})(Z_i-m_{\hat{{\bm{\beta}}}}(\mathbf{X}_i))\bigg]^2w(\mathbf{x})d\mathbf{x}\\&=&n\abs{\mathbf{H}}^{1/2}\int \bigg[e_1'\left(
		\begin{array}{ll}
			\frac{1}{n}\sum_{i=1}^{n} K_{\mathbf{H}}(\mathbf{X}_i-\mathbf{x}) & \frac{1}{n}\sum_{i=1}^{n} K_{\mathbf{H}}(\mathbf{X}_i-\mathbf{x})(\mathbf{X}_i-\mathbf{x})'  \\
			\frac{1}{n}\sum_{i=1}^{n} K_{\mathbf{H}}(\mathbf{X}_i-\mathbf{x})(\mathbf{X}_i-\mathbf{x})  & \frac{1}{n}\sum_{i=1}^{n} K_{\mathbf{H}}(\mathbf{X}_i-\mathbf{x})(\mathbf{X}_i-\mathbf{x}) (\mathbf{X}_i-\mathbf{x}) '
		\end{array}
		\right)^{-1}\\&\cdot&\bigg(
		\begin{array}{ll}
			\frac{1}{n}\sum_{i=1}^{n} K_{\mathbf{H}}(\mathbf{X}_i-\mathbf{x})(Z_i-m_{\hat{{\bm{\beta}}}}(\mathbf{X}_i))  \\
			\frac{1}{n}\sum_{i=1}^{n} K_{\mathbf{H}}(\mathbf{X}_i-\mathbf{x})(\mathbf{X}_i-\mathbf{x}) (Z_i-m_{\hat{{\bm{\beta}}}}(\mathbf{X}_i))
		\end{array}
		\bigg)\bigg]^2w(\mathbf{x})d\mathbf{x}.
	\end{eqnarray*}

According to \cite{liu2001kernel} and taking into account that for every $\eta>0$, $\hat{f}_{\mathbf{H}}(\mathbf{x})=\frac{1}{n}\sum_{i=1}^{n} K_{\mathbf{H}}(\mathbf{X}_i-\mathbf{x})=f(\mathbf{x})+ O_p(n^{-2/(4+d)+\eta})$ uniformly in $\mathbf{x}$ \citep[see]{hardle1993comparing}, it follows that
\begin{eqnarray*}
	\frac{1}{n}X_x'W_xX_x&=&\left(
	\begin{array}{ll}
		\frac{1}{n}\sum_{i=1}^{n} K_{\mathbf{H}}(\mathbf{X}_i-\mathbf{x}) & \frac{1}{n}\sum_{i=1}^{n} K_{\mathbf{H}}(\mathbf{X}_i-\mathbf{x})(\mathbf{X}_i-\mathbf{x})'  \\
		\frac{1}{n}\sum_{i=1}^{n} K_{\mathbf{H}}(\mathbf{X}_i-\mathbf{x})(\mathbf{X}_i-\mathbf{x})  & \frac{1}{n}\sum_{i=1}^{n} K_{\mathbf{H}}(\mathbf{X}_i-\mathbf{x})(\mathbf{X}_i-\mathbf{x}) (\mathbf{X}_i-\mathbf{x}) '
	\end{array}
	\right)\\&=&\left(
	\begin{array}{ll}
		f(\mathbf{x})+O_p(n^{-2/(4+d)+\eta}) & \mu_2(K)\nabla f(\mathbf{x})'\mathbf{H}^2+O_p(n^{-2/(4+d)+\eta}\mathbf{H}^2) \\
		\mu_2(K)\mathbf{H}^2\nabla f(\mathbf{x})+O_p(n^{-2/(4+d)+\eta}\mathbf{H}^2)  & \mu_2(K)f(\mathbf{x})\mathbf{H}^2+O_p(\mathbf{H}\mathbf{1}_{d\times d}\mathbf{H})
	\end{array}
	\right),\end{eqnarray*}
where $\nabla f(\mathbf{x})$ denotes the $d \times 1$ vector of first-order partial derivatives of $f$ (and $\nabla f(\mathbf{x})'$ its transpose). 

Therefore,
\begin{eqnarray}\label{Tn}
	T_n\nonumber&=& n\abs{\mathbf{H}}^{1/2}\\&\cdot&\int \bigg[e_1'\bigg(
	\begin{array}{ll}
		f(\mathbf{x})+O_p(n^{-2/(4+d)+\eta}) & \mu_2(K)\nabla f(\mathbf{x})'\mathbf{H}^2+O_p(n^{-2/(4+d)+\eta}\mathbf{H}^2) \\
		\mu_2(K)\mathbf{H}^2\nabla f(\mathbf{x})+O_p(n^{-2/(4+d)+\eta}\mathbf{H}^2)  & \mu_2(K)f(\mathbf{x})\mathbf{H}^2+O_p(\mathbf{H}\mathbf{1}_{d\times d}\mathbf{H})
	\end{array}
	\bigg)^{-1}\nonumber\\&\cdot&\bigg(
	\begin{array}{ll}
		\frac{1}{n}\sum_{i=1}^{n} K_{\mathbf{H}}(\mathbf{X}_i-\mathbf{x})(Z_i-m_{\hat{{\bm{\beta}}}}(\mathbf{X}_i))  \\
		\frac{1}{n}\sum_{i=1}^{n} K_{\mathbf{H}}(\mathbf{X}_i-\mathbf{x})(\mathbf{X}_i-\mathbf{x}) (Z_i-m_{\hat{{\bm{\beta}}}}(\mathbf{X}_i))
	\end{array}
	\bigg)\bigg]^2w(\mathbf{x})d\mathbf{x}\nonumber \\&=&\nonumber n\abs{\mathbf{H}}^{1/2}\\&\cdot&\int \bigg[e_1'\bigg(
	\begin{array}{ll}
	f^{-1}(\mathbf{x})+O_p(n^{-2/(4+d)+\eta}) & -f^{-2}(\mathbf{x})\nabla f(\mathbf{x})'+O_p(n^{-2/(4+d)+\eta}\mathbf{1}_d') \\
	-f^{-2}(\mathbf{x})\nabla f(\mathbf{x})+O_p(n^{-2/(4+d)+\eta}\mathbf{1})  & \{\mu_2(K)f(\mathbf{x})\mathbf{H}^2\}^{-1}+O_p(n^{-2/(4+d)+\eta}\mathbf{H}\mathbf{1}_{d\times d}\mathbf{H})	\end{array}
	\bigg)\nonumber\\&\cdot&\bigg(
	\begin{array}{ll}
		\frac{1}{n}\sum_{i=1}^{n} K_{\mathbf{H}}(\mathbf{X}_i-\mathbf{x})(Z_i-m_{\hat{{\bm{\beta}}}}(\mathbf{X}_i))  \\
		\frac{1}{n}\sum_{i=1}^{n} K_{\mathbf{H}}(\mathbf{X}_i-\mathbf{x})(\mathbf{X}_i-\mathbf{x}) (Z_i-m_{\hat{{\bm{\beta}}}}(\mathbf{X}_i))
	\end{array}
	\bigg)\bigg]^2w(\mathbf{x})d\mathbf{x}\nonumber \\&=&\nonumber n\abs{\mathbf{H}}^{1/2}\int \bigg[		\frac{1}{nf(\mathbf{x})}\sum_{i=1}^{n} K_{\mathbf{H}}(\mathbf{X}_i-\mathbf{x})(Z_i-m_{\hat{{\bm{\beta}}}}(\mathbf{X}_i))\\\nonumber &-&\nabla f(\mathbf{x})		\frac{1}{nf^{2}(\mathbf{x})}\sum_{i=1}^{n} K_{\mathbf{H}}(\mathbf{X}_i-\mathbf{x})(\mathbf{X}_i-\mathbf{x}) (Z_i-m_{\hat{{\bm{\beta}}}}(\mathbf{X}_i))\bigg]^2w(\mathbf{x})d\mathbf{x}+O_p(n^{-2/(4+d)+\eta})\\&=&T_{n1}+T_{n2}+2T_{n12}+O_p(n^{-2/(4+d)+\eta}),
\end{eqnarray}
with
	\begin{eqnarray*}
	T_{n1}&=&n\abs{\mathbf{H}}^{1/2}\int \bigg[		\frac{1}{nf(\mathbf{x})}\sum_{i=1}^{n} K_{\mathbf{H}}(\mathbf{X}_i-\mathbf{x})(Z_i-m_{\hat{{\bm{\beta}}}}(\mathbf{X}_i))\bigg]^2w(\mathbf{x})d\mathbf{x},\\
	T_{n2}&=&n\abs{\mathbf{H}}^{1/2}\int \bigg[		\nabla f(\mathbf{x})		\frac{1}{nf^{2}(\mathbf{x})}\sum_{i=1}^{n} K_{\mathbf{H}}(\mathbf{X}_i-\mathbf{x})(\mathbf{X}_i-\mathbf{x}) (Z_i-m_{\hat{{\bm{\beta}}}}(\mathbf{X}_i))\bigg]^2w(\mathbf{x})d\mathbf{x},\\
\end{eqnarray*}
and the  $T_{n12}$ term is the integral of the cross product.

Regarding $T_{n1}$, taking into account that the regression functions considered are of the form $m  =m_{{\bm{\beta}}_0}  + n^{-1/2}\abs{\mathbf{H}}^{-1/4}g  $, one gets
	\begin{eqnarray*}
			T_{n1}&=&n\abs{\mathbf{H}}^{1/2}\int \bigg[		\frac{1}{nf(\mathbf{x})}\sum_{i=1}^{n} K_{\mathbf{H}}(\mathbf{X}_i-\mathbf{x})(Z_i-m_{\hat{{\bm{\beta}}}}(\mathbf{X}_i))\bigg]^2w(\mathbf{x})d\mathbf{x}\\&=&n\abs{\mathbf{H}}^{1/2}\int \bigg[		\frac{1}{nf(\mathbf{x})}\sum_{i=1}^{n} K_{\mathbf{H}}(\mathbf{X}_i-\mathbf{x})(m(\mathbf{X}_i)+\varepsilon_i-m_{\hat{{\bm{\beta}}}}(\mathbf{X}_i))\bigg]^2w(\mathbf{x})d\mathbf{x}\\&=&n\abs{\mathbf{H}}^{1/2}\int \bigg[		\frac{1}{nf(\mathbf{x})}\sum_{i=1}^{n} K_{\mathbf{H}}(\mathbf{X}_i-\mathbf{x})(m_{{\bm{\beta}}_0}(\mathbf{X}_i)+n^{-1/2}\abs{\mathbf{H}}^{-1/4}g(\mathbf{X}_i)+\varepsilon_i-m_{\hat{{\bm{\beta}}}}(\mathbf{X}_i))\bigg]^2w(\mathbf{x})d\mathbf{x}\\&=&n\abs{\mathbf{H}}^{1/2}\int 	\frac{1}{f^2(\mathbf{x})}(I_1(\mathbf{x})+I_2(\mathbf{x})+I_3(\mathbf{x}))^2w(\mathbf{x})d\mathbf{x},
	\end{eqnarray*}
where
	\begin{eqnarray*}
	I_1(\mathbf{x})&=&\dfrac{1}{n}\sum_{i=1}^n K_{\mathbf{H}}\left({\mathbf{X}_i-\mathbf{x}}\right)(m_{{\bm{\beta}}_0}(\mathbf{X}_i)-m_{\hat{{\bm{\beta}}}}(\mathbf{X}_i)),\\
	I_2(\mathbf{x})&=&\dfrac{1}{n}\sum_{i=1}^n K_{\mathbf{H}}\left({\mathbf{X}_i-\mathbf{x}}\right)n^{-1/2}\abs{\mathbf{H}}^{-1/4}g(\mathbf{X}_i),\\
	I_3(\mathbf{x})&=&\dfrac{1}{n}\sum_{i=1}^n K_{\mathbf{H}}\left({\mathbf{X}_i-\mathbf{x}}\right)\varepsilon_i.
\end{eqnarray*}

With respect to the term $I_1(\mathbf{x})$, using assumptions (A1)--(A3) and (A7), and given that the difference  $m_{\hat{{\bm{\beta}}}}(\mathbf{x})-m_{{{\bm{\beta}}}_0}(\mathbf{x})=O_p(n^{-1/2})$ uniformly in $\mathbf{x}$, it is obtained that 

\begin{eqnarray}
&&n\abs{\mathbf{H}}^{1/2}\int 	\frac{1}{f^2(\mathbf{x})}I_1^2(\mathbf{x})w(\mathbf{x})d\mathbf{x}\nonumber\\&=&n\abs{\mathbf{H}}^{1/2}\int 	\frac{1}{f^2(\mathbf{x})}\left[\dfrac{1}{n}\sum_{i=1}^n K_{\mathbf{H}}\left({\mathbf{X}_i-\mathbf{x}}\right)(m_{{\bm{\beta}}_0}(\mathbf{X}_i)-m_{\hat{{\bm{\beta}}}}(\mathbf{X}_i))\right]^2w(\mathbf{x})d\mathbf{x}\nonumber\\&=&O_p(\abs{\mathbf{H}}^{1/2}).
\label{i_1}
\end{eqnarray}

As for the  term $I_2(\mathbf{x})$, taking into account Lemma \ref{L1}, it follows that
	\begin{eqnarray}\label{b1H}
&&n\abs{\mathbf{H}}^{1/2}\int 	\frac{1}{f^2(\mathbf{x})}I_2^2(\mathbf{x})w(\mathbf{x})d\mathbf{x}\nonumber\\&=&n\abs{\mathbf{H}}^{1/2}\int 	\frac{1}{f^2(\mathbf{x})}\left[\dfrac{1}{n}\sum_{i=1}^n K_{\mathbf{H}}\left({\mathbf{X}_i-\mathbf{x}}\right)n^{-1/2}\abs{\mathbf{H}}^{-1/4}g(\mathbf{X}_i)\right]^2w(\mathbf{x})d\mathbf{x}\nonumber\\&=& \int \frac{1}{f^2(\mathbf{x})} \bigg[\int K\left(\mathbf{p}\right)g(\mathbf{x}+\mathbf{Hp})f(\mathbf{x}+\mathbf{Hp})d\mathbf{p}+o_p(1)\bigg]^2 w(\mathbf{x})d\mathbf{x} \nonumber\\&=& \int \frac{1}{f^2(\mathbf{x})} \bigg[\int K\left(\mathbf{p}\right)g(\mathbf{x}+\mathbf{Hp})\{f(\mathbf{x})+o(1)\}d\mathbf{p}\bigg]^2w(\mathbf{x})d\mathbf{x} \cdot \{1+o_p(1)\}\nonumber\\&=& \int \bigg[\int K_\mathbf{H}\left({\mathbf{u}-\mathbf{x}}\right)g(\mathbf{u})d\mathbf{u}\bigg]^2w(\mathbf{x})d\mathbf{x}\cdot\{1+o_p(1)\}\nonumber\\&=& \int (K_\mathbf{H}\ast g)^2(\mathbf{x})w(\mathbf{x})d\mathbf{x}\cdot\{1+o_p(1)\}.
	\end{eqnarray}
	
The leading term of (\ref{b1H}) is the term $b_{1\mathbf{H}}$ in Theorem 1. Finally, the term $I_3(\mathbf{x})$, associated with the error component of the model, can be split as 
			\begin{eqnarray*}\label{decom}
	n\abs{\mathbf{H}}^{1/2}\int 	\frac{1}{f^2(\mathbf{x})}I_3^2(\mathbf{x})w(\mathbf{x})d\mathbf{x}&=&n\abs{\mathbf{H}}^{1/2}\int 	\frac{1}{f^2(\mathbf{x})}\left[\dfrac{1}{n}\sum_{i=1}^n K_{\mathbf{H}}\left({\mathbf{X}_i-\mathbf{x}}\right)\varepsilon_i\right]^2w(\mathbf{x})d\mathbf{x}\nonumber\\&=& n\abs{\mathbf{H}}^{1/2}\int \frac{1}{f^2(\mathbf{x})}\dfrac{1}{n^2}\sum_{i=1}^n K^2_{\mathbf{H}}\left({\mathbf{X}_i-\mathbf{x}}\right)\varepsilon_i^2 w(\mathbf{x})d\mathbf{x}\nonumber\\&+& n\abs{\mathbf{H}}^{1/2}\int \frac{1}{f^2(\mathbf{x})}\dfrac{1}{n^2}\sum_{i\neq j} K_{\mathbf{H}}\left({\mathbf{X}_i-\mathbf{x}}\right)K_{\mathbf{H}}\left({\mathbf{X}_j-\mathbf{x}}\right)\varepsilon_i\varepsilon_jw(\mathbf{x})d\mathbf{x}\nonumber\\&=&I_{31}+I_{32}.
	\end{eqnarray*}
	
Close expressions of $I_{31}$ and $I_{32}$ can be obtained computing the expectation and the variance of these terms. For doing so, general results on the conditional expectation and conditional variance can be used. Specifically, given two random variables $X$ and $Y$, it is known that  $\mathbb{E}(X)=\mathbb{E}(\mathbb{E}(X|Y))$ and $\mbox{Var}(X)=\mathbb{E}(\mbox{Var}(X|Y))+\mbox{Var}(\mathbb{E}(X|Y))$.
	
	For $I_{31}$, using the result for the conditional mean, it follows that $\mathbb{E}(I_{31})=	\mathbb{E}(	\mathbb{E}(I_{31}|\mathbf{X}_1,\dots,\mathbf{X}_n))$. Firstly,  
	\begin{eqnarray}\label{expI31}
		\mathbb{E}(\abs{\mathbf{H}}^{1/2}I_{31}|\mathbf{X}_1,\dots,\mathbf{X}_n)\nonumber&=&\mathbb{E}\bigg[n\abs{\mathbf{H}}\int \frac{1}{f^2(\mathbf{x})}\frac{1}{n^2}\sum_{i=1}^n K^2_{\mathbf{H}}\left({\mathbf{X}_i-\mathbf{x}}\right)\varepsilon_i^2 w(\mathbf{x})d\mathbf{x}|\mathbf{X}_1,\dots,\mathbf{X}_n\bigg]\nonumber\\&=&\sigma^2n\abs{\mathbf{H}}\int \frac{1}{f^2(\mathbf{x})}\frac{1}{n^2}\sum_{i=1}^n K^2_{\mathbf{H}}\left({\mathbf{X}_i-\mathbf{x}}\right)w(\mathbf{x})d\mathbf{x}.
	\end{eqnarray}
	Considering the first part of the proof of Lemma \ref{L2}, one gets that,
	\begin{eqnarray}\label{exp1}
	\mathbb{E}(\abs{\mathbf{H}}^{1/2}I_{31})=	\mathbb{E}(	\mathbb{E}(\abs{\mathbf{H}}^{1/2}I_{31}|\mathbf{X}_1,\dots,\mathbf{X}_n))\nonumber&=&\mathbb{E}\bigg[\sigma^2n\abs{\mathbf{H}}\int \frac{1}{f^2(\mathbf{x})}\frac{1}{n^2}\sum_{i=1}^n K^2_{\mathbf{H}}\left({\mathbf{X}_i-\mathbf{x}}\right)w(\mathbf{x})d\mathbf{x}\bigg]\nonumber\\&=&\sigma^2\abs{\mathbf{H}}\int \frac{1}{f^2(\mathbf{x})}\abs{\mathbf{H}}^{-1}K^{(2)}(\mathbf{0})\{f(\mathbf{x})+o(1)\}w(\mathbf{x})d\mathbf{x} \nonumber\\&=&\sigma^2 {}K^{(2)}(\mathbf{0})\int \dfrac{w(\mathbf{x})}{f(\mathbf{x})}d\mathbf{x}\cdot\{1+o(1)\}.
\end{eqnarray} 

On the other hand,
\begin{equation}
\mbox{Var}(I_{31})=\mathbb{E}(\mbox{Var}(I_{31}|\mathbf{X}_1,\dots,\mathbf{X}_n)) + \mbox{Var}(	\mathbb{E}(I_{31}|\mathbf{X}_1,\dots,\mathbf{X}_n)).
\label{var_i31}
\end{equation}

Using assumption (A5), it is obtained that  
\begin{eqnarray*}
		&&\mbox{Var}(\abs{\mathbf{H}}^{1/2}I_{31}|\mathbf{X}_1,\dots,\mathbf{X}_n)\\&=&{\mbox{Var}}\bigg[n\abs{\mathbf{H}}\int \frac{1}{f^2(\mathbf{x})}\frac{1}{n^2}\sum_{i=1}^n K^2_{\mathbf{H}}\left({\mathbf{X}_i-\mathbf{x}}\right)\varepsilon_i^2 w(\mathbf{x})d\mathbf{x}|\mathbf{X}_1,\dots,\mathbf{X}_n\bigg]\\&=&\frac{1}{n^2}\abs{\mathbf{H}}^2\sum_{i=1}^n\sum_{j=1}^n\int \int \dfrac{1}{f^2(\mathbf{x})f^2(\mathbf{t})}K^2_\mathbf{H}\left({\mathbf{X}_i-\mathbf{x}}\right)K^2_\mathbf{H}\left({\mathbf{X}_j-\mathbf{t}}\right)w(\mathbf{x})w(\mathbf{t})d\mathbf{x}d\mathbf{t}\\&\cdot&{\mbox{Cov}}(\varepsilon_i^2,\varepsilon_j^2)\\&=&\frac{1}{n^2}\abs{\mathbf{H}}^2\sum_{i=1}^n\sum_{j=1}^n\int \int \dfrac{1}{f^2(\mathbf{x})f^2(\mathbf{t})}K^2_\mathbf{H}\left({\mathbf{X}_i-\mathbf{x}}\right)K^2_\mathbf{H}\left({\mathbf{X}_j-\mathbf{t}}\right)w(\mathbf{x})w(\mathbf{t})d\mathbf{x}d\mathbf{t}\\&\cdot&2({\mbox{Cov}}(\varepsilon_i,\varepsilon_j))^2\\&=&\frac{2\sigma^4}{n^2}\abs{\mathbf{H}}^2\sum_{i=1}^n\sum_{j=1}^n\int \int \dfrac{1}{f^2(\mathbf{x})f^2(\mathbf{t})}K^2_\mathbf{H}\left({\mathbf{X}_i-\mathbf{x}}\right)K^2_\mathbf{H}\left({\mathbf{X}_j-\mathbf{t}}\right)w(\mathbf{x})w(\mathbf{t})d\mathbf{x}d\mathbf{t}\\&\cdot&\rho^2_n(\mathbf{X}_i-\mathbf{X}_j)
	\end{eqnarray*}
and, therefore by using and Lemma \ref{L3},
	\begin{eqnarray*}
	&&\mathbb{E}(\mbox{Var}(\abs{\mathbf{H}}^{1/2}I_{31}|\mathbf{X}_1,\dots,\mathbf{X}_n))\\&=&\mathbb{E}\bigg[\frac{2\sigma^4}{n^2}\abs{\mathbf{H}}^2\sum_{i=1}^n\sum_{j=1}^n\int \int \dfrac{1}{f^2(\mathbf{x})f^2(\mathbf{t})}K^2_\mathbf{H}\left({\mathbf{X}_i-\mathbf{x}}\right)K^2_\mathbf{H}\left({\mathbf{X}_j-\mathbf{t}}\right)w(\mathbf{x})w(\mathbf{t})d\mathbf{x}d\mathbf{t}\\&\cdot&\rho^2_n(\mathbf{X}_i-\mathbf{X}_j)\bigg]\\&=&2\sigma^4\abs{\mathbf{H}}^2\int \int \dfrac{1}{f^2(\mathbf{x})f^2(\mathbf{t})}\abs{\mathbf{H}}^{-2}\int \int K^2(\mathbf{p})K^2(\mathbf{q})\rho^2_n(\mathbf{x}-\mathbf{t}+\mathbf{H}(\mathbf{p}-\mathbf{q}))d\mathbf{p}d\mathbf{q}\\&\cdot& f(\mathbf{x})f(\mathbf{t})w(\mathbf{x})w(\mathbf{t})d\mathbf{x}d\mathbf{t}\cdot\{1+o(1)\}\\&=&2\sigma^4\int \int \int \int \dfrac{K^2(\mathbf{p})K^2(\mathbf{q})}{f(\mathbf{x})f(\mathbf{t})}w(\mathbf{x})w(\mathbf{t})\rho^2_n(\mathbf{x}-\mathbf{t}+\mathbf{H}(\mathbf{p}-\mathbf{q}))d\mathbf{p}d\mathbf{q}d\mathbf{x}d\mathbf{t}\\&\cdot&\{1+o(1)\}\\&=&2\sigma^4\abs{\mathbf{H}}\int \int \int \int \dfrac{K^2(\mathbf{p})K^2(\mathbf{q})}{f(\mathbf{x})f(\mathbf{x}+\mathbf{Hu})}w(\mathbf{x})w(\mathbf{x}+\mathbf{Hu})\rho^2_n(\mathbf{H}(\mathbf{p}-\mathbf{q}+\mathbf{u}))d\mathbf{p}d\mathbf{q}d\mathbf{x}d\mathbf{u}\\&\cdot&\{1+o(1)\}\\&=&2\sigma^4\abs{\mathbf{H}}\int \int \int \int \dfrac{K^2(\mathbf{p})K^2(\mathbf{q})}{f^2(\mathbf{x})}w^2(\mathbf{x})\rho^2_n(\mathbf{H}(\mathbf{p}-\mathbf{q}+\mathbf{u}))d\mathbf{p}d\mathbf{q}d\mathbf{x}d\mathbf{u}\\&\cdot&\{1+o(1)\}.
\end{eqnarray*}

Let \begin{eqnarray*}
		j_n(\mathbf{p},\mathbf{u})&=&n\abs{\mathbf{H}}\int K^2(\mathbf{q})\rho^2_n(\mathbf{H}(\mathbf{p}-\mathbf{q}+\mathbf{u}))d\mathbf{q}.
	\end{eqnarray*}
	Notice that, using assumption (A4),
	\begin{eqnarray*}\abs{j_n(\mathbf{p},\mathbf{u})}&\le& K^2_M(n\abs{\mathbf{H}}\int \abs{\rho^2_n(\mathbf{H}(\mathbf{p}-\mathbf{q}+\mathbf{u}))}d\mathbf{q})\\&\le&K^2_M(n\int \abs{\rho_n(\mathbf{t})}d\mathbf{t})\\&\le&K_M^2\rho_{M},\end{eqnarray*}
where $K_M\equiv\displaystyle \max_{\mathbf{x}}(K(\mathbf{x}))$ and $\rho_M\equiv\displaystyle \max_{\mathbf{x}}(\rho(\mathbf{x}))$, and using assumptions (A2), (A3), (A7) and (A9), one gets that
	\begin{equation}
			\mathbb{E}(\mbox{Var}(I_{31}|\mathbf{X}_1,\dots,\mathbf{X}_n))=o_p(1).
			\label{e_var_i31}
	\end{equation}
	
On the other hand, using expression (\ref{expI31}), the second part of Lemma \ref{L2} and assumption (A9), it follows  that
	\begin{eqnarray}
	&&\mbox{Var}(	\mathbb{E}(\abs{\mathbf{H}}^{1/2}I_{31}|\mathbf{X}_1,\dots,\mathbf{X}_n))\nonumber\\&=&\mbox{Var}\bigg[\sigma^2n\abs{\mathbf{H}}\int \frac{1}{f^2(\mathbf{x})}\frac{1}{n^2}\sum_{i=1}^n K^2_{\mathbf{H}}\left({\mathbf{X}_i-\mathbf{x}}\right)w(\mathbf{x})d\mathbf{x}\bigg]\nonumber\\&=&\sum_{i=1}^n\mbox{Var}\bigg[\sigma^2n\abs{\mathbf{H}}\int \frac{1}{f^2(\mathbf{x})}\frac{1}{n^2} K^2_{\mathbf{H}}\left({\mathbf{X}_i-\mathbf{x}}\right)w(\mathbf{x})d\mathbf{x}\bigg]\nonumber\\&\le&\sigma^4\abs{\mathbf{H}}^2\sum_{i=1}^n\mathbb{E}\bigg[\int \int\frac{1}{f^2(\mathbf{x})f^2(\mathbf{t})}\frac{1}{n^2} K^2_{\mathbf{H}}\left({\mathbf{X}_i-\mathbf{x}}\right)K^2_{\mathbf{H}}\left({\mathbf{X}_i-\mathbf{t}}\right)w(\mathbf{x})w(\mathbf{t})d\mathbf{x}d\mathbf{t}\bigg]\nonumber\\&=&o_p(1) \label{var_e_i31}.
\end{eqnarray} 

Now, considering (\ref{var_i31}), (\ref{e_var_i31}) and (\ref{var_e_i31}), it is obtained that
\begin{equation}
\mbox{Var}(\abs{\mathbf{H}}^{1/2}I_{31})=o_p(1),
\label{var_i31_t}
\end{equation}
and considering (\ref{exp1}) and (\ref{var_i31_t}),
\begin{equation}\label{b0H}
I_{31}=\sigma^2 \abs{\mathbf{H}}^{-1/2}{}K^{(2)}(\mathbf{0})\int \dfrac{w(\mathbf{x})}{f(\mathbf{x})}d\mathbf{x}\cdot\{1+o_p(1))\}.
\end{equation}

Taking into account assumption (A9), the leading term of (\ref{b0H}) corresponds to the first term of $b_{0\mathbf{H}}$ in Theorem 1.
	 
Now, consider the term   
	\begin{eqnarray*}       
		I_{32}&=&n\abs{\mathbf{H}}^{1/2}\int \frac{1}{f^2(\mathbf{x})}\dfrac{1}{n^2}\sum_{i\neq j} K_{\mathbf{H}}\left({\mathbf{X}_i-\mathbf{x}}\right)K_{\mathbf{H}}\left({\mathbf{X}_j-\mathbf{x}}\right)\varepsilon_i\varepsilon_jw(\mathbf{x})d\mathbf{x}.\\
	\end{eqnarray*}
	Let 
	\begin{eqnarray*}
		\kappa_{ij}&=&n\abs{\mathbf{H}}^{1/2}\int \frac{1}{f^2(\mathbf{x})}\dfrac{1}{n^2} K_{\mathbf{H}}\left({\mathbf{X}_i-\mathbf{x}}\right)K_{\mathbf{H}}\left({\mathbf{X}_j-\mathbf{x}}\right)\varepsilon_i\varepsilon_jw(\mathbf{x})d\mathbf{x},  
	\end{eqnarray*}
thus,
	$$I_{32}=\sum_{i\neq j}\kappa_{ij},$$
	and this can be seen as a U-statistic with degenerate kernel. To obtain the asymptotic normality of $I_{32}$, Theorem 2 given in \cite{kim2013central} will be applied. In this work, the central limit theorem for degenerate reduced U-statistics under $\alpha-$mixing is derived. The assumptions of this result hold (specifically, assumption (A6)) and the expectation and the variance of $I_{32}$ should be computed.

Proceeding as for $I_{31}$, it follows that $\mathbb{E}(I_{32})=\mathbb{E}(	\mathbb{E}(I_{32}|\mathbf{X}_1,\dots,\mathbf{X}_n))$. Taking into account the first part of   Lemma \ref{L4}, one gets that
		\begin{eqnarray}
		\mathbb{E}(I_{32}|\mathbf{X}_1,\dots,\mathbf{X}_n)\nonumber&=&\mathbb{E}\bigg[n\abs{\mathbf{H}}^{1/2}\dfrac{1}{n^2}\sum_{i\neq j}\displaystyle\int {\frac{1}{f^2(\mathbf{x})}K_{\mathbf{H}}\left({\mathbf{X}_i-\mathbf{x}}\right)K_{\mathbf{H}}\left({\mathbf{X}_j-\mathbf{x}}\right)}w(\mathbf{x})d\mathbf{x}\cdot{\varepsilon}_i{\varepsilon}_j|\mathbf{X}_1,\dots,\mathbf{X}_n\bigg]\nonumber\\&=&n\abs{\mathbf{H}}^{1/2}\dfrac{1}{n^2}\int \frac{1}{f^2(\mathbf{x})}\sum_{i\neq j}\mathbb{E}({\varepsilon}_i{\varepsilon}_j)\displaystyle{K_{\mathbf{H}}\left({\mathbf{X}_i-\mathbf{x}}\right)K_{\mathbf{H}}\left({\mathbf{X}_j-\mathbf{x}}\right)}w(\mathbf{x})d\mathbf{x} \nonumber\\&=&n\abs{\mathbf{H}}^{1/2}\int \frac{1}{f^2(\mathbf{x})}\dfrac{1}{n^2}\sum_{i\neq j}\mbox{Cov}({\varepsilon}_i,{\varepsilon}_j){K_{\mathbf{H}}\left({\mathbf{X}_i-\mathbf{x}}\right)K_{\mathbf{H}}\left({\mathbf{X}_j-\mathbf{x}}\right)}w(\mathbf{x})d\mathbf{x} \nonumber\\&=&\abs{\mathbf{H}}^{1/2}\sigma^2\int \frac{1}{f^2(\mathbf{x})}\dfrac{1}{n}\sum_{i\neq j}\rho_n(\mathbf{X}_i-\mathbf{X}_j){K_{\mathbf{H}}\left({\mathbf{X}_i-\mathbf{x}}\right)K_{\mathbf{H}}\left({\mathbf{X}_j-\mathbf{x}}\right)}w(\mathbf{x})d\mathbf{x},
		\label{E_i32}
	\end{eqnarray}
	and, therefore,
	\begin{eqnarray*}\label{biasT3}
&&\mathbb{E}(	\mathbb{E}(I_{32}|\mathbf{X}_1,\dots,\mathbf{X}_n))\nonumber\\&=&\mathbb{E}\bigg[n\abs{\mathbf{H}}^{1/2}\sigma^2\int \frac{1}{f^2(\mathbf{x})}\dfrac{1}{n^2}\sum_{i\neq j}\rho_n(\mathbf{X}_i-\mathbf{X}_j){K_{\mathbf{H}}\left({\mathbf{X}_i-\mathbf{x}}\right)K_{\mathbf{H}}\left({\mathbf{X}_j-\mathbf{x}}\right)}w(\mathbf{x})d\mathbf{x}\bigg]\nonumber\\&=&n\abs{\mathbf{H}}^{1/2}\sigma^2\int \frac{1}{f^2(\mathbf{x})}\left(\dfrac{n-1}{n}{f^2(\mathbf{x})}\int \int K\left(\mathbf{p}\right)K\left(\mathbf{q}\right)\rho_n(\mathbf{H}(\mathbf{p}-\mathbf{q}))d\mathbf{p}d\mathbf{q}+o(1)\right)\\&\cdot&w(\mathbf{x})d\mathbf{x}\nonumber\\&=&\dfrac{n-1}{n}\abs{\mathbf{H}}^{-1/2}\sigma^2\int \left(n\abs{\mathbf{H}}\int \int K\left(\mathbf{p}\right)K\left(\mathbf{q}\right)\rho_n(\mathbf{H}(\mathbf{p}-\mathbf{q}))d\mathbf{p}d\mathbf{q}+o(1)\right)\\&\cdot&w(\mathbf{x})d\mathbf{x}.
	\end{eqnarray*} 

Under the assumptions (A4), (A7), (A8) and (A9), as shown in \cite{liu2001kernel}, 
$$\lim_{n\to\infty}n\abs{\mathbf{H}}\int K(\mathbf{p})K(\mathbf{q})\rho_n(\mathbf{H}(\mathbf{p}-\mathbf{q}))d\mathbf{p}d\mathbf{q}=K^{(2)}(0)\rho_c,$$
and, therefore, 
\begin{equation}\label{biasI32}
				\mathbb{E}\left(I_{32}\right)=\abs{\mathbf{H}}^{-1/2}\sigma^2K^{(2)}(0)\rho_{c}\int w(\mathbf{x})d\mathbf{x}\cdot\{1+o(1)\},
				\end{equation}
corresponding to the second term of $b_{0\mathbf{H}}$ in Theorem 1.

The variance of $I_{32}$ can be computed considering that:
	\begin{eqnarray}
	\mbox{Var}(I_{32})&=&\mathbb{E}(\mbox{Var}(I_{32}|\mathbf{X}_1,\dots,\mathbf{X}_n))+\mbox{Var}(\mathbb{E}(I_{32}|\mathbf{X}_1,\dots,\mathbf{X}_n)).
	\label{var_i32}
\end{eqnarray}
Let
\begin{eqnarray*}
	W_{ij}&=&\displaystyle\int \frac{1}{f^2(\mathbf{x})}{K_{\mathbf{H}}\left({\mathbf{X}_i-\mathbf{x}}\right)K_{\mathbf{H}}\left({\mathbf{X}_j-\mathbf{x}}\right)}w(\mathbf{x})d\mathbf{x},
\end{eqnarray*}
thus, 
	\begin{eqnarray}
	\mbox{Var}(I_{32}|\mathbf{X}_1,\dots,\mathbf{X}_n)&=&\mbox{Var}\left(n^{-1}\abs{\mathbf{H}}^{1/2}\sum_{i\neq j}W_{ij}{\varepsilon}_i{\varepsilon}_j|\mathbf{X}_1,\dots,\mathbf{X}_n\right)\nonumber\\&=&4n^{-2}\abs{\mathbf{H}}\sum_{i=1}^{n-1}\sum_{j=i+1}^{n}\sum_{k=1}^{n-1}\sum_{l=k+1}^{n}W_{ij}W_{kl}\mbox{Cov}({\varepsilon}_i{\varepsilon}_j,{\varepsilon}_k{\varepsilon}_l )\nonumber	\\&=&T_{31}+T_{32}+T_{33},
	\label{var_i32_con}
\end{eqnarray}
	where
	\begin{eqnarray*}
		T_{31}&=&4n^{-2}\abs{\mathbf{H}}\sum_{i=1}^{n-1}\sum_{j=i+1}^{n}W^2_{ij}\mbox{Cov}({\varepsilon}_i{\varepsilon}_j,{\varepsilon}_i{\varepsilon}_j),\\
		T_{32}&=&4n^{-2}\abs{\mathbf{H}}\sum_{i=1}^{n-2}\sum_{j=i+1}^{n-1}\sum_{l=i+2}^{n}W_{ij}W_{il}\mbox{Cov}({\varepsilon}_i{\varepsilon}_j,{\varepsilon}_i{\varepsilon}_l ),\\
		T_{33}&=&4n^{-2}\abs{\mathbf{H}}\sum_{\text{\scriptsize all different indices  $i$, $j$, $k$, $l$}}W_{ij}W_{kl}\mbox{Cov}({\varepsilon}_i{\varepsilon}_j,{\varepsilon}_k{\varepsilon}_l ).
	\end{eqnarray*}

	First, when $i = k$ and $j = l$, the total number of terms is  $n(n-1)/2$. Second, when one of the $i$ and $j$ is equal to one of the $k$ and $l$ (without loss of generality, assume $i = k$ and $j\neq l$), the total number of terms can be bounded by $n^3$. Finally, when $i, j, k$, and $l$ are all different, the total number of terms can be bounded by $n^4$.
	
The expected value of $\mbox{Var}(I_{32}|\mathbf{X}_1,\dots,\mathbf{X}_n)$ is computed, calculating the mean of the terms $T_{31}$,  $T_{32}$, and $T_{33}$,
\begin{equation}
\mathbb{E}(\mbox{Var}(I_{32}|\mathbf{X}_1,\dots,\mathbf{X}_n))=\mathbb{E}(T_{31})+\mathbb{E}(T_{32})+\mathbb{E}(T_{33}).
\label{E_var_i32}
\end{equation}

As for $T_{31}$, using assumption (A5), this term can be split as 	\begin{eqnarray*}T_{31}&=&4n^{-2}\abs{\mathbf{H}}\sum_{i=1}^{n-1}\sum_{j=i+1}^{n}W^2_{ij}\mbox{Cov}({\varepsilon}_i{\varepsilon}_j,{\varepsilon}_i{\varepsilon}_j )\\&=&4n^{-2}\abs{\mathbf{H}}\sum_{i=1}^{n-1}\sum_{j=i+1}^{n}W^2_{ij}[\sigma^4+\mbox{Cov}^2({\varepsilon}_i,{\varepsilon}_j)]\\&=&T_{311}+T_{312},\end{eqnarray*}	
	where
	\begin{eqnarray*}T_{311}&=&4\sigma^4 n^{-2}\abs{\mathbf{H}}\int \int \frac{1}{f^2(\mathbf{x})f^2(\mathbf{t})}\sum_{i=1}^{n-1}\sum_{j=i+1}^{n} K_{\mathbf{H}}\left({\mathbf{X}_i-\mathbf{x}}\right)K_{\mathbf{H}}\left({\mathbf{X}_j-\mathbf{x}}\right)K_{\mathbf{H}}({\mathbf{X}_i-\mathbf{t}})K_{\mathbf{H}}({\mathbf{X}_j-\mathbf{t}})\\&\cdot&w(\mathbf{x})w(\mathbf{t})d\mathbf{x}d\mathbf{t},\end{eqnarray*}	
	and 
	\begin{eqnarray*}T_{312}&=&4\sigma^4 n^{-2}\abs{\mathbf{H}}\int \int \frac{1}{f^2(\mathbf{x})f^2(\mathbf{t})}\sum_{i=1}^{n-1}\sum_{j=i+1}^{n} K_{\mathbf{H}}\left({\mathbf{X}_i-\mathbf{x}}\right)K_{\mathbf{H}}\left({\mathbf{X}_j-\mathbf{x}}\right)K_{\mathbf{H}}({\mathbf{X}_i-\mathbf{t}})K_{\mathbf{H}}({\mathbf{X}_j-\mathbf{t}})\\&\cdot&\rho^2_n(\mathbf{X}_i-\mathbf{X}_j)w(\mathbf{x})w(\mathbf{t})d\mathbf{x}d\mathbf{t}.\end{eqnarray*}	
	
Taking into account the first part of Lemma \ref{L2},
	\begin{eqnarray}\mathbb{E}(T_{311})&=&4\sigma^4 \abs{\mathbf{H}}\dfrac{n-1}{2n}\int \int \frac{1}{f^2(\mathbf{x})f^2(\mathbf{t})}\bigg[\abs{\mathbf{H}}^{-1}(K^{(2)}(\mathbf{H}^{-1}({\mathbf{x}}-\mathbf{t}))\{f(\mathbf{t})+o(1)\}\bigg]^2w(\mathbf{x})w(\mathbf{t})d\mathbf{x}d\mathbf{t}\nonumber\\&=&2\sigma^4 \abs{\mathbf{H}}\dfrac{n-1}{n}\int \int \frac{1}{f^2(\mathbf{x})f^2(\mathbf{t})}\abs{\mathbf{H}}^{-2}(K^{(2)}(\mathbf{H}^{-1}({\mathbf{x}}-\mathbf{t}))^2f^2(\mathbf{t})w(\mathbf{x})w(\mathbf{t})d\mathbf{x}d\mathbf{t}\cdot\{1+o(1)\}\nonumber\\&=&2\sigma^4 \dfrac{n-1}{n}\int \int \frac{1}{f^2(\mathbf{x})} (K^{(2)}(\mathbf{p}))^2w(\mathbf{x})w(\mathbf{x}+\mathbf{Hp})d\mathbf{x}d\mathbf{t}\cdot\{1+o(1)\}\nonumber\\&=&2\dfrac{n-1}{n}\sigma^4 K^{(4)}(0)\int  \dfrac{w^2(\mathbf{x})}{f^2(\mathbf{x})}d\mathbf{x}\cdot\{1+o(1)\}\nonumber\\&=&2\sigma^4 K^{(4)}(0)\int  \dfrac{w^2(\mathbf{x})}{f^2(\mathbf{x})}d\mathbf{x}\cdot\{1+o(1)\}.
\label{E_t311}	
\end{eqnarray}

	Similarly for $T_{312}$,  using assumptions (A2), (A3) and (A7), and taking into account Lemma \ref{L5}, this term becomes
		\begin{eqnarray*}\label{T31''}\mathbb{E}(T_{312})&=&4\sigma^4 \abs{\mathbf{H}}\dfrac{n-1}{2n}\int \int \frac{1}{f^2(\mathbf{x})f^2(\mathbf{t})}\bigg[\abs{\mathbf{H}}^{-2}\bigg(f^2(\mathbf{t})\int\int K(-\mathbf{p}+\mathbf{H}^{-1}({\mathbf{x}-\mathbf{t}}))\\&\cdot&K(-\mathbf{q}+\mathbf{H}^{-1}({\mathbf{x}-\mathbf{t}}))K(\mathbf{p})K(\mathbf{q})\rho^2_n(\mathbf{H}(\mathbf{p}-\mathbf{q}))d\mathbf{p}d\mathbf{q}\cdot\{1+o(1)\}\bigg)\bigg]w(\mathbf{x})w(\mathbf{t})d\mathbf{x}d\mathbf{t}\\&=&2\dfrac{n-1}{n}\sigma^4 \int \int \int \int \frac{1}{f^2(\mathbf{x})} K\left(-\mathbf{p}+\mathbf{u}\right)K\left(-\mathbf{q}+\mathbf{u}\right)K\left(\mathbf{p}\right)K\left(\mathbf{q}\right)w(\mathbf{x})w(\mathbf{x}+\mathbf{Hu})\\&\cdot&\rho^2_n(\mathbf{H}(\mathbf{p}-\mathbf{q}))d\mathbf{p}d\mathbf{q}d\mathbf{x}d\mathbf{u}\cdot\{1+o(1)\}
		\\&\le&2\dfrac{n-1}{n^2\abs{\mathbf{H}}}\dfrac{\sigma^4K_M^4w_M^2}{f_M^2}\int \{n\abs{\mathbf{H}}\int \rho^2_n(\mathbf{H}(\mathbf{p}-\mathbf{q}))d\mathbf{p}\}d\mathbf{q}\cdot\{1+o(1)\},
		\end{eqnarray*}	
	where $f_M$ denotes the lower bound of $f$ (assumption (A3)).
		
		Since 
			$$n\abs{\mathbf{H}}\int \rho^2_n(\mathbf{H}(\mathbf{p}-\mathbf{q}))d\mathbf{p}\le n\int \abs{\rho_n(\mathbf{t})}d\mathbf{t}\le C_1,$$
	it is obtained that
		 \begin{eqnarray}
		 \mathbb{E}(T_{312})&\le& 2\dfrac{\sigma^4K_M^4w_M^2}{f_M^2}\dfrac{C_1}{n\abs{\mathbf{H}}}\dfrac{n-1}{n}\nonumber\\&=& O_p(n^{-1}\abs{\mathbf{H}}^{-1}).
\label{E_t312}		 
\end{eqnarray}

Then, from (\ref{E_t311}) and (\ref{E_t312}), it follows that
\begin{equation}
\mathbb{E}(T_{31})=2\sigma^4 K^{(4)}(0)\int  \dfrac{w^2(\mathbf{x})}{f^2(\mathbf{x})}d\mathbf{x}\cdot\{1+o(1)\}+O_p(n^{-1}\abs{\mathbf{H}}^{-1}).
\label{E_t31}
\end{equation}

	With respect to the term $T_{32}$ (corresponding to the case with $i = k$ and $j\neq l$ in (\ref{var_i32_con})), using assumption (A5), it follows that
	\begin{eqnarray*}T_{32}&=&4n^{-2}\abs{\mathbf{H}}\sum_{i=1}^{n-2}\sum_{j=i+1}^{n-1}\sum_{j=i+2}^{n}W_{ij}W_{il}\mbox{Cov}({\varepsilon}_i{\varepsilon}_j,{\varepsilon}_i{\varepsilon}_l )\\&=&4n^{-2}\abs{\mathbf{H}}\sum_{i=1}^{n-2}\sum_{j=i+1}^{n-1}\sum_{j=i+2}^{n}W_{ij}W_{il}[\mbox{Var}({\varepsilon}_i)\mbox{Cov}({\varepsilon}_j,{\varepsilon}_l)+\mbox{Cov}({\varepsilon}_i,{\varepsilon}_l)\mbox{Cov}({\varepsilon}_j,{\varepsilon}_i)]\\&=&T_{321}+T_{322},\end{eqnarray*}	
	where
	\begin{eqnarray*}T_{321}&=&4\sigma^4n^{-2}\abs{\mathbf{H}}\int \int \frac{1}{f^2(\mathbf{x})f^2(\mathbf{t})}\sum_{i=1}^{n-2}\sum_{j=i+1}^{n-1}\sum_{l=i+2}^{n} K_{\mathbf{H}}\left({\mathbf{X}_i-\mathbf{x}}\right)K_{\mathbf{H}}\left({\mathbf{X}_j-\mathbf{x}}\right)K_{\mathbf{H}}({\mathbf{X}_i-\mathbf{t}})K_{\mathbf{H}}({\mathbf{X}_l-\mathbf{t}})\\&\cdot&\rho_n(\mathbf{X}_j-\mathbf{X}_l)w(\mathbf{x})w(\mathbf{t})d\mathbf{x}d\mathbf{t},\end{eqnarray*}	
	and 
	\begin{eqnarray*}T_{322}&=&4\sigma^4n^{-2}\abs{\mathbf{H}}\int \int \frac{1}{f^2(\mathbf{x})f^2(\mathbf{t})}\sum_{i=1}^{n-2}\sum_{j=i+1}^{n-1}\sum_{l=i+2}^{n} K_{\mathbf{H}}\left({\mathbf{X}_i-\mathbf{x}}\right)K_{\mathbf{H}}\left({\mathbf{X}_j-\mathbf{x}}\right)K_{\mathbf{H}}({\mathbf{X}_i-\mathbf{t}})K_{\mathbf{H}}({\mathbf{X}_l-\mathbf{t}})\\&\cdot&\rho_n(\mathbf{X}_i-\mathbf{X}_l)\rho_n(\mathbf{X}_j-\mathbf{X}_i)w(\mathbf{x})w(\mathbf{t})d\mathbf{x}d\mathbf{t}.\end{eqnarray*}	
	
Using  the assumption (A4) and the first part of Lemma \ref{L2} and of Lemma \ref{L4},  one gets
	\begin{eqnarray*}\mathbb{E}(T_{321})&=&4\sigma^4n^{}\abs{\mathbf{H}}\int \int \frac{1}{f^2(\mathbf{x})f^2(\mathbf{t})}\dfrac{1}{\abs{\mathbf{H}}^{}}(K^{(2)}(\mathbf{H}^{-1}({\mathbf{x}}-\mathbf{t}))f(\mathbf{t})\cdot\{1+o
		(1)\}\\&\cdot&\left(\int \int K\left(\mathbf{p}\right)K\left(\mathbf{q}\right)\rho_n(\mathbf{x}-\mathbf{t}+\mathbf{H}(\mathbf{p}-\mathbf{q}))d\mathbf{p}d\mathbf{q}f(\mathbf{x})f(\mathbf{t})\cdot\{1+o
		(1)\}\right)w(\mathbf{x})w(\mathbf{t})d\mathbf{x}d\mathbf{t}\\&=&4\sigma^4n\int \int \frac{1}{f^2(\mathbf{x})f^2(\mathbf{t})}\bigg(K^{(2)}(\mathbf{H}^{-1}({\mathbf{x}}-\mathbf{t}))f(\mathbf{t})\int \int K\left(\mathbf{p}\right)K\left(\mathbf{q}\right)\rho_n(\mathbf{x}-\mathbf{t}+\mathbf{H}(\mathbf{p}-\mathbf{q}))\\& &d\mathbf{p}d\mathbf{q}f(\mathbf{x})f(\mathbf{t})\bigg)w(\mathbf{x})w(\mathbf{t})d\mathbf{x}d\mathbf{t}\cdot\{1+o
		(1)\}\\&=&4\sigma^4n^{}\int \int \dfrac{1}{f(\mathbf{x})}K^{(2)}(\mathbf{H}^{-1}({\mathbf{x}}-\mathbf{t}))\int \int K\left(\mathbf{p}\right)K\left(\mathbf{q}\right)\rho_n(\mathbf{x}-\mathbf{t}+\mathbf{H}(\mathbf{p}-\mathbf{q}))d\mathbf{p}d\mathbf{q}\\&\cdot&w(\mathbf{x})w(\mathbf{t})d\mathbf{x}d\mathbf{t}\{1+o
		(1)\}\\&=&4\sigma^4 n\abs{\mathbf{H}}^{}\int \int \int \int \frac{1}{f(\mathbf{x})} K^{(2)}\left(\mathbf{r}\right)K\left(\mathbf{p}\right)K\left(\mathbf{q}\right)w(\mathbf{x})w(\mathbf{x}-\mathbf{Hr})\rho_n(\mathbf{H}(\mathbf{p}-\mathbf{q}+\mathbf{r}))d\mathbf{p}d\mathbf{q}d\mathbf{x}d\mathbf{r}\\&\cdot&\{1+o
		(1)\}\\&=&4\sigma^4 \int \int \int \dfrac{1}{f(\mathbf{x})} K^{(2)}\left(\mathbf{r}\right)K\left(\mathbf{q}\right)w^2(\mathbf{x})\{n\abs{\mathbf{H}}^{}\int K\left(\mathbf{p}\right)\rho_n(\mathbf{H}(\mathbf{p}-\mathbf{q}+\mathbf{r}))d\mathbf{p}\}d\mathbf{q}d\mathbf{x}d\mathbf{r}\\&\cdot&\{1+o
		(1)\}. \end{eqnarray*}
	As it was shown in \cite{liu2001kernel}, 	
	$$\lim_{n\to\infty}n\abs{\mathbf{H}}\int K\left(\mathbf{p}\right)\rho_n(\mathbf{H}(\mathbf{p}-\mathbf{q}+\mathbf{r}))d\mathbf{p}=K(\mathbf{q}-\mathbf{r})\rho_c,$$
and, therefore,
		\begin{eqnarray}
		\mathbb{E}(T_{321})&=&4 \sigma^4 \rho_c\int \int \int \frac{1}{f(\mathbf{x})} K^{(2)}\left(\mathbf{r}\right)K\left(\mathbf{q}\right)K(\mathbf{r}-\mathbf{q})w^2(\mathbf{x})d\mathbf{q}d\mathbf{x}d\mathbf{r}\cdot\{1+o
			(1)\}\nonumber\\&=&4\sigma^4 \rho_c\int \int \frac{1}{f(\mathbf{x})} (K^{(2)}\left(\mathbf{r}\right))^2w^2(\mathbf{x})d\mathbf{r}d\mathbf{x}\cdot\{1+o
			(1)\}\nonumber\\&=&4\sigma^4K^{(4)}(0) \rho_c\int \frac{w^2(\mathbf{x})}{f(\mathbf{x})} d\mathbf{x}\cdot\{1+o
			(1)\}.
			\label{E_t321}
			\end{eqnarray}
	
	Similarly,  taking into account that $K$ is bounded, assumption (A4) and Lemma \ref{L6}, the expected value of $T_{322}$ becomes
	\begin{eqnarray*}\label{T32''}\mathbb{E}(T_{322})\nonumber&=&4\sigma^4n\abs{\mathbf{H}}\int \int \frac{1}{f^2(\mathbf{x})f^2(\mathbf{t})}\abs{\mathbf{H}}^{-1}\bigg(f^2(\mathbf{x})f(\mathbf{t})\int\int \int K(\mathbf{p}) K\left(\mathbf{q}\right)K\left(-\mathbf{p}+\mathbf{H}^{-1}({\mathbf{x}-\mathbf{t}})\right)K\left(\mathbf{r}\right)\\&\nonumber\cdot&\rho_n(\mathbf{H}(\mathbf{p}-\mathbf{q}))\rho_n(\mathbf{x}-\mathbf{t}+\mathbf{H}(\mathbf{p}-\mathbf{r}))d\mathbf{p}d\mathbf{q}d\mathbf{r}\cdot\{1+o
		(1)\}\bigg)w(\mathbf{x})w(\mathbf{t})d\mathbf{x}d\mathbf{t}\\&=&4\sigma^4 n\int \int \int \int \int \frac{1}{f(\mathbf{t})} K\left(\mathbf{p}\right)K\left(-\mathbf{p}+\mathbf{H}^{-1}({\mathbf{x}-\mathbf{t}})\right)K\left(\mathbf{q}\right)K\left(\mathbf{r}\right)w(\mathbf{x})w(\mathbf{t})\\&\cdot&\rho_n(\mathbf{x}-\mathbf{t}+\mathbf{H}(\mathbf{p}-\mathbf{r}))\rho_n(\mathbf{H}(\mathbf{p}-\mathbf{q}))d\mathbf{p}d\mathbf{q}d\mathbf{r}d\mathbf{x}d\mathbf{t}\cdot\{1+o
		(1)\}\\&=&4\sigma^4 n\abs{\mathbf{H}}\int \int \int \int \int \frac{1}{f(\mathbf{t})} K\left(\mathbf{p}\right)K\left(-\mathbf{p}+\mathbf{u})\right)K\left(\mathbf{q}\right)K\left(\mathbf{r}\right)w(\mathbf{t}+\mathbf{Hu})w(\mathbf{t})\\&\cdot&\rho_n(\mathbf{H}(\mathbf{p}-\mathbf{r}+\mathbf{u}))\rho_n(\mathbf{H}(\mathbf{p}-\mathbf{q}))d\mathbf{p}d\mathbf{q}d\mathbf{r}d\mathbf{u}d\mathbf{t}\cdot\{1+o
		(1)\}\nonumber\\&=&4\sigma^4 n^{-1}\abs{\mathbf{H}}^{-1}\int \int \int \frac{1}{f(\mathbf{t})} K\left(\mathbf{p}\right)K\left(-\mathbf{p}+\mathbf{u}\right)w^2(\mathbf{t})\{n\abs{\mathbf{H}}\int K\left(\mathbf{r}\right)\rho_n(\mathbf{H}(\mathbf{p}-\mathbf{r}+\mathbf{u}))d\mathbf{r}\}\\&\cdot&\{n\abs{\mathbf{H}}\int K\left(\mathbf{q}\right)\rho_n(\mathbf{H}(\mathbf{p}-\mathbf{q}))d\mathbf{q}\}d\mathbf{p}d\mathbf{u}d\mathbf{t}\cdot\{1+o
		(1)\}.
	\end{eqnarray*}

	Since 	
$$\lim_{n\to\infty}n\abs{\mathbf{H}}\int K(\mathbf{r})\rho_n(\mathbf{H}(\mathbf{p}-\mathbf{r}+\mathbf{u}))d\mathbf{r}=K(\mathbf{p}+\mathbf{u})\rho_c,$$
$$\lim_{n\to\infty}n\abs{\mathbf{H}}\int K(\mathbf{q})\rho_n(\mathbf{H}(\mathbf{p}-\mathbf{q}))d\mathbf{q}=K(\mathbf{p})\rho_c,$$
and taking into account that the functions $K$, $w$ are bounded, and $f$ is bounded away from zero, it follows that 
\begin{equation}
\mathbb{E}(T_{322})=O_p(n^{-1}\abs{\mathbf{H}}^{-1}).
\label{E_t322}
\end{equation}

Then, from (\ref{E_t321}) and (\ref{E_t322}), one gets that
\begin{equation}
\mathbb{E}(T_{32})=4\sigma^4K^{(4)}(0) \rho_c\int \frac{w^2(\mathbf{x})}{f(\mathbf{x})} d\mathbf{x}\cdot\{1+o
			(1)\}+O_p(n^{-1}\abs{\mathbf{H}}^{-1}).
\label{E_t32}
\end{equation}

	Regarding the term $T_{33}$ (when all $i$, $j$, $k$, $l$ are different in (\ref{var_i32_con})), using assumption (A5), it follows that
	\begin{eqnarray*}T_{33}&=&4n^{-2}\abs{\mathbf{H}}\sum_{\text{\scriptsize all different indices  $i$, $j$, $k$, $l$}}W_{ij}W_{kl}\mbox{Cov}({\varepsilon}_i{\varepsilon}_j,{\varepsilon}_k{\varepsilon}_l )\\&=&4n^{-2}\abs{\mathbf{H}}\sum_{\text{\scriptsize all different indices  $i$, $j$, $k$, $l$}}W_{ij}W_{kl}[\mbox{Cov}({\varepsilon}_i,{\varepsilon}_k)\mbox{Cov}({\varepsilon}_j,{\varepsilon}_l)\\&+&\mbox{Cov}({\varepsilon}_i,{\varepsilon}_l)\mbox{Cov}({\varepsilon}_j,{\varepsilon}_k)]\\&=&T_{331}+T_{332},\end{eqnarray*}	
	where
	\begin{eqnarray*}T_{331}&=&4\sigma^4n^{-2}\abs{\mathbf{H}}\sum_{\text{\scriptsize all different indices  $i$, $j$, $k$, $l$}}\int \int \frac{1}{f^2(\mathbf{x})f^2(\mathbf{t})} K_{\mathbf{H}}\left({\mathbf{X}_i-\mathbf{x}}\right)K_{\mathbf{H}}\left({\mathbf{X}_j-\mathbf{x}}\right)\\&\cdot&K_{\mathbf{H}}({\mathbf{X}_k-\mathbf{t}})K_{\mathbf{H}}({\mathbf{X}_l-\mathbf{t}})w(\mathbf{x})w(\mathbf{t})d\mathbf{x}d\mathbf{t}\rho_n(\mathbf{X}_i-\mathbf{X}_k)\rho_n(\mathbf{X}_j-\mathbf{X}_l),\end{eqnarray*}	
	and 
	\begin{eqnarray*}T_{332}&=&4\sigma^4n^{-2}\abs{\mathbf{H}}\sum_{\text{\scriptsize all different indices  $i$, $j$, $k$, $l$}}\int \int \frac{1}{f^2(\mathbf{x})f^2(\mathbf{t})} K_{\mathbf{H}}\left({\mathbf{X}_i-\mathbf{x}}\right)K_{\mathbf{H}}\left({\mathbf{X}_j-\mathbf{x}}\right)\\&\cdot&K_{\mathbf{H}}({\mathbf{X}_k-\mathbf{t}})K_{\mathbf{H}}({\mathbf{X}_l-\mathbf{t}})w(\mathbf{x})w(\mathbf{t})d\mathbf{x}d\mathbf{t}\rho_n(\mathbf{X}_i-\mathbf{X}_l)\rho_n(\mathbf{X}_j-\mathbf{X}_k).\end{eqnarray*}

Using the assumption (A4) and Lemma \ref{L4},
	\begin{eqnarray*}\mathbb{E}(T_{331})&=&4\sigma^4n^{2}\abs{\mathbf{H}}\int \int \frac{1}{f^2(\mathbf{x})f^2(\mathbf{t})}\bigg[	f(\mathbf{x})f(\mathbf{t})\int \int K\left(\mathbf{p}\right)K\left(\mathbf{q}\right)\rho_n(\mathbf{x}-\mathbf{t}+\mathbf{H}(\mathbf{p}-\mathbf{q}))d\mathbf{p}d\mathbf{q}\\&\cdot&\{1+o
		(1)\}\bigg]^2w(\mathbf{x})w(\mathbf{t})d\mathbf{x}d\mathbf{t}\\&=&4\sigma^4n^{2}\abs{\mathbf{H}}\int \int \bigg[\int \int K\left(\mathbf{p}\right)K\left(\mathbf{q}\right)\rho_n(\mathbf{x}-\mathbf{t}+\mathbf{H}(\mathbf{p}-\mathbf{q}))d\mathbf{p}d\mathbf{q}\\&\cdot&\int \int K\left(\mathbf{m}\right)K\left(\mathbf{r}\right)\rho_n(\mathbf{x}-\mathbf{t}+\mathbf{H}(\mathbf{m}-\mathbf{r}))d\mathbf{m}d\mathbf{r}\bigg]w(\mathbf{x})w(\mathbf{t})d\mathbf{x}d\mathbf{t}\cdot\{1+o
		(1)\}\\\nonumber&=&4\sigma^4 n^2\abs{\mathbf{H}}^{}\int \int \int \int \int \int  K\left(\mathbf{p}\right)K\left(\mathbf{q}\right)K\left(\mathbf{m}\right)K\left(\mathbf{r}\right)w(\mathbf{x})w(\mathbf{t})\\&\nonumber\cdot&\rho_n(\mathbf{x}-\mathbf{t}+\mathbf{H}(\mathbf{p}-\mathbf{q}))\rho_n(\mathbf{x}-\mathbf{t}+\mathbf{H}(\mathbf{m}-\mathbf{r}))d\mathbf{p}d\mathbf{q}d\mathbf{m}d\mathbf{r}d\mathbf{x}d\mathbf{t}\cdot\{1+o
		(1)\}\\\nonumber&=&4\sigma^4 n^2\abs{\mathbf{H}}^{2}\int \int \int \int \int \int  K\left(\mathbf{p}\right)K\left(\mathbf{q}\right)K\left(\mathbf{m}\right)K\left(\mathbf{r}\right)w(\mathbf{x})w(\mathbf{x}-\mathbf{Hu})\\&\nonumber\cdot&\rho_n(\mathbf{H}(\mathbf{p}-\mathbf{q}+\mathbf{u}))\rho_n(\mathbf{H}(\mathbf{m}-\mathbf{r}+\mathbf{u}))d\mathbf{p}d\mathbf{q}d\mathbf{m}d\mathbf{r}d\mathbf{x}d\mathbf{u}\cdot\{1+o
		(1)\}\\\nonumber&=&4\sigma^4 \int \int \int \int K\left(\mathbf{q}\right)K\left(\mathbf{r}\right)w^2(\mathbf{x})\{n\abs{\mathbf{H}}^{}\int K\left(\mathbf{m}\right)\rho_n(\mathbf{H}(\mathbf{m}-\mathbf{r}+\mathbf{u}))d\mathbf{m}\}\\&\nonumber\cdot&\{n\abs{\mathbf{H}}^{}\int K\left(\mathbf{p}\right)\rho_n(\mathbf{H}(\mathbf{p}-\mathbf{q}+\mathbf{u}))d\mathbf{p}\} d\mathbf{q}d\mathbf{r}d\mathbf{x}d\mathbf{u}\cdot\{1+o
		(1)\}.
	\end{eqnarray*}

	Since 	
$$\lim_{n\to\infty}n\abs{\mathbf{H}}\int K(\mathbf{p})\rho_n(\mathbf{H}(\mathbf{p}-\mathbf{q}+\mathbf{u}))d\mathbf{p}=K(\mathbf{q}-\mathbf{u})\rho_c,$$
and	
$$\lim_{n\to\infty}n\abs{\mathbf{H}}\int K(\mathbf{m})\rho_n(\mathbf{H}(\mathbf{m}-\mathbf{r}+\mathbf{u}))d\mathbf{m}=K(\mathbf{r}-\mathbf{u})\rho_c,$$ 	it follows that  
	\begin{eqnarray}
	\mathbb{E}(T_{331})\nonumber&=&4\sigma^4\rho^2_c\int \int \int \int K\left(\mathbf{q}\right)K\left(\mathbf{u}-\mathbf{q}\right)K\left(\mathbf{r}\right)K\left(\mathbf{u}-\mathbf{r}\right)w^2(\mathbf{x})d\mathbf{q}d\mathbf{r}d\mathbf{x}d\mathbf{u}\cdot\{1+o
		(1)\}\nonumber\\&=&4\sigma^4\rho^2_c\int \int (K^{(2)}(\mathbf{u}))^2w^2(\mathbf{x})d\mathbf{x}d\mathbf{u}\cdot\{1+o
		(1)\}\nonumber\\&=&4\sigma^4\rho^2_cK^{(4)}(0)\int w^2(\mathbf{x})d\mathbf{x}\cdot\{1+o
		(1)\}.
		\label{E_t331}
		\end{eqnarray}
		
For symmetry, $\mathbb{E}(T_{332})=\mathbb{E}(T_{331})$ and, therefore, using (\ref{E_t331}), it follows that
\begin{equation}
\mathbb{E}(T_{33})=	8\sigma^4K^{(4)}(0)\rho^2_c\int w^2(\mathbf{x})d\mathbf{x}\cdot\{1+o
	(1)\}.	
\label{E_t33}
\end{equation}
		
	So, from (\ref{E_var_i32}), (\ref{E_t31}), (\ref{E_t32}) and (\ref{E_t33}), it is obtained that
\begin{eqnarray}
\mathbb{E}(\mbox{Var}(I_{32}|\mathbf{X}_1,\dots,\mathbf{X}_n))&=&2\sigma^4 K^{(4)}(0)\int  \dfrac{w^2(\mathbf{x})}{f^2(\mathbf{x})}d\mathbf{x}\cdot\{1+o(1)\}+O_p(n^{-1}\abs{\mathbf{H}}^{-1})\nonumber\\&+&4\sigma^4K^{(4)}(0) \rho_c\int \frac{w^2(\mathbf{x})}{f(\mathbf{x})} d\mathbf{x}\cdot\{1+o
	(1)\}+O_p(n^{-1}\abs{\mathbf{H}}^{-1})\nonumber\\&+&8\sigma^4K^{(4)}(0)\rho^2_c\int w^2(\mathbf{x})d\mathbf{x}\cdot\{1+o
	(1)\}.
	\label{E_var_i32b}
	\end{eqnarray}
	
	With respect to the $\mbox{Var}(\mathbb{E}(I_{32}|\mathbf{X}_1,\dots,\mathbf{X}_n))$, the second term in equation (\ref{var_i32}), denoting by
	$$\phi_{ij}=\int \frac{1}{f^2(\mathbf{x})}{K_{\mathbf{H}}\left({\mathbf{X}_i-\mathbf{x}}\right)K_{\mathbf{H}}\left({\mathbf{X}_j-\mathbf{x}}\right)}\rho_n(\mathbf{X}_i-\mathbf{X}_j)w(\mathbf{x})d\mathbf{x},$$and using the expression of the $\mathbb{E}(I_{32}|\mathbf{X}_1,\dots,\mathbf{X}_n)$, given in (\ref{E_i32}), it can be split as:
	\begin{eqnarray}
	\mbox{Var}(\mathbb{E}(I_{32}|\mathbf{X}_1,\dots,\mathbf{X}_n))&=&\mbox{Var}\left(\abs{\mathbf{H}}^{1/2}\sigma^2\dfrac{1}{n}\sum_{i\neq j}\phi_{ij}\right)\nonumber\\&=&4\sigma^4n^{-2}\abs{\mathbf{H}}\sum_{i=1}^{n-1}\sum_{j=i+1}^{n}\sum_{k=1}^{n-1}\sum_{l=k+1}^{n}\mbox{Cov}(\phi_{ij},\phi_{kl}).
	\label{var_e_i32}
	\end{eqnarray}
	
	Now, consider the value of $\mbox{Cov}(\phi_{ij},\phi_{kl})$ according to the following three exclusive cases. 	First, when $i = k$ and $j = l$, the total number of such terms is $n(n-1)/2$. In this case, using Lemma \ref{L5}, one gets
	\begin{eqnarray*}
\mbox{Cov}(\phi_{ij},\phi_{ij})&\le&\mathbb{E}\bigg(\int\int \frac{1}{f^2(\mathbf{x})f^2(\mathbf{t})}K_{\mathbf{H}}\left({\mathbf{X}_i-\mathbf{x}}\right)K_{\mathbf{H}}\left({\mathbf{X}_i-\mathbf{t}}\right)K_{\mathbf{H}}({\mathbf{X}_j-\mathbf{x}})K_{\mathbf{H}}\left(\mathbf{X}_j-\mathbf{t}\right)\\&\cdot&\rho^2_n(\mathbf{X}_i-\mathbf{X}_j)w(\mathbf{x})w(\mathbf{t})d\mathbf{x}d\mathbf{t}\bigg)\\&=&\int\int \frac{1}{f^2(\mathbf{x})f^2(\mathbf{t})}\abs{\mathbf{H}}^{-2}\bigg(f^2(\mathbf{t})\int\int K\left(-\mathbf{p}+\mathbf{H}^{-1}({\mathbf{x}-\mathbf{t}})\right)K\left(-\mathbf{q}+\mathbf{H}^{-1}({\mathbf{x}-\mathbf{t}})\right)\\&\cdot&K\left(\mathbf{p}\right)K\left(\mathbf{q}\right)\rho^2_n(\mathbf{H}(\mathbf{p}-\mathbf{q}))d\mathbf{p}d\mathbf{q}+o(1)\bigg)w(\mathbf{x})w(\mathbf{t})d\mathbf{x}d\mathbf{t}
	\\&=& \abs{\mathbf{H}}^{-1}\int \int \int \int \frac{1}{f^2(\mathbf{x})} K\left(-\mathbf{p}+\mathbf{u}\right)K\left(-\mathbf{q}+\mathbf{u}\right)K\left(\mathbf{p}\right)K\left(\mathbf{q}\right)w(\mathbf{x})w(\mathbf{x}-\mathbf{Hu})\\&\cdot&\rho^2_n(\mathbf{H}(\mathbf{p}-\mathbf{q}))d\mathbf{p}d\mathbf{q}d\mathbf{x}d\mathbf{u}\cdot\{1+o(1)\}\\&\le&\dfrac{K_M^4w_M^2}{f_M^2n\abs{\mathbf{H}}^2}\int \{n\abs{\mathbf{H}}\int \rho^2_n(\mathbf{H}(\mathbf{p}-\mathbf{q}))d\mathbf{p}\}d\mathbf{q}\cdot\{1+o(1)\}.\end{eqnarray*}	
Since 
$$n\abs{\mathbf{H}}\int \rho^2_n(\mathbf{H}(\mathbf{p}-\mathbf{q}))d\mathbf{p}\le n\int \abs{\rho_n(\mathbf{t})}d\mathbf{t}\le C_2,$$
then
\begin{eqnarray}
\mbox{Cov}(\phi_{ij},\phi_{ij})&\le& \dfrac{K_M^4w_M^2}{f_M^2}\dfrac{C_2}{n\abs{\mathbf{H}}^2}.
\label{cov_1}
\end{eqnarray}

Second, when $i = k$ and $j\neq l$ in (\ref{var_e_i32}). In this case, the total number of such terms can be bounded by $n^3$. Using Lemma \ref{L6}, it follows that
\begin{eqnarray*}
\mbox{Cov}(\phi_{ij},\phi_{il})&=&\mathbb{E}(\phi_{ij},\phi_{il})-\mathbb{E}(\phi_{ij})\mathbb{E}(\phi_{il})\\&=&\mathbb{E}(\phi_{ij},\phi_{il})-(\mathbb{E}(\phi_{ij}))^2\\&\le&\mathbb{E}(\phi_{ij},\phi_{il})\\&=&\mathbb{E}\bigg(\int\int \frac{1}{f^2(\mathbf{x})f^2(\mathbf{t})}K_{\mathbf{H}}\left({\mathbf{X}_i-\mathbf{x}}\right)K_{\mathbf{H}}\left({\mathbf{X}_i-\mathbf{t}}\right)K_{\mathbf{H}}({\mathbf{X}_j-\mathbf{x}})K_{\mathbf{H}}\left(\mathbf{X}_l-\mathbf{t}\right)\\&\cdot&\rho_n(\mathbf{X}_i-\mathbf{X}_j)\rho_n(\mathbf{X}_i-\mathbf{X}_l)w(\mathbf{x})w(\mathbf{t})d\mathbf{x}d\mathbf{t}\bigg)\\&=&\int \int \frac{1}{f^2(\mathbf{x})f^2(\mathbf{t})}\abs{\mathbf{H}}^{-1}\bigg(f^2(\mathbf{x})f(\mathbf{t})\int\int \int K(\mathbf{p}) K\left(\mathbf{q}\right)K\left(-\mathbf{p}+\mathbf{H}^{-1}({\mathbf{x}-\mathbf{t}})\right)K\left(\mathbf{r}\right)\\&\nonumber\cdot&\rho_n(\mathbf{H}(\mathbf{p}-\mathbf{q}))\rho_n(\mathbf{x}-\mathbf{t}+\mathbf{H}(\mathbf{p}-\mathbf{r}))d\mathbf{p}d\mathbf{q}d\mathbf{r}\cdot\{1+o
(1)\}\bigg)w(\mathbf{x})w(\mathbf{t})d\mathbf{x}d\mathbf{t}\\&=&\int \int \int \int \int \frac{1}{f(\mathbf{t})} K\left(\mathbf{p}\right)K\left(-\mathbf{p}+\mathbf{u})\right)K\left(\mathbf{q}\right)K\left(\mathbf{r}\right)w(\mathbf{t}+\mathbf{Hu})w(\mathbf{t})\\&\cdot&\rho_n(\mathbf{H}(\mathbf{p}-\mathbf{r}+\mathbf{u}))\rho_n(\mathbf{H}(\mathbf{p}-\mathbf{q}))d\mathbf{p}d\mathbf{q}d\mathbf{r}d\mathbf{u}d\mathbf{t}\cdot\{1+o
(1)\}\nonumber\\&=&n^{-2}\abs{\mathbf{H}}^{-2}\int \int \int \frac{1}{f(\mathbf{t})} K\left(\mathbf{p}\right)K\left(-\mathbf{p}+\mathbf{u}\right)w^2(\mathbf{t})\{n\abs{\mathbf{H}}\int K\left(\mathbf{r}\right)\rho_n(\mathbf{H}(\mathbf{p}-\mathbf{r}+\mathbf{u}))d\mathbf{r}\}\\&\cdot&\{n\abs{\mathbf{H}}\int K\left(\mathbf{q}\right)\rho_n(\mathbf{H}(\mathbf{p}-\mathbf{q}))d\mathbf{q}\}d\mathbf{p}d\mathbf{u}d\mathbf{t}\cdot\{1+o
(1)\}.
\end{eqnarray*}

Since 	
$$\lim_{n\to\infty}n\abs{\mathbf{H}}\int K(\mathbf{r})\rho_n(\mathbf{H}(\mathbf{p}-\mathbf{r}+\mathbf{u}))d\mathbf{r}=K(\mathbf{p}+\mathbf{u})\rho_c,$$
$$\lim_{n\to\infty}n\abs{\mathbf{H}}\int K(\mathbf{q})\rho_n(\mathbf{H}(\mathbf{p}-\mathbf{q}))d\mathbf{q}=K(\mathbf{p})\rho_c,$$
and taking into account that the functions $K$, $w$ are bounded, and $f$ is bounded away from zero, it is obtained that
\begin{equation}
\mbox{Cov}(\phi_{ij},\phi_{il})\le \dfrac{C_3}{ n^{2}\abs{\mathbf{H}}^2}.
\label{cov_2}
\end{equation}

Finally, when $i,j,k,l$ are all distinct in (\ref{var_e_i32}), as $\phi_{ij}$ and $\phi_{kl}$ are independent,
\begin{equation}
\mbox{Cov}(\phi_{ij},\phi_{kl})=0,
\label{cov_3}
\end{equation}

Then, considering (\ref{var_e_i32}), (\ref{cov_1}), (\ref{cov_2}) and (\ref{cov_3}), it follows that
	\begin{eqnarray}
\mbox{Var}(\mathbb{E}(I_{32}|\mathbf{X}_1,\dots,\mathbf{X}_n))&=&\mbox{Var}\left(\abs{\mathbf{H}}^{1/2}\sigma^2\dfrac{1}{n}\sum_{i\neq j}\phi_{ij}\right)\nonumber\\&=&4\sigma^4n^{-2}\abs{\mathbf{H}}\left(\dfrac{n^2-n}{2}\dfrac{C_2}{n\abs{\mathbf{H}}^2}+n^3\dfrac{C_3}{ n^{2}\abs{\mathbf{H}}^2}\right)\nonumber\\&=&O_p(n^{-1}\abs{\mathbf{H}}^{-1}).
\label{var_e_i32b}
\end{eqnarray}

Now, from (\ref{var_i32}), (\ref{E_var_i32b}) and (\ref{var_e_i32b}), the leading term of the variance of $I_{32}$ is given by:
	\begin{eqnarray}\label{vart3}	V&=&2\sigma^4 K^{(4)}(0)\bigg[\int \dfrac{w^2(\mathbf{x})}{f^2(\mathbf{x})}d\mathbf{x}+2\rho_{c}\int \dfrac{w^2(\mathbf{x})}{f(\mathbf{x})}d\mathbf{x}+4\rho^2_{c}\int {w^2(\mathbf{x})}d\mathbf{x}\bigg].
	\end{eqnarray}
	 
Therefore, using the central limit theorem for degenerate reduced U-statistics under $\alpha-$mixing conditions, given by \cite{kim2013central}, it is obtained that the term $I_{32}$ converges, in distribution, to a normal distribution with mean the leading term of (\ref{biasT3}) and variance given by (\ref{vart3}).

On the other hand, in virtue of the Cauchy-Schwarz inequality, the  cross terms in $T_{n1}$ resulting from the products of $I_1(\mathbf{x})$, $I_2(\mathbf{x})$ and $I_3(\mathbf{x})$ are all of smaller order. Therefore, combining the results in (\ref{i_1}), (\ref{b1H}) and (\ref{b0H}), and the asymptotic normality of $I_{32}$ (with bias the leading term of (\ref{biasI32}) and variance (\ref{vart3})), one gets  
	\begin{equation}\label{Tn1}
	V^{-1/2}(T_{n1}-b_{0\mathbf{H}}-b_{1\mathbf{H}})\to_{\mathcal{L}} N(0,1) \text{ as } n\to\infty,
	\end{equation}
	where 
	\begin{eqnarray*}
	b_{0\mathbf{H}}&=& \abs{\mathbf{H}}^{-1/2}\sigma^2K^{(2)}(\bm{0})\bigg[\int \dfrac{w(\mathbf{x})}{f(\mathbf{x})}d\mathbf{x}+\rho_{c}\int {w(\mathbf{x})}d\mathbf{x}\bigg],\\
	b_{1\mathbf{H}}&=& \int (K_{\mathbf{H}}\ast g(\mathbf{x}))^2w(\mathbf{x})d\mathbf{x},\\
\end{eqnarray*}
and
$$
V=2\sigma^4 K^{(4)}(0)\bigg[\int \dfrac{w^2(\mathbf{x})}{f^2(\mathbf{x})}d\mathbf{x}+2\rho_{c}\int \dfrac{w^2(\mathbf{x})}{f(\mathbf{x})}d\mathbf{x}+4\rho^2_{c}\int {w^2(\mathbf{x})}d\mathbf{x}\bigg].
$$

	The term $T_{n2}$ in $T_n$ is of smaller order than $T_{n1}$ (specifically, $T_{n2}=O_p(\tr(\mathbf{H}^2)T_{n1})$), and by the Cauchy-Schwarz inequality, the cross term $T_{n12}$ is of smaller order as well. Therefore, from (\ref{Tn}), it follows that
	$$T_n=T_{n1}+O_p(tr(\mathbf{H}^2))+O_p(n^{-2/(4+d)+\eta}).$$
Taking into account (\ref{Tn1}), it follows that
			\begin{equation*}\label{T1}
	V^{-1/2}(T_n-b_{0\mathbf{H}}-b_{1\mathbf{H}})\to_{\mathcal{L}} N(0,1) \text{ as } n\to\infty,
	\end{equation*}
	with $b_{0\mathbf{H}}$, $b_{1\mathbf{H}}$ and $V$ given above.
		\end{proof}

\section*{Appendix B. Additional simulations results}
In this appendix, additional simulations complementing the study presented in Section 4 are presented. It is organized as follows. First, the asymptotic distribution of the test is illustrated with a particular example. The next subsections present an extension of the simulation results, considering the use of non-scalar bandwidth matrices, employing a different regression function, assuming a random design, and including a nugget effect in dependence structure.

\subsection*{B.1. Asymptotic distribution of the test}
Asymptotic distribution of test statistics are usually employed for test calibration in practice. However, the convergence of $T_n$ to its limit distribution, as it happens with other smooth-based test, is too slow. This issue is pointed out in Section 3.2: the asymptotic distribution obtained in Theorem 1 could not be sufficiently precise when the sample size is small or medium. This was also noted in other nonparametric testing contexts \citep[see][for example]{hardle1993comparing}. Moreover, the limit distribution of the test statistic depends on unknown quantities such as the design density and the error variance that, in a practical situation, must be estimated from the data. For these reasons, resampling methods are considered as an alternative to the asymptotic distribution. As previously shown, the bootstrap approach designed to be used in this context provides satisfactory results. Nevertheless, and for the sake of illustration, in this section, a brief simulation experiment is presented to study the performance of the asymptotic distribution of the test under the null hypothesis. Specifically, we consider the simple case of assuming $f$ and $\sigma^2$ known, and the density estimator of  $V^{-1/2}(T_n-b_{0\mathbf{H}})$ and the standard normal density function are compared.

  A linear parametric regression family is chosen,
$
m_{\bm{\beta}}(X_1,X_2)=\beta_0+\beta_1 X_1+\beta_2 X_2,
$ being $\mathbf{X}=(X_1,X_2)$, and the regression function considered is:
\begin{equation}
m(X_1,X_2)=2+X_1+X_2.
\label{trend_sim}
\end{equation}
500 samples of sizes $n=400$, $2500$ and $10000$ are generated from a regression model with explanatory variables drawn from  a bivariate uniform distribution in the unit square, regression function (\ref{trend_sim}), and 
random errors $\varepsilon_i$ normally distributed with zero mean and with isotropic exponential covariance function:  
\begin{equation}
	\mbox{Cov}({\varepsilon}_i,{\varepsilon}_j)=   
	\sigma^2\{\exp(-\lambda n\norm{\mathbf{X}_i-\mathbf{X}_j})\},
	\label{expo2}
	\end{equation}
	with values of $\sigma^2=0.4$ and  $\lambda=0.0005$. Note that with this selection $\lambda$, the values for the practical range are 5, 0.8 and 0.2, for $n=400$, $2500$ and $10000$, respectively.  The parametric fit  was computed using the iterative least squares procedure described in Section 2.2, considering a linear model. The nonparametric fit was obtained using the multivariate local linear estimator with  a multivariate Gaussian kernel and a scalar bandwidth matrix.  With this kernel, the quantities $K^{(2)}(0)$ and $K^{(4)}(0)$  in the asymptotic bias and variance of $T_n$ can be easily calculated. Additionally, considering (\ref{expo2}), it is straightforward to prove that $\rho_c=1/\lambda$.  For simplicity, we also take $w(\mathbf{x})=f(\mathbf{x})$,  $\forall\mathbf{x}\in D\in\mathbb{R}^d$. For each sample and in every scenario, the statistic $V^{-1/2}(T_n-b_{0\mathbf{H}})$ is computed.
	
\begin{figure*}[htb]
	\centering
		\includegraphics[width=0.45\textwidth]{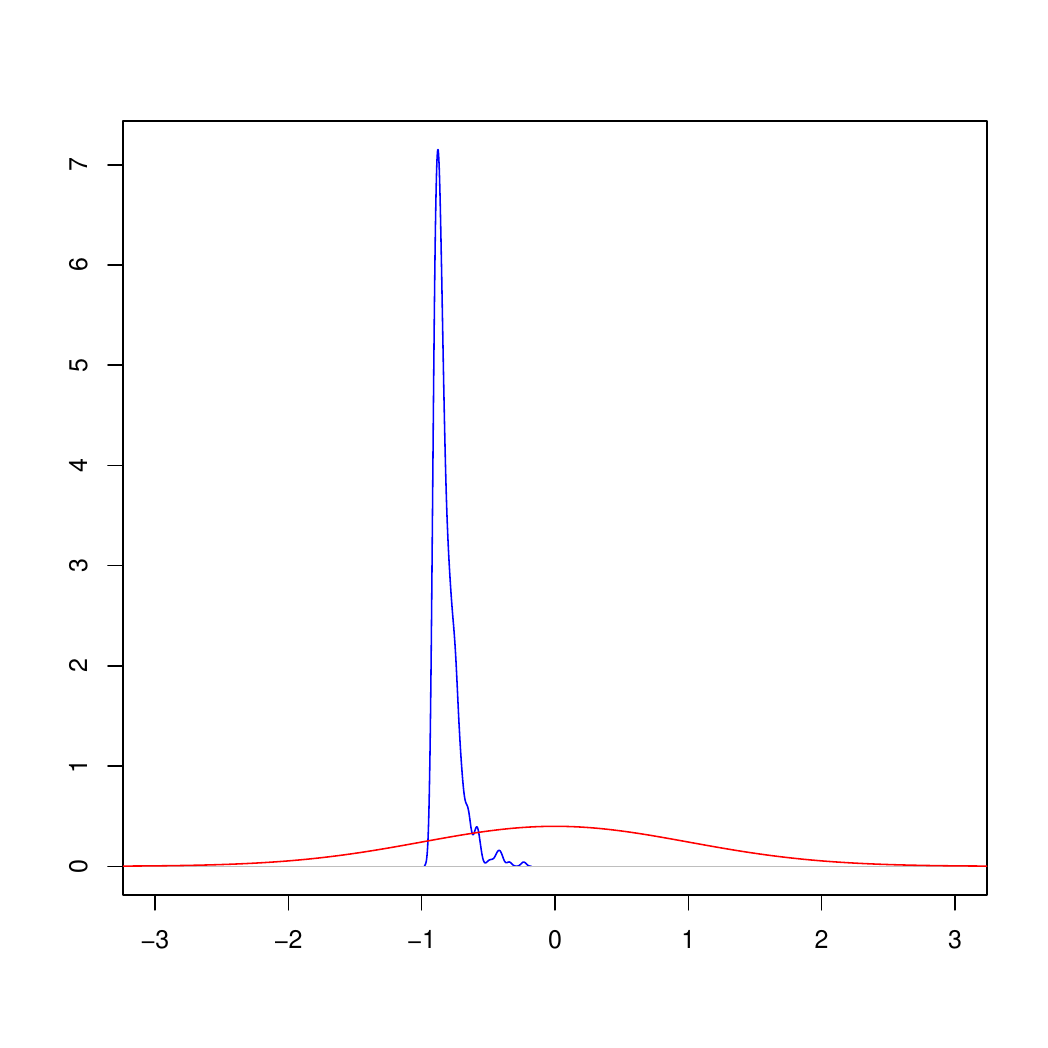}\hspace{0.5cm}
			\includegraphics[width=0.45\textwidth]{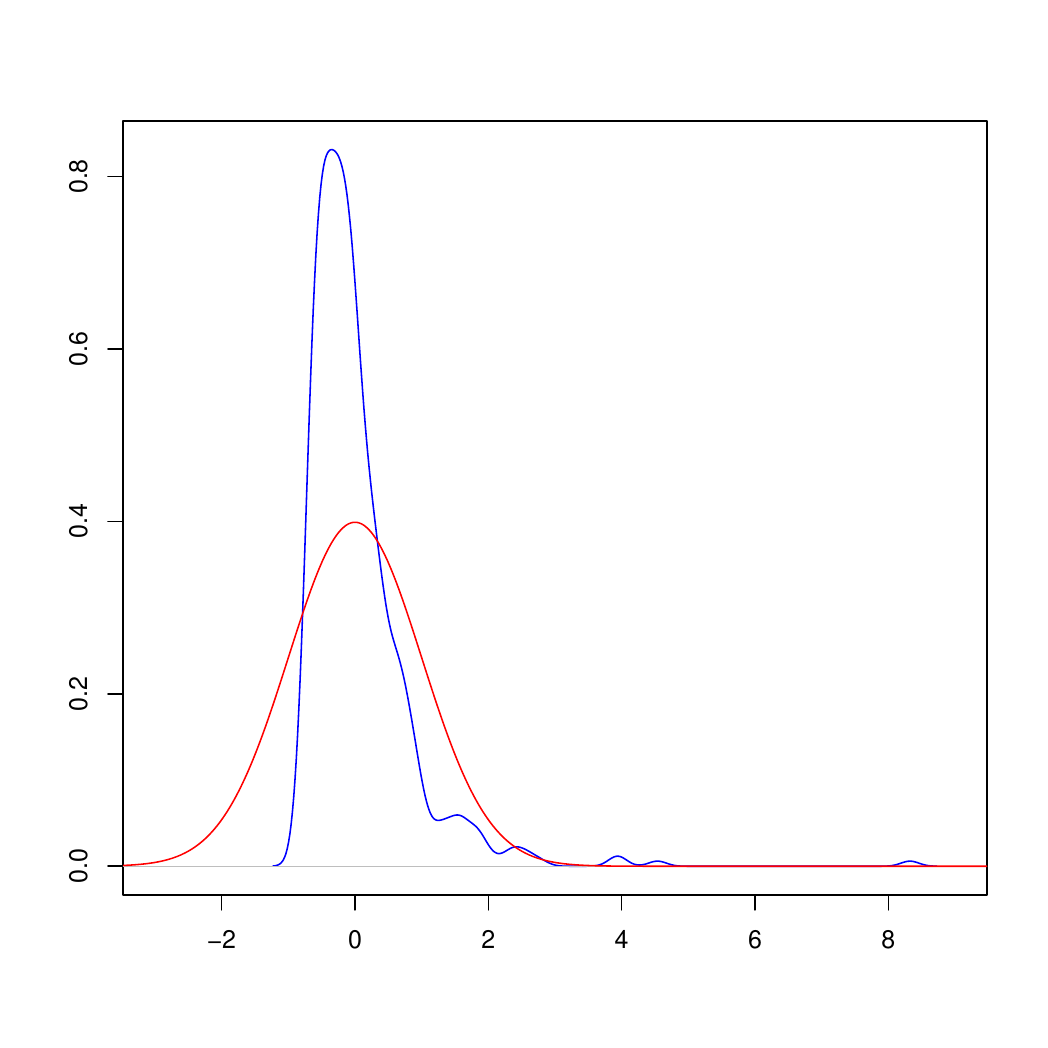}
	\caption{Density estimates of  $V^{-1/2}(T_n-b_{0\mathbf{H}})$ (blue lines) and  normal standard densities (red lines),  considering $n=2500$ (left panel) and $n=10000$ (right panel).}
	\label{asy}
\end{figure*}

Figure \ref{asy} shows  density estimates of  $V^{-1/2}(T_n-b_{0\mathbf{H}})$ (blue lines), computed with a Gaussian kernel and the rule-of-thumb bandwidth selector, and the  standard normal densities (red lines). The plot in the left panel corresponds to $n=2500$ and the one in the right panel to $n=10000$. When $n=400$, the asymptotic distribution of $V^{-1/2}(T_n-b_{0\mathbf{H}})$  is very far from the standard normal distribution and it is not shown here.  Only when the sample size is very large, the sampling distribution of the test statistic seems to approximate rea\-so\-na\-bly well the Gaussian limit distribution. It is expected that this approximation will be better for larger sample sizes. That means that to obtain reliable results with the asymptotic distribution of the test, it would be necessary to consider a huge sample size (ignoring $f$ and $\sigma^2$, which should be estimated). In this situation, the application of the test will take an enormous computing time.
In such scenarios, the use of binning techniques or big data methods could be of special interest to accelerate the running time when applying the test. These approaches are out of the scope of the present paper, but can be an interesting issue of research in future.

\subsection*{B.2. Non-scalar bandwidths} 
This section contains additional simulations similar to those presented in Section 4, but taking a different type of bandwidth matrices to compute the nonparametric estimation of the regression function. While in Section 4, scalar matrix bandwidths (diagonal matrix  with equal values in the main diagonal)  were considered, here, diagonal bandwidths with different elements are used. A linear model $m_{\bm{\beta}}(X_1,X_2)=\beta_0+\beta_1 X_1+\beta_2 X_2$ is chosen, and for different values of $c$ (specifically, 0, 3 and 5) the regression function
\begin{equation}
m(X_1,X_2)=2+X_1+X_2+cX_1^3
\label{trend_xim}
\end{equation}
is considered. For each value of $c$, 500 samples of sizes  $n=225$ and $400$ are generated on a bidimensional regular grid in the unit square, with regression function (\ref{trend_xim}) and 
random errors $\varepsilon_i$ normally distributed with  zero mean and isotropic exponential covariance function:  
\begin{equation}
\mbox{Cov}({\varepsilon}_i,{\varepsilon}_j)=   
\sigma^2\{\exp(-\norm{\mathbf{X}_i-\mathbf{X}_j}/a_e)\},
\label{expo}
\end{equation}
with $\sigma=0.4$, $0.6$, and $0.8$. Different values of parameter $a_e$ are considered: $a_e=0.1$ (weak correlation), $a_e=0.2$ (medium correlation) and $a_e=0.4$ (strong correlation).  No nugget effect is considered in this scenario.

\begin{figure*}[htb]
	\centering
	\includegraphics[width=0.9\textwidth]{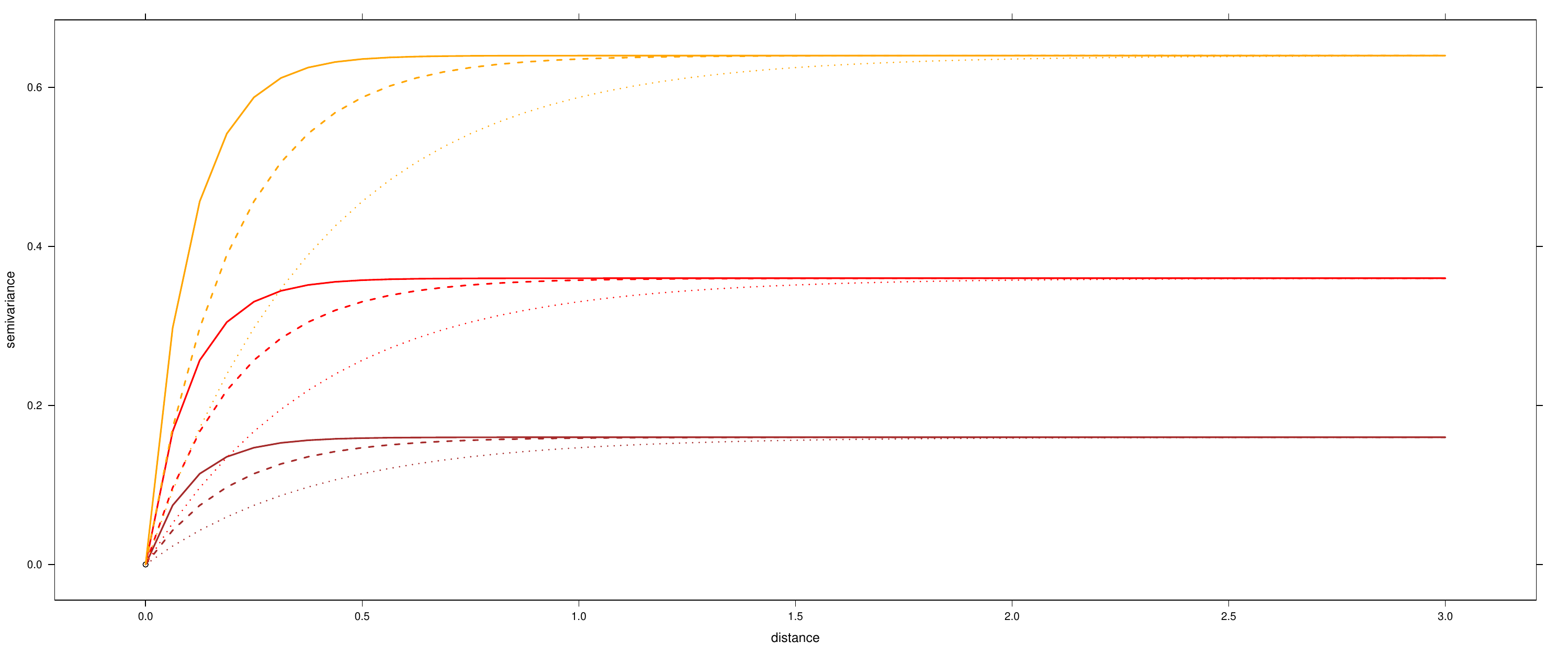}\hspace{0.5cm}
	\caption{Exponential variogram models for the simulation scenario trying different bandwidth matrices.}
	\label{vario}
\end{figure*}

Figure \ref{vario} shows the different exponential variogram models considered (brown lines for $\sigma=0.4$, red lines for $\sigma=0.6$, and orange lines for $\sigma=0.8$. For each value of $\sigma$, solid, dashed and dotted lines for $a_e=0.1,0.2$ and $0.4$, respectively).

\begin{figure*}[htb]
	\centering
	\includegraphics[width=0.48\textwidth]{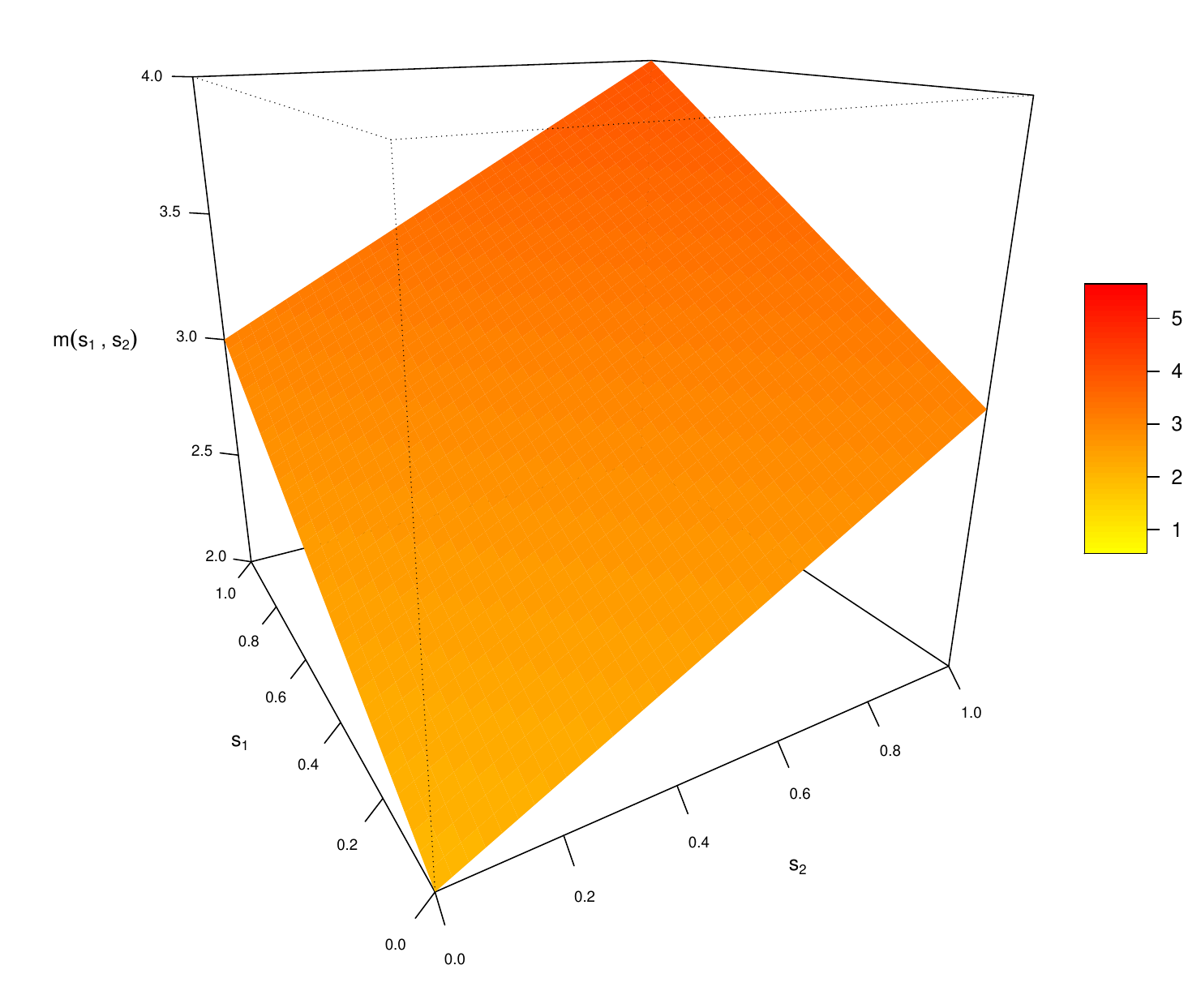}\hspace{0.5cm}
	\includegraphics[width=0.48\textwidth]{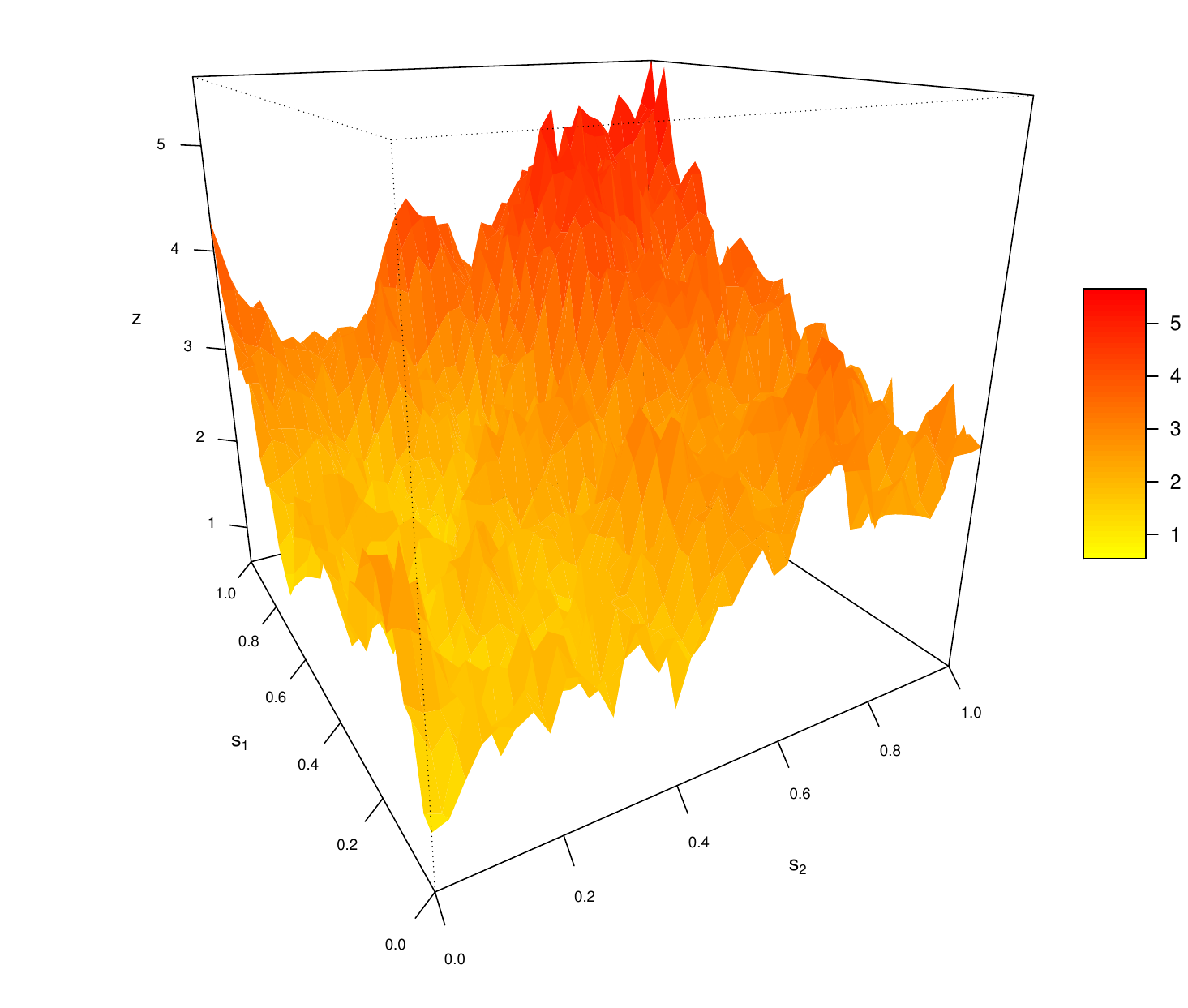}
	\caption{Regression model (\ref{trend_xim}) for $c=0$ (left panel) and a realization of the spatial process  (right panel). The dependence structure of the errors is explained by an exponential covariogram  with parameters $\sigma=0.4$ and $a_e=0.2$. }
	\label{surface}
\end{figure*}

Figure \ref{surface}  shows, for $c=0$, in the left panel, the regression function function (\ref{trend_xim}) and, in the right panel, a simulated spatial process, considering $\sigma=0.6$ and $a_e=0.2$ in (\ref{expo}).

\begin{figure*}[htb]
	\centering
	\includegraphics[width=0.48\textwidth]{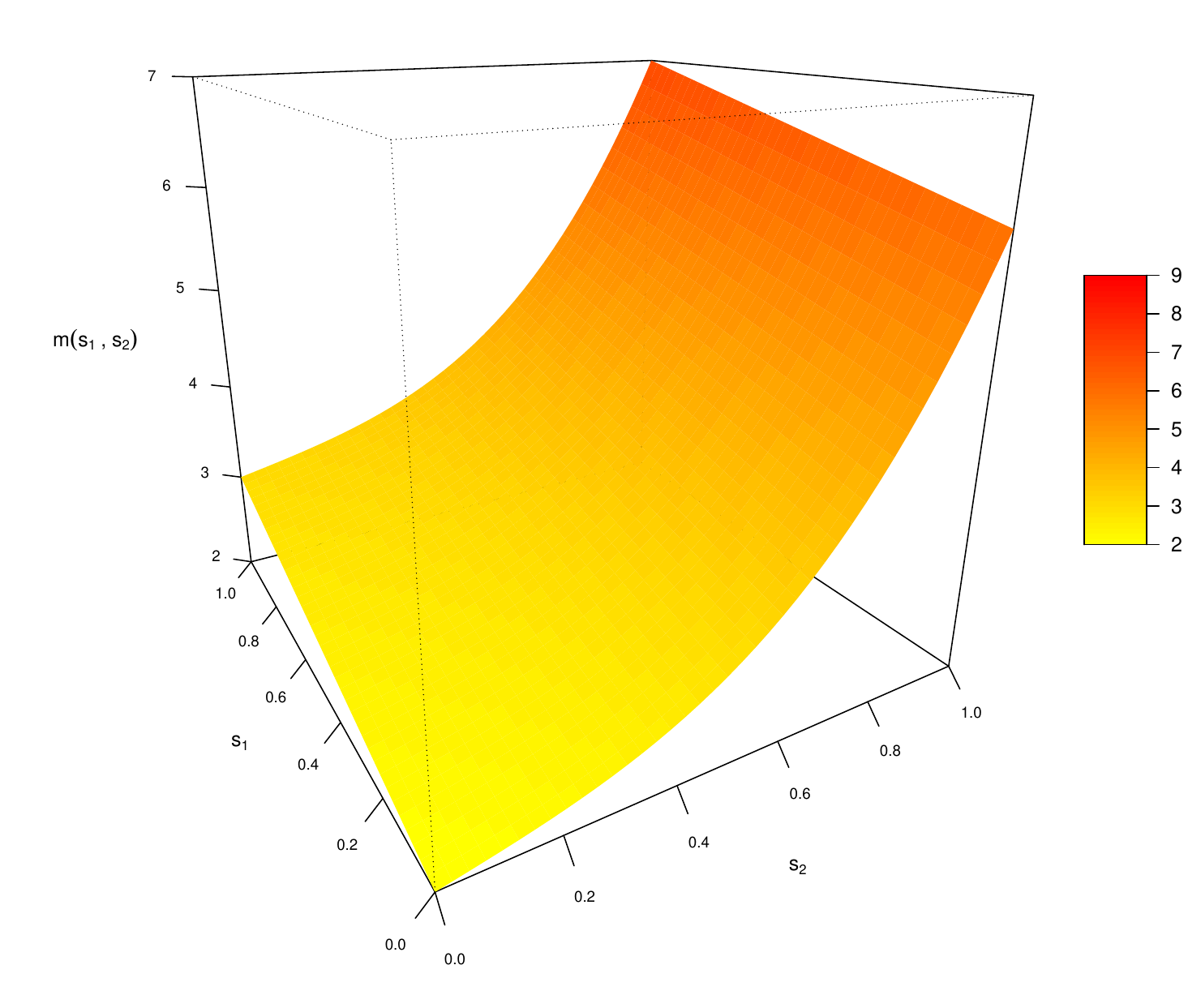}\hspace{0.5cm}
	\includegraphics[width=0.48\textwidth]{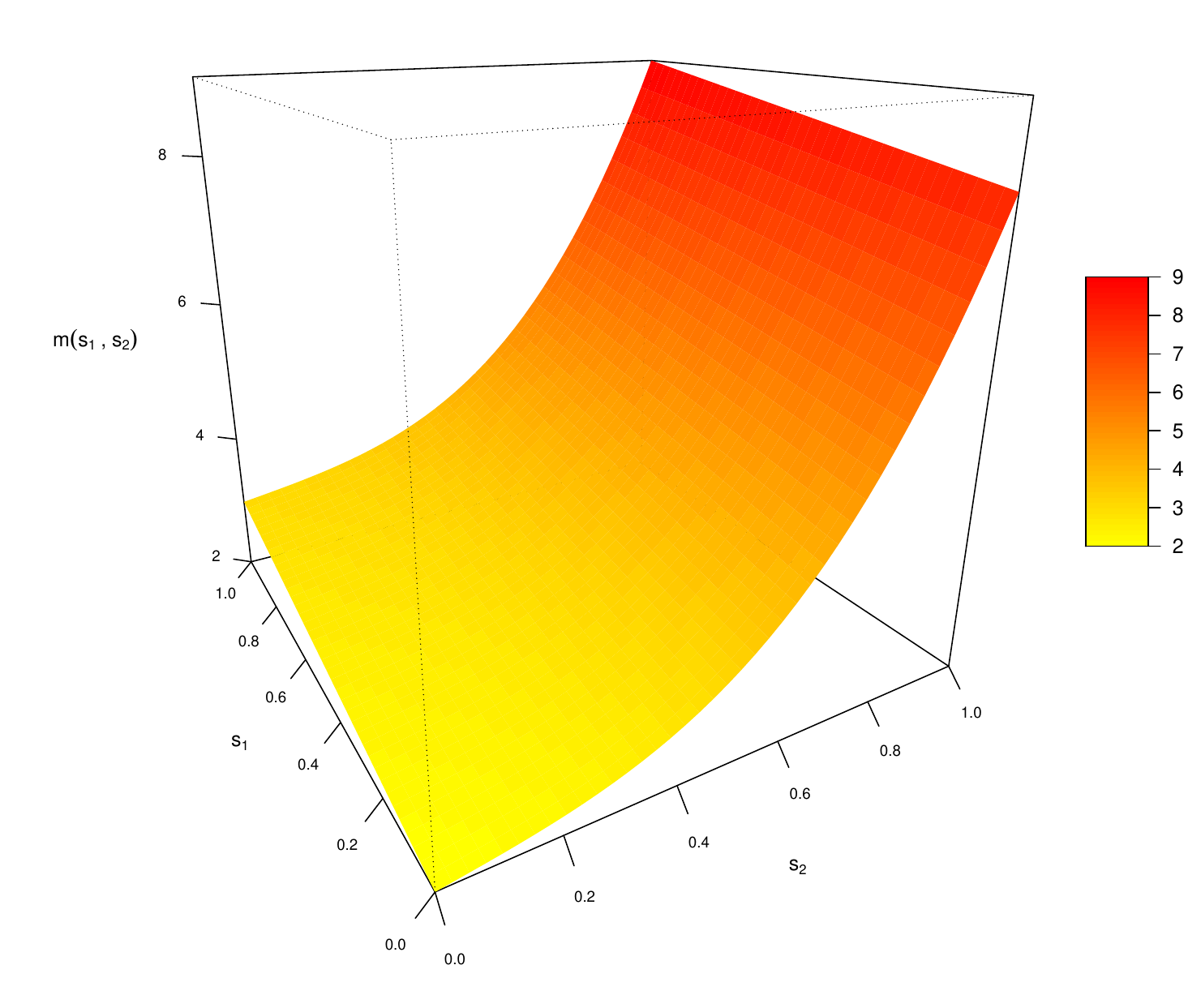}
	\caption{Regression model (\ref{trend_xim}) for   $c=3$ (left panel)  and  $c=5$ (right panel).}
	\label{trend2}
\end{figure*}

The regression functions, using (\ref{trend_xim}), for $c=3$ (left panel) and for $c=5$ (right panel) are shown in Figure \ref{trend2}.

The bootstrap procedure described in Section 3.2 was applied, using  $B=500$ replicates. The weight function  was taken constant with value 1. The parametric fit used for constructing (\ref{statistic}) was computed using the iterative least squares procedure, considering a linear model, while the nonparametric fit was obtained using the multivariate local linear estimator estimator with a multiplicative triweight kernel. The bandwidth is taken as a diagonal matrix $\mathbf{H}=\text{diag}(h_1,h_2)$, being the values of $h_1$ and $h_2$ different.

Results are presented in Table \ref{m1db}, where the rejection proportions of the null hypothesis, for $\alpha=0.05$, are displayed. Similarly to the results shown in Section 4, it can be observed that the test has a reasonable behavior. In particular, if $c = 0$ (under the null hypothesis) the rejection proportions are similar to the
theoretical level,  for the different values of $h_1$ and $h_2$ considered. For the alternative hypothesis ($c=3$ and $c=5$),  the power of the test becomes larger as the value of $c$ increases.  On the other hand, the power of the test decreases with the point variance $\sigma^2$. In all  scenarios, it can be seen that the rejection proportions depend on the bandwidth $\mathbf{H}$, especially, under the alternative hypothesis.

For example, for a $15\times 15$ grid, with $\sigma=0.4$ and $a_e=0.2$, it follows that, under the null hypothesis, the rejection proportions obtained are not significantly different from the theoretical level, considering both bandwidth matrices $\mathbf{H}=\text{diag}(1,0.6)$ and $\mathbf{H}=\text{diag}(0.6,1)$. However, the power of the test shows a different behavior. It is 
significantly larger when $\mathbf{H}=\text{diag}(0.6,1)$ is considered. Then, under the alternative hypothesis, the rejection proportion depends on the values of $h_1$ and $h_2$. Note that, a comparison between Table \ref{Simu} and Table \ref{m1db} reveals that there are not relevant differences in terms of rejection proportions if $\mathbf{H}=\text{diag}(h,h)$ or $\mathbf{H}=\text{diag}(h_1,h_2)$ (with $h_1\neq h_2$) is considered, for this particular scenario. 

\subsection*{B.3. Alternative regression function}

The second framework considered is similar to the previous regression scenario, but with mean function 
\begin{equation}
m(X_1,X_2)=3+2X_1+X_2+cx_{1}^3.
\label{alt_reg}
\end{equation}
The errors of the model are also normally distributed with an exponential dependence structure, and the same parameters for $c$, $\sigma$, $a_e$, $B$, and $n$ as in the previous framework are considered in this case. Table \ref{m3eb} shows the rejection proportions of the null hypothesis, for $\alpha=0.05$, considering that the bandwidth is taken as a diagonal matrix $\mathbf{H}=\text{diag}(h,h)$, and different  values of $h$ are chosen, $h=0.6, 0.7, 0.8, 0.9, 1.$ Table \ref{m3db} shows the results when the bandwidth is taken as a diagonal matrix $\mathbf{H}=\text{diag}(h_1,h_2)$, being the values of $h_1$ and $h_2$ different. It can be observed that considering different regression parameters ($\beta_0=3$, $\beta_1=2$ for the first coordinate and $\beta_2=1$ for the second one), the rejection proportions (under the null and the alternative hypothesis) are really similar to those obtained in the first setting (where $\beta_0=2$ and $\beta_1=\beta_2=1$) and analogous conclusions can be deduced.

\begin{table}[H]
	\scriptsize
	\centering
	\begin{tabular}{ccccccccccc}
		&&	&&&&	& &$\mathbf{H}$&&  \\

		\cline{6-11}
		
		\vspace{0.2cm}	
		$\sigma$	&$a_e$&$c$&$n$&&$\begin{pmatrix}  0.8 & 0 \\ 0 &  0.6 \end{pmatrix}$&$\begin{pmatrix}  1 & 0 \\ 0 &  0.6 \end{pmatrix}$&$\begin{pmatrix}  0.6 & 0 \\ 0 &  0.8 \end{pmatrix}$&$\begin{pmatrix}  1 & 0 \\ 0 &  0.8 \end{pmatrix}$&$\begin{pmatrix}  0.6 & 0 \\ 0 &  1 \end{pmatrix}$&$\begin{pmatrix}  0.8 & 0 \\ 0 &  1 \end{pmatrix}$    \\		
		\hline
		$0.4$ &0.1&0&225&&0.074& 0.054& 0.066& 0.038& 0.052& 0.036 \\
		&&&400&&  0.034 &0.028& 0.032& 0.022& 0.028& 0.024 \\  
		$0.4$ &0.1&3&225&&0.394 &0.356 &0.576 &0.416 &0.592 &0.478 \\
		&&&400&& 0.298 &0.236& 0.502& 0.322& 0.530& 0.404  \\			
		$0.4$ &0.1&5&225&&0.998 &0.994& 1.000& 0.998& 1.000& 1.000 \\
		&&&400&& 0.998& 0.994& 1.000& 0.998& 1.000& 1.000  \\
		\hline
		$0.4$ &0.2&0&225&& 0.062& 0.050& 0.060& 0.036& 0.050 &0.036  \\
		&&&400&& 0.054 &0.038& 0.054& 0.024& 0.038& 0.022
		\\  
		$0.4$ &0.2&3&225&&0.780& 0.726& 0.870& 0.786& 0.876& 0.822  \\
		&&&400&& 0.796& 0.726& 0.912& 0.772& 0.914 &0.846  \\			
		$0.4$ &0.2&5&225&& 1.000& 1.000&1.000& 1.000&1.000& 1.000  \\
		&&&400&& 0.998 &1.000& 1.000& 1.000& 1.000& 1.000  \\
		\hline
		$0.4$ &0.4&0&225&&0.126 &0.106& 0.132& 0.070& 0.092& 0.080\\
		&&&400&&  0.200& 0.164& 0.126& 0.098& 0.076& 0.058  \\  
		$0.4$ &0.4&3&225&&0.978& 0.970& 0.988& 0.978& 0.990& 0.984   \\
		&&&400&&0.980 &0.974& 0.992& 0.984& 0.992& 0.988\\			
		$0.4$ &0.4&5&225&&1.000& 1.000&1.000& 1.000&1.000& 1.000  \\
		&&&400&& 1.000& 1.000&1.000& 1.000&1.000& 1.000  \\
		\hline	
		$0.6$ &0.1&0&225&& 0.074 &0.062 &0.070 &0.040 &0.056 &0.036 \\
		&&&400&&0.034 &0.028& 0.032& 0.022& 0.028& 0.022 \\  
		$0.6$ &0.1&3&225&&0.060 &0.050 &0.154 &0.062 &0.166 &0.086  \\
		&&&400&&  0.026 &0.016& 0.106& 0.032& 0.106& 0.044 \\			
		$0.6$ &0.1&5&225&&0.552& 0.508 &0.740& 0.562& 0.756& 0.656   \\
		&&&400&&  0.476& 0.434& 0.700& 0.494& 0.708& 0.570 \\
		\hline
		$0.6$ &0.2&0&225&& 0.062 &0.050& 0.062& 0.036& 0.050& 0.036   \\
		&&&400&& 0.054 &0.038& 0.054& 0.024& 0.036& 0.022   \\  
		$0.6$ &0.2&3&225&&0.304& 0.230& 0.498& 0.286& 0.506& 0.382  \\
		&&&400&&0.322 &0.268& 0.502& 0.314& 0.504& 0.380   \\			
		$0.6$ &0.2&5&225&&0.888 &0.860& 0.944& 0.882& 0.942& 0.924  \\
		&&&400&&0.888 &0.858& 0.938& 0.880& 0.940& 0.908  \\
		\hline
		$0.6$ &0.4&0&225&& 0.204 &0.172& 0.128& 0.096& 0.076& 0.058  \\
		&&&400&& 0.198& 0.164& 0.126& 0.100& 0.076& 0.058 \\  
		$0.6$ &0.4&3&225&&0.734 &0.658 &0.844 &0.700 &0.838 &0.752\\
		&&&400&& 0.718 &0.656& 0.830&0.698& 0.822& 0.752  \\			
		$0.6$ &0.4&5&225&& 0.996 &0.996& 0.998& 0.996& 0.998& 0.998  \\
		&&&400&&  0.994& 0.994& 0.996& 0.994& 0.996& 0.994 \\
		\hline
		$0.8$ & 0.1&0&225&& 0.072& 0.060& 0.068& 0.038& 0.052& 0.036 \\ 
		&&&400&&0.034& 0.028& 0.032& 0.022& 0.028& 0.022 \\  
		$0.8$ & 0.1&3&225&& 0.034 &0.024& 0.086& 0.026& 0.096& 0.042   \\
		&&&400&&  0.008& 0.006& 0.054& 0.004& 0.052& 0.010 \\			
		$0.8$ & 0.1&5&225&& 0.154 &0.132& 0.328& 0.180& 0.360& 0.230   \\
		&&&400&& 0.086 &0.070& 0.240& 0.096& 0.248& 0.162 \\
		\hline
		$0.8$ & 0.2&0&225&& 0.064 &0.050& 0.062& 0.036& 0.050& 0.036 \\ 
		&&&400&& 0.052 &0.038& 0.050& 0.024& 0.036& 0.022 \\  
		$0.8$ & 0.2&3&225&&0.144 &0.112& 0.280& 0.122& 0.268& 0.166  \\
		&&&400&& 0.158 &0.118 &0.276& 0.134& 0.270& 0.168  \\			
		$0.8$ & 0.2&5&225&&0.534& 0.468& 0.710& 0.524& 0.710& 0.608   \\
		&&&400&&0.556 &0.472 &0.722 &0.546 &0.722 &0.628 \\
		\hline
		$0.8$ & 0.4&0&225&& 0.126 &0.110& 0.134& 0.068& 0.094& 0.082  \\ 
		&&&400&& 0.196 &0.164& 0.126& 0.100& 0.074& 0.058 \\  
		$0.8$ & 0.4&3&225&&0.462 &0.390 &0.610& 0.414& 0.596& 0.496  \\
		&&&400&& 0.472& 0.414& 0.598& 0.426& 0.592& 0.480 \\			
		$0.8$ & 0.4&5&225&& 0.902& 0.880& 0.956& 0.898& 0.956& 0.922  \\
		&&&400&& 0.914 &0.868& 0.960& 0.906& 0.958& 0.930 \\
		\hline
	\end{tabular}
	\vspace{0.3cm}
	\caption{ \small Rejection proportions of the null hypothesis for $\alpha=0.05$. Non-scalar bandwidths.}
\label{m1db}
\end{table}

\begin{table}[H]
	\scriptsize
	\centering
	\begin{tabular}{cccccccccc}
		&&	&&&&	 &$h$&&  \\
		\cline{6-10}
		$\sigma$	&$a_e$&$c$&$n$&&0.6&$0.7$&0.8&0.9&1    \\
		\hline
		$0.4$ &0.1&0&225&& 0.060& 0.042& 0.042& 0.030& 0.024\\
		&&&400&&  0.044& 0.040& 0.030& 0.022& 0.016\\  
		$0.4$ &0.1&3&225&& 0.454& 0.408& 0.394 &0.392& 0.398 \\
		&&&400& & 0.420 &0.368& 0.324& 0.316& 0.324 \\			
		$0.4$ &0.1&5&225&&1.000 &0.998& 0.998& 0.998& 0.998  \\
		&&&400&& 1.000& 0.998& 0.998& 0.996& 0.994 \\
		\hline
		$0.4$ &0.2&0&225&& 0.086 &0.058& 0.048& 0.032& 0.024  \\
		&&&400&&0.104 &0.072& 0.034& 0.024& 0.020 \\  
		$0.4$ &0.2&3&225&&0.852 &0.838& 0.809 &0.794& 0.790 \\
		&&&400& &0.886& 0.862& 0.834& 0.822& 0.816\\			
		$0.4$ &0.2&5&225&&1.000&1.000&1.000&1.000&1.000  \\
		&&&400&&1.000&1.000&1.000&1.000&1.000 \\
		\hline
		$0.4$ &0.4&0&225&&0.164 &0.126& 0.084& 0.074& 0.068 \\
		&&&400&& 0.172 &0.130& 0.096& 0.080& 0.064 \\  
		$0.4$ &0.4&3&225&&0.978 &0.976& 0.974& 0.970& 0.970  \\
		&&&400&&0.994& 0.994& 0.992& 0.992& 0.988 \\			
		$0.4$ &0.4&5&225&&1.000&1.000&1.000&1.000&1.000 \\
		&&&400&& 1.000&1.000&1.000&1.000&1.000 \\
		\hline	
		$0.6$ &0.1&0&225&&0.060& 0.042& 0.042& 0.034& 0.024\\
		&&&400&&  0.044 &0.038& 0.030& 0.022& 0.016 \\  
		$0.6$ &0.1&3&225& &0.096& 0.084& 0.062& 0.056& 0.066\\
		&&&400& &0.072 &0.046 &0.036& 0.030& 0.030\\			
		$0.6$ &0.1&5&225&& 0.640& 0.606& 0.574& 0.558& 0.568 \\
		&&&400&& 0.604& 0.554& 0.532& 0.508& 0.522\\
		\hline
		$0.6$ &0.2&0&225&&  0.086& 0.060& 0.048& 0.032& 0.026   \\
		&&&400& &0.104& 0.066& 0.032& 0.022& 0.016  \\  
		$0.6$ &0.2&3&225&&0.418& 0.362& 0.314& 0.282& 0.272 \\
		&&&400& &0.460 &0.400& 0.346& 0.306& 0.300 \\			
		$0.6$ &0.2&5&225&&0.938 &0.920& 0.916& 0.894& 0.890  \\
		&&&400& &0.944 &0.942& 0.926& 0.916& 0.916\\
		\hline
		$0.6$ &0.4&0&225&&0.158& 0.126& 0.084& 0.074& 0.068\\
		&&&400&&  0.168& 0.124& 0.090& 0.076& 0.058\\  
		$0.6$ &0.4&3&225& &0.766& 0.742& 0.716& 0.694& 0.684 \\
		&&&400&& 0.810 &0.776& 0.740& 0.716& 0.708 \\			
		$0.6$ &0.4&5&225& &1.000 &0.998& 0.998& 0.998& 0.998\\
		&&&400&&  1.000 &1.000& 0.996& 0.996& 0.996 \\
		\hline
		$0.8$ & 0.1&0&225&&0.060& 0.040& 0.042& 0.036& 0.026  \\ 
		&&&400& &0.044& 0.038& 0.030& 0.022& 0.016\\  
		$0.8$ & 0.1&3&225&& 0.062& 0.028& 0.016& 0.012& 0.018  \\
		&&&400& &0.040 &0.024& 0.014& 0.010& 0.010 \\			
		$0.8$ & 0.1&5&225&& 0.218 &0.196& 0.180& 0.172& 0.166 \\
		&&&400&& 0.164& 0.128& 0.102& 0.110& 0.124\\
		\hline
		$0.8$ & 0.2&0&225&&0.086& 0.060& 0.048& 0.034& 0.026 \\ 
		&&&400&& 0.104& 0.064& 0.030& 0.022& 0.016 \\  
		$0.8$ & 0.2&3&225&& 0.234& 0.180& 0.1386& 0.118& 0.116  \\
		&&&400&&0.278 &0.222& 0.176& 0.140& 0.136 \\			
		$0.8$ & 0.2&5&225&& 0.654& 0.612& 0.578& 0.560& 0.550 \\
		&&&400&& 0.698& 0.644& 0.586& 0.568& 0.560\\
		\hline
		$0.8$ & 0.4&0&225&& 0.158& 0.124& 0.084& 0.074& 0.068
		\\ 
		&&&400& &0.168& 0.120& 0.092& 0.078& 0.060\\  
		$0.8$ & 0.4&3&225&&0.556& 0.496& 0.458& 0.434& 0.426\\
		&&&400& &0.572 &0.542& 0.494& 0.474& 0.460 \\			
		$0.8$ & 0.4&5&225& &0.928 &0.920& 0.906& 0.888& 0.874 \\
		&&&400& &0.952& 0.946& 0.928& 0.928& 0.918 \\
		\hline
	\end{tabular}
	\vspace{0.3cm}
	\caption{\small Rejection proportions of the null hypothesis for $\alpha=0.05$. Regression function (\ref{alt_reg}).}
	\label{m3eb}
\end{table}

\begin{table}[H]
	\scriptsize
	\centering
	\begin{tabular}{ccccccccccc}
		&&	&&&&	& &$\mathbf{H}$&&  \\

	\cline{6-11}
	
	\vspace{0.2cm}	
	$\sigma$	&$a_e$&$c$&$n$&&$\begin{pmatrix}  0.8 & 0 \\ 0 &  0.6 \end{pmatrix}$&$\begin{pmatrix}  1 & 0 \\ 0 &  0.6 \end{pmatrix}$&$\begin{pmatrix}  0.6 & 0 \\ 0 &  0.8 \end{pmatrix}$&$\begin{pmatrix}  1 & 0 \\ 0 &  0.8 \end{pmatrix}$&$\begin{pmatrix}  0.6 & 0 \\ 0 &  1 \end{pmatrix}$&$\begin{pmatrix}  0.8 & 0 \\ 0 &  1 \end{pmatrix}$    \\		
		\hline
		$0.4$ &0.1&0&225&& 0.050& 0.038& 0.048& 0.036& 0.036& 0.034\\
		&&&400&&  0.038& 0.032& 0.034& 0.026& 0.030& 0.024\\  
		$0.4$ &0.1&3&225&& 0.328& 0.272& 0.520& 0.354& 0.538& 0.422\\
		&&&400& &0.238& 0.196& 0.462& 0.278& 0.484& 0.364  \\			
		$0.4$ &0.1&5&225&&0.996 &0.988& 1.000& 0.998& 1.000& 0.998 \\
		&&&400&&0.992 &0.986& 1.000& 0.990& 1.000& 0.998 \\
		\hline
		$0.4$ &0.2&0&225&& 0.062 &0.042& 0.064& 0.032& 0.054& 0.038 \\
		&&&400&&0.072& 0.048& 0.064& 0.030& 0.030& 0.020 \\  
		$0.4$ &0.2&3&225&&0.770& 0.706& 0.874& 0.774& 0.876& 0.824 \\
		&&&400& &0.798 &0.728& 0.904& 0.796& 0.912& 0.848 \\			
		$0.4$ &0.2&5&225&&1.000& 1.000& 1.000& 1.000& 1.000& 1.000 \\
		&&&400&&1.000& 1.000& 1.000& 1.000& 1.000& 1.000  \\
		\hline
		$0.4$ &0.4&0&225&&0.118 &0.104& 0.110& 0.080& 0.084& 0.072 \\
		&&&400&& 0.136& 0.114& 0.136& 0.084& 0.110& 0.072 \\  
		$0.4$ &0.4&3&225&&0.970 &0.952& 0.982& 0.966& 0.984& 0.972  \\
		&&&400&&0.986 &0.974 &0.994& 0.984& 0.996& 0.992  \\			
		$0.4$ &0.4&5&225&&1.000& 1.000& 1.000& 1.000& 1.000& 1.000 \\
		&&&400&& 1.000& 1.000& 1.000& 1.000& 1.000& 1.000 \\
		\hline	
		$0.6$ &0.1&0&225&&0.048& 0.042& 0.046& 0.036& 0.042& 0.034\\
		&&&400&& 0.038 &0.032& 0.034& 0.026& 0.030& 0.022\\  
		$0.6$ &0.1&3&225& &0.052& 0.044& 0.110& 0.054& 0.124& 0.072 \\
		&&&400& &0.026& 0.014& 0.096& 0.026& 0.108& 0.040 \\			
		$0.6$ &0.1&5&225&& 0.496 &0.442& 0.684& 0.530& 0.708& 0.612  \\
		&&&400&& 0.452&0.378& 0.680& 0.470& 0.700& 0.562\\
		\hline
		$0.6$ &0.2&0&225&& 0.062& 0.042& 0.066& 0.032& 0.054& 0.038   \\
		&&&400& &0.070&0.046& 0.064& 0.028& 0.030& 0.016  \\  
		$0.6$ &0.2&3&225&&0.278 &0.214& 0.456& 0.258& 0.466& 0.332 \\
		&&&400& &0.308& 0.240& 0.486& 0.278& 0.486& 0.356 \\			
		$0.6$ &0.2&5&225&&0.878& 0.846& 0.946& 0.892& 0.948& 0.916 \\
		&&&400& &0.900 &0.856& 0.954& 0.900& 0.954& 0.934\\
		\hline
		$0.6$ &0.4&0&225&&0.116& 0.102& 0.108& 0.080& 0.084& 0.072 \\
		&&&400&&  0.134& 0.110& 0.132& 0.078& 0.106& 0.070  \\  
		$0.6$ &0.4&3&225& &0.664 &0.606& 0.806& 0.654& 0.798& 0.732\\
		&&&400&& 0.706& 0.646& 0.828& 0.672& 0.826 &0.750  \\			
		$0.6$ &0.4&5&225& &0.994& 0.988& 0.998& 0.994& 0.998& 0.998\\
		&&&400&& 0.996& 0.996& 1.000& 0.996& 1.000& 1.000 \\
		\hline
		$0.8$ & 0.1&0&225&&  0.050 &0.042& 0.048& 0.036& 0.042& 0.036  \\ 
		&&&400& & 0.038& 0.032& 0.034& 0.026& 0.030& 0.022\\  
		$0.8$ & 0.1&3&225&&  0.010 &0.010& 0.078& 0.012& 0.080& 0.018  \\
		&&&400&& 0.008& 0.006& 0.050& 0.008& 0.054& 0.018 \\			
		$0.8$ & 0.1&5&225&& 0.134 &0.118 &0.262& 0.158& 0.274& 0.200\\
		&&&400&& 0.066 &0.054& 0.222& 0.086& 0.232& 0.138\\
		\hline
		$0.8$ & 0.2&0&225&&0.062& 0.040& 0.064& 0.032& 0.054& 0.040 \\ 
		&&&400&& 0.068 &0.042& 0.064& 0.028& 0.030& 0.016\\  
		$0.8$ & 0.2&3&225&&  0.132 &0.096& 0.262& 0.108& 0.252& 0.148  \\
		&&&400&& 0.168&0.110& 0.302& 0.124& 0.282& 0.182 \\			
		$0.8$ & 0.2&5&225&& 0.520 &0.456& 0.684& 0.518& 0.684& 0.610\\
		&&&400&&0.546 &0.450& 0.720& 0.518 &0.728 &0.612 \\
		\hline
		$0.8$ & 0.4&0&225&& 0.114 &0.102& 0.110& 0.080& 0.084& 0.074
		\\ 
		&&&400& &0.134&0.112& 0.130& 0.080& 0.106& 0.072\\  
		$0.8$ & 0.4&3&225&&0.432 &0.354& 0.596& 0.406& 0.570& 0.464\\
		&&&400& &0.492 &0.418& 0.590& 0.440& 0.588& 0.516 \\			
		$0.8$ & 0.4&5&225&& 0.884 &0.834& 0.934& 0.870& 0.932& 0.908
		\\
		&&&400& &0.918 &0.872& 0.958& 0.904& 0.956& 0.930 \\
		\hline
	\end{tabular}
	\vspace{0.3cm}
	\caption{\small Rejection proportions of the null hypothesis for $\alpha=0.05$. Regression function (\ref{alt_reg}).}
	\label{m3db}
\end{table}

\subsection*{B.4. Random design}
The methodology is now illustrated with covariate variables generated from a random design. As in Section 4, the regression function 
$
m(X_1,X_2)=2+X_1+X_2+cX_1^3
$
is considered. In this case, for each value of $c$ (being $c$ equal to 0 or 5), 500 samples of sizes  $n=225$ and $400$ are uniformly sampled in the unit square. The 
random errors $\varepsilon_i$ are normally distributed with  zero mean and isotropic exponential covariance function (\ref{expo}),
with $\sigma=0.4$, $0.8$, and $a_e=0.1$, $0.4$.  No nugget effect is considered. Table \ref{m3} shows the rejection proportions of the null hypothesis, for $\alpha=0.05$, considering that the bandwidth is taken as a diagonal matrix $\mathbf{H}=\text{diag}(h,h)$, and different  values of $h$ are chosen, $h=0.6, 0.7, 0.8, 0.9, 1.$ Similar conclusions as in the case of considering a fixed design can be deduced. 

\begin{table}[thb]
	\scriptsize
	\centering
	\begin{tabular}{cccccccccc}
		&&	&&&&	 &$h$&&  \\
		\cline{6-10}
		$\sigma$	&$a_e$&$c$&$n$&&0.6&$0.7$&0.8&0.9&1    \\
		\hline
		$0.4$ &0.1&0&225&&0.066& 0.056& 0.036& 0.028& 0.022\\
		&&&400&& 0.080 &0.068& 0.058& 0.048& 0.042\\  
		$0.4$ &0.1&5&225&& 1.000 &1.000&1.000&1.000 &1.000\\
		&&&400&& 1.000 &1.000&1.000&1.000 &1.000 \\
		\hline
		$0.4$ &0.4&0&225&&0.144& 0.100& 0.082& 0.060& 0.052\\
		&&&400&& 0.146 &0.118& 0.086& 0.068& 0.056\\  
		$0.4$ &0.4&5&225&&  1.000 &1.000&1.000&1.000 &1.000 \\
		&&&400&&  1.000 &1.000&1.000&1.000 &1.000\\
		\hline
		$0.8$ & 0.1&0&225&& 0.072 &0.056& 0.036 &0.030& 0.024 \\ 
		&&&400& & 0.080 &0.068& 0.058& 0.048& 0.044\\  
		$0.8$ & 0.1&5&225&& 0.916& 0.890& 0.870& 0.860& 0.858\\
		&&&400&& 0.954& 0.946& 0.944& 0.944& 0.948\\
		\hline
		$0.8$ & 0.4&0&225&& 0.142&0.112&0.100&0.090&0.076
		\\ 
		&&&400& &0.160 &0.122&0.086&0.062&0.054\\  
		$0.8$ & 0.4&5&225&&1.000 &1.000&1.000&1.000 &1.000\\
		&&&400& &   1.000 &1.000&1.000&1.000 &1.000\\
		\hline
	\end{tabular}
	\vspace{0.3cm}
	\caption{\small Rejection proportions of the null hypothesis for $\alpha=0.05$. Random design.}
	\label{m3}
\end{table}

\subsection*{B.5. Nugget effect}
Finally, a nugget effect is included in the dependence model.  Recall that in the previous frameworks the nugget effect was zero. In this case, the model considered is similar to the one of Section 4: the regression function is the same, $m(X_1,X_2)=2+X_1+X_2+cX_1^3$ (the data are generated on a bidimensional regular grid in the unit square, and $c$ is considered equal to 0 or 5). However, a nugget effect is included in the dependence structure. Then, the 
random errors $\varepsilon_i$ are normally distributed with  zero mean and isotropic exponential covariance function:  
$
\mbox{Cov}({\varepsilon}_i,{\varepsilon}_j)=   
c_e\{\exp(-\norm{\mathbf{X}_i-\mathbf{X}_j}/a_e)\},
$ if $\norm{\mathbf{X}_i-\mathbf{X}_j}\neq 0$, where $c_e=\sigma^2-c_0$ is the partial sill, with $\sigma=0.4$ and nugget effect $c_0$ being  $20\%$ and  $50\%$ of the total variance $\sigma^2$.  Two values for the practical range are considered, $a_e=0.1$ and $0.4$.  Table \ref{m4} shows the rejection proportions of the null hypothesis, for $\alpha=0.05$, considering that the bandwidth is taken as a diagonal matrix $\mathbf{H}=\text{diag}(h,h)$, and different  values of $h$ are chosen, $h=0.6, 0.7, 0.8, 0.9, 1.$ It can be observed that the performance of the test is satisfactory, with similar results to those in the previous scenarios. As the nugget is larger, the bandwidth value should be smaller.

\begin{table}[thb]
	\scriptsize
	\centering
	\begin{tabular}{ccccccccccc}
		&&&	&&&&	 &$h$&&  \\
		\cline{7-11}
		$c_0$&	$\sigma$	&$a_e$&$c$&$n$&&0.6&$0.7$&0.8&0.9&1    \\
		\hline
		&	$0.4$ &0.1&0&225&&0.078 &0.060& 0.042& 0.030& 0.026
		\\
		&	&&&400&&  0.052& 0.038& 0.028& 0.016& 0.010\\  
		$20\%$ &	$0.4$ &0.1&5&225&& 1.000 &1.000&1.000&1.000 &1.000 \\
		&	&&&400&& 1.000 &1.000&1.000&1.000 &1.000\\
		\hline
		&	$0.4$ &0.1&0&225&& 0.074 &0.056& 0.030& 0.020& 0.020\\
		&	&&&400&& 0.028 &0.016& 0.014& 0.012 &0.012 \\  
	$50\%$		&	$0.4$ &0.1&5&225&& 1.000 &1.000&1.000&1.000 &1.000\\
		&	&&&400&&  1.000 &1.000&1.000&1.000 &1.000\\
		\hline
		&	$0.4$ & 0.4&0&225&& 0.052 &0.048& 0.036& 0.036& 0.030\\ 
		&	&&&400& & 0.044 &0.040& 0.032& 0.026& 0.020\\  
		$20\%$&	$0.4$ & 0.4&5&225&& 1.000 &1.000&1.000&1.000 &1.000\\
		&	&&&400&&1.000 &1.000&1.000&1.000 &1.000\\
		\hline
		&	$0.4$ & 0.4&0&225&& 0.062& 0.050& 0.044& 0.038& 0.026
		\\ 
		&	&&&400& & 0.024 &0.024 &0.020 &0.020& 0.014\\  
		$50\%$	&	$0.4$ & 0.4&5&225& &1.000 &1.000&1.000&1.000 &1.000\\
		&	&&&400& &1.000 &1.000&1.000&1.000 &1.000\\
		\hline
	\end{tabular}
	\vspace{0.3cm}
	\caption{\small Rejection proportions of the null hypothesis for $\alpha=0.05$. Nugget effect.}
	\label{m4}
\end{table}

\end{document}